\documentclass[a4paper,12pt,onecolumn,article]{memoir}

\usepackage{times}
\usepackage{graphicx}
\usepackage{bm}
\usepackage{amsmath}
\usepackage{amssymb,latexsym,pifont}
\usepackage{chngcntr}
\usepackage{pdfpages}
\usepackage{epstopdf}
\usepackage{color,calc,soul}
\usepackage{subfig}
\usepackage{upgreek}
\usepackage{listings}
\usepackage{memhfixc}
\usepackage{bibentry}
\usepackage[numbers]{natbib}

\nobibliography*

\definecolor{gp}{RGB}{52,165,51}
\definecolor{rp}{RGB}{192,32,32}

\newcommand{\be}{\begin{equation}}
\newcommand{\ee}{\end{equation}}
\newcommand{\e}{\mathrm{e}}
\newcommand{\im}{\mathrm{i}}
\newcommand{\md}{\mathrm{d}}
\newcommand{\nn}{\nonumber}
\newcommand{\eF}{\epsilon_\mathrm{_F}}
\newcommand{\Fref}[1]{Fig.~\ref{#1}}
\newcommand{\Eqref}[1]{Eq.~(\ref{#1})}

\newcommand\ddfrac[2]{\frac{\displaystyle #1}{\displaystyle #2}}
\newcommand{\oone}{\text{\small 1} \!\! 1}

\newcommand{\REM}[1]{}

\usepackage{soul}
\sethlcolor{yellow}
\newcommand{\hlc}[2][yellow]{{%
                \colorlet{foo}{#1}%
                \sethlcolor{foo}\hl{#2}}%
}

\usepackage[%
	unicode=true
	,colorlinks=true
	,linkcolor=blue
	,anchorcolor=blue
	,breaklinks=true
	,citecolor=blue
	,filecolor=cyan
	,menucolor=blue
	,breaklinks=true
	,urlcolor=blue
			]{hyperref}

\definecolor{nicered}{rgb}{.647,.129,.149}
\makeatletter
\newlength\dlf@normtxtw
\newsavebox{\feline@chapter}
\newcommand\feline@chapter@marker[1][4cm]{%
  \sbox\feline@chapter{%
    \resizebox{!}{#1}{\fboxsep=1pt%
      \colorbox{nicered}{\color{white}\bfseries\sffamily\thechapter}%
  }}%
  \rotatebox{90}{%
    \resizebox{%
      \heightof{\usebox{\feline@chapter}}+\depthof{\usebox{\feline@chapter}}}%
	      {!}{\scshape\so\@chapapp}}\quad%
  \raisebox{\depthof{\usebox{\feline@chapter}}}{\usebox{\feline@chapter}}%
}
\newcommand\feline@chm[1][4cm]{%
  \sbox\feline@chapter{\feline@chapter@marker[#1]}%
  \makebox[0pt][l]{
    \makebox[1cm][r]{\usebox\feline@chapter}%
}}
\makechapterstyle{daleif1}{

  \renewcommand\printchapternum{\null\hfill\feline@chm[2.5cm]\par}

}
\makeatother
\chapterstyle{daleif1}

\begin{document}

\frontmatter

\title{
{\Large\textbf{The Influence of \textit{s-d} Exchange Interaction \\
on the Gap Anisotropy  and Anisotropy of the Lifetime of 
Normal State Charge Carriers of Layered Cuprates}}
}

\author{Todor M. Mishonov\thanks{mishonov@gmail.com},
Nedeltcho I. Zahariev\thanks{zahariev@issp.bas.bg},
Albert M. Varonov\thanks{varonov@issp.bas.bg}
\\ \\
\textit{Georgi Nadjakov Institute of Solid State Physics, Bulgarian Academy of Sciences}\\
\textit{72 Tzarigradsko Chaussee Blvd., BG-1784 Sofia, Bulgaria}
}



\date{{\normalsize 2 December 2022}}

\maketitle

\begin{abstract}
The anisotropy of the electron scattering rate and lifetime $\Gamma_\mathbf{p}=1/\tau_\mathbf{p}$
observed by Angle Resolved Photoemission Spectroscopy (ARPES) 
is evaluated using \textit{s-d} Kondo-Zener exchange Hamiltonian used previously to describe superconducting properties of high-$T_c$ cuprates;
for correlation between critical temperature $T_c$ and BCS coupling constant, for example.
The performed qualitative analysis reveals that
``cold spots'' correspond to nodal regions of the superconducting phase 
where the superconducting gap is zero, because the exchange interaction is annulled.
Vice versa, ``hot spots'' and intensive scattering in the normal state 
correspond to the regions with maximal gap in the superconducting phase.
We have obtained that separable kernel postulated in the Fermi liquid approach to the
normal phase is the same kernel which is exactly calculated 
in the framework of the \textit{s-d} approach in the 
Linear Combination of Atomic Orbital (LCAO) approximation for CuO$_2$ plane.
In this sense, at least on the qualitative level, the superconducting 
cuprates are described by one and the same Hamiltonian applied to their 
superconducting and normal properties.
For the overdopedcuprates having conventional Fermi liquid behavior
we derived the corresponding Stoss-integral for the \textit{s-d} interaction
in the framework of LCAO approximation which is the central result of the present study.
\textcolor{black}{
Repeating the thermodynamics we observed that for optimally doped cuprates
$2\Delta_\mathrm{max}/T_c$ is in good agreement with Pokrovsky theory for
anisotropic gap superconductors.}
\end{abstract}


%
%
%
%
%
%
%
%
%
%
%
%
%
%
%
%
%
%


\thispagestyle{empty}

\maxsecnumdepth{subsubsection}
\setsecnumdepth{subsubsection}
\settocdepth{subsection}

\tableofcontents


\mainmatter
\setsecnumdepth{none}
\maxsecnumdepth{none}
\maxsecnumdepth{subsubsection}
\setsecnumdepth{subsubsection}
\chapter{Introduction}

It took almost 50 years between the experimental discovery of superconductivity to
its theoretical explanation.
Nowadays history repeats itself with high-temperature superconductivity;
it's been more than 35 years since its discovery 
with no convincing theo retical candidate in sight
despite that qualitative understanding has been already achieved~\cite{Kivelson:15}.
One of the first proposed mechanism for cuprates was the extended Hubbard model between
Cu3$d$ and O2$p$ states in the CuO$_2$ plane~\cite{Emery:87,Zhang:88}, 
but it is still an open question whether the Hubbard model can quantitatively describe high temperature superconductivity~\cite{Arovas:22}.
Currently the $t$-$J$~\cite{Spalek:07} and $t$-$J$-$U$~\cite{Spalek:17} models are on the agenda of the theoretical studies~\cite{Spalek:22}
both of them descendants of the Hubbard model.

In this work the  Shubin-Kondo-Zener \textit{s-d} exchange interaction~\cite{Shubin,Zener1,Kondo}
with the Linear Combination of Atomic Orbitals (LCAO) Hamiltonian
in the CuO$_2$ plane of optimally and overdoped cuprates
is studied.
R\"ohler for the first time noted that 
the Cu$4s$-3$d_{x^2-y^2}$ hybridization seems to be the crucial quantum chemical parameter controlling related electronic degrees of freedom~\cite{Roehler:00,Roehler:00a}.
The obtained results with \textit{s-d} exchange interaction described here
convincingly support this observation
as they qualitatively agree with contemporary calculations and experiments.

For optimally doped and overdoped cuprates LCAO approximation for the electron bands~\cite{MishonovPenev:11} agrees with Local Density Approximation (LDA) band 
calculations~\cite{Andersen:95,Andersen:96}.
A remarkable property of the high-temperature superconducting hole doped cuprates
is the correlation between the critical temperature $T_c$ and the Cu$4s$ energy level
discovered with LDA band calculations~\cite{Pavarini:01}.
One of the results from this study qualitatively agrees with this correlation using one and the same calculated \textit{s-d} exchange interaction.

This interaction is also used to calculate
the anisotropy of the electron scattering rate and lifetime
which again qualitatively agrees with the one
observed by Angle Resolved Photoemission Spectroscopy (ARPES) experiments.
This experimental technique has been widely used to study cuprates
~\cite{Shen:95,Randeria:97,Feng:01,Takeuchi:01,Armitage:02,Feng:02,Damascelli:03,
Kaminski:05,Inosov:07,Vishik:10,He:11,Hank:17,Yu:20,Yu:21,Shen:21,Zonno:21}
and for the observed scattering rate anisotropy
the phenomenology of ``hot spots'' by
Hlubina and Rice~\cite{Hlubina:95} and
``cold spots'' by Ioffe and Millis~\cite{Ioffe:98} on the Fermi surface was introduced.
The regions with strong scattering and short lifetime are coined hot spots
and the cold spots are the regions with the longest lifetime or the weakest scattering.
This hot/cold spot phenomenology is applicable roughly speaking when cuprates demonstrate Fermi liquid behavior for optimally and overdoped regime.

Last but not least, 
for a ferromagnetic sign of the \textit{s-d} exchange interaction
a propagation of zero sound is predicted.
Already for half a century different kinds of zero sound
have been theoretically studied extensively and
this topic continues to attract a great deal of interest within the scientific community.
For instance,
recent studies include  two-dimensional zero-sound~\cite{Khoo:19,Gochan:20}
and shear~\cite{Marel:21} zero sound for $p$-type interaction~\cite{Ding:19}.
We can finally conclude that except for $^3$He thin films,
two dimensional structures with large exchange interaction with ferromagnetic sign
will soon become an interesting object for the realization of the old idea of Landau~\cite{Landau_theory,Landau_zero_sound}.

As it is well known the basic superconducting state for overdoped cuprates is 
well described by a simple mean-field theory with a $d$-wave superconducting gap~\cite{Lee:16,Kivelson:21}.
The purpose of the present study is to describe the anysotropy of the 
lifetime of the normal charge carriers using
the \textit{s-d} exchange interaction which naturally explains gap 
anisotropy of the superconducting phase.
The use of one and the same Hamiltonian, for both the normal and superconducting phase, is only the first step on the right track.

To summarize, using the LCAO method with the \textit{s-d} exchange interaction
two properties of the high-$T_c$ cuprates are described:
1) the correlation of the Cu4$s$ energy level and $T_c$ for the superconducting regime and
2) the anisotropy of the lifetime of charge carriers on the Fermi surface in the normal phase measured by ARPES experiments.
In addition, the microscopic Hamiltonian leads to the prediction of zero sound
propagation  along the diagonals (``through the cold spots''),
but only for ferromagnetic sign of the \textit{s-d} exchange interaction,
i.e. for non-superconducting cuprates.


After this extended review of the Hamiltonian used to explain the superconducting properties, 
in Sec.~\ref{Fermi liquid reduction} we perform
Fermi liquid reduction of the exchange \textit{s-d} Hamiltonian
and suggest a possible explanation of the 
phenomenology of ``hot'' and ``cold'' spots used to 
describe the normal properties of high-$T_c$ cuprates.
A short version of this study is given in Ref.~\cite{hotspot2022}
For a lateral illustration of the Fermi liquid theory we 
analyze in Sec.~\ref{Zero sound for ferromagnetic sign of s-d exchange interaction}
the imaginary case of a layered perovskite which is not superconducting, but has
ferromagnetic sign of the exchange amplitude $J_{sd}$.
For such perovskites we predict propagation of zero sound;
a short version of this study can be found in Ref.~\cite{sound2022}.
In Sec.~\ref{Fermi liquid behavior of overdoped cuprates}
using the same Hamiltonian and notions we analyze
Fermi liquid behavior of overdoped cuprates where Ohmic resistance is quadratic with temperature $\varrho_{_\Omega}=A_\Omega T^2$.
We calculate the coefficient $A_\Omega$ which is determined by \textit{s-d}
exchange interaction.
The explicite formula for the scattering rate \Eqref{centrum} 
in the Landau Fermi liquid theory for the Stoss-integral is the central result of
the present study. We reproduced here the results of our short 
communications \cite{hotspot2022} and \cite{sound2022}
in order to represent a complete and coherent picture of the normal and supeconducting 
properties of high-$T_c$ cuprates and to introduce a 
common system of notions and notations.

The main qualitative conclusion of the work is that 
the phenomenology of the normal properties can be derived 
from the \textit{s-d} Hamiltonian used tho describe the superconducting properties.
In the discussion and conclusion Chap.~\ref{Discussion and conclusions}
we analyze: 
A) the motivation of the phenomenology,
B) what compromises are necessary to build a coherent picture
and 
C) we try to mention some seminal papers which in our opinion are important 
to create a complete mosaic.
For a general review of the physics of cuprates we recommend the monograph 
by Plakida~\cite{Plakida:10}.
For the physics of metal-insulator transitions in transitional-metal oxides
see the classical monograph by Mott~\cite[Chap.~6]{Mott:90}.
However, we wish to emphasise that our attention is concentrated on the overdoped cuprates with doping level far from the Mott transition.

\section{Basic notions of the elementary kinetics}
\label{Basic notions of the elementary kinetics}

\subsection{Transport cross-section of two dimensional coulomb scattering}
Let us consider scattering by two dimensional Coulomb potential in a text-book style
\be
U(r)=\frac{Ze^2}{r},\quad r=|\mathbf{r}|=\sqrt{x^2+y^2},\quad
e^2\equiv\frac{q_e^2}{4\pi\varepsilon_0}.
\ee
Our first step is to calculate the matrix elements between normalized plane waves
\begin{align}
& \psi_\mathrm{i}({\mathbf{r}})
=\frac{1}{\sqrt{S}}\mathrm{e}
^{\im \mathbf{P}_\mathrm{i}\cdot\mathbf{r}/\hbar}, \qquad
 \psi_\mathrm{f}({\mathbf{r}})
=\frac{1}{\sqrt{S}}\mathrm{e}
^{\im \mathbf{P}_\mathrm{f}\cdot\mathbf{r}/\hbar},\\
& P=P_\mathrm{i}=P_\mathrm{f},\qquad
\mathbf{P}_\mathrm{f}=\mathbf{P}_\mathrm{i}+\hbar\mathbf{K},\qquad
\hbar K=2P\sin(\theta/2), \nn
\end{align}
where $\theta$ is the angle between the initial $p_\mathrm{i}$ and
final $p_\mathrm{f}$ momentum.
For the distances $L_x$ and $L_y$ we suppose periodic boundary conditions 
and $S=L_xL_y$.
Using the well-known integral
\be
\int_0^{2\pi}
\frac{\md \varphi}{a+b\cos(\varphi)}
=\frac{2\pi}{\sqrt{a^2-b^2}},
\ee
after some regularization and analytical continuation
for the Fourier transform we obtain
\be
\left(\frac1{r}\right)_\mathbf{\!K}
=\int \frac{1}{r}\, \mathrm{e}^{-\im \mathbf{K}\cdot\mathbf{r}}\,
\md x\,\md y=\frac{2\pi}{K},
\ee
and for the matrix elements between the initial and final states we have
\be
U_\mathrm{f,i}=\int \psi_\mathrm{f}^*({\mathbf{r}})U(r)
\psi_\mathrm{i}({\mathbf{r}})\,\md x\md y
=\frac{2\pi Ze^2}{2(P/\hbar)\sin(\theta/2)S}.
\ee
Then for the density of final states per unit angle for free particles
$P_E \equiv \sqrt{2mE}$ we have
\be
 \rho_f (E)
=\frac{1}{2\pi}\sum_{\mathbf{P}}
\delta(E-E_\mathbf{P}) 
 =\frac{1}{2\pi}\frac{S}{(2\pi\hbar)^2}
\int\limits_0^\infty
\delta\left(\frac{P^2}{2m}-\frac{P_E^2}{2m}\right)\mathrm{d}(\pi P^2)
=\frac{mS}{(2\pi\hbar)^2}.
\ee
And for the flux of the probability of incoming electron we have the product of the velocity
$v$ and the density of the probability $1/S$ of a plane wave
\be
j_\mathrm{i}=\frac{V_\mathrm{i}}{S}, \qquad V_\mathrm{i}=\frac{P}{m}.
\ee

According to the second Fermi golden rule for the cross-section 
with dimension length in 2D we derive
\be
\sigma(\theta)
=\frac{2\pi}{\hbar}\left\vert U_\mathrm{f,i}\right\vert^2\frac{\rho_f}{j_\mathrm{i}}
=\frac{\pi}{4\hbar} \frac{(Ze^2)^2}{VE \sin^2(\theta/2)}, \;\;\;
E=\frac{P^2}{2m}
\label{differential}
\ee
and using 
$\sin^2(\theta/2)=\frac12\left(1-\cos(\theta)\right)$ 
one can easily calculate the transport section
\be
\sigma_\mathrm{tr}
=\int_0^\pi \sigma(\theta)\left(1-\cos\theta\right)\mathrm{d}\theta
=\frac{\pi^2}{2\hbar} \frac{(Ze^2)^2}{VE}.
\label{transport}
\ee
For the applicability of the Born approximation the effective charge $|Z|\ll1.$
In the next subsection we incorporate this cross-section in the formula for the
temperature dependence of the resistivity.

\subsection{Linear temperature dependence of the in-plane resistivity}
\label{Plane_capacitor_model}

The mean free path $l$, 
impurity concentration $n_\mathrm{imp}$ 
and transport section $\sigma_\mathrm{tr}$
are involved in the well-known relation
\be
ln_\mathrm{imp}\sigma_\mathrm{tr}=1
\ee
which determines the electrical conductivity in the Drude formula which we apply to the
2D case
\be
\label{Drude}
\frac{1}{\varrho}=\sigma_\mathrm{Drude}=\frac{n_eq_e^2\tau}{m},
\qquad \tau=\frac{l}{V}, \qquad
\Gamma_\mathrm{\! C}\equiv\frac1{\tau}=n_\mathrm{imp}\sigma_\mathrm{tr}V
=\frac{\pi^2(Ze^2)^2\,n_\mathrm{imp}}{2\hbar E},
\ee
where $\Gamma_\mathrm{\! C}$ is the Columb scattering rate, 
$\tau$ is the mean free time, and $\varrho$ is the resistivity
of the 2D conductor with dimension $\Omega$ in SI units.
For a general introduction of kinetics of metals, see 
Refs.~\cite{LL10,AbrGorDzya,Abrikosov,LifAzbKag}.

High-$T_c$ cuprates are layered materials, but in order to evaluate the contribution
of the classical fluctuation of the electric field between conducting 2D layers
in Ref.~\cite{Mishonov:00} was analyzed a plane capacitor model for a (CuO$_2$)$_2$
bi-layer.
Imagine that a 2D plane is divided in small squares (plaquettes)
with a side equal to the Cu-Cu
distance, the in-plane lattice constant $a_0$
and the distance between the planes (double or single) is $d_0$.
The capacity of the considered small capacitor
\be
C=\varepsilon_0\frac{a_0^2}{d_0}.
\ee
For the square of the fluctuation charge $Q=Zq_e$ of this plaquette 
the equipartition theorem~\cite{Herapath:21,Waterston:51,Waterston:92}
with temperature in energy units 
\be
\frac{\left<Q^2\right>_T}{2C}=\frac{T}{2}
\label{PlaneCapacitor}
\ee
gives
\be
(Zq_e)^2=\left<Q^2\right>=CT=\varepsilon_0\frac{a_0^2}{d_0}T,
\ee
where for brevity from now on we omit the brackets $\left<\,\right>_T$
 denoting thermal averaging.
The used here and in Ref.~\cite{Mishonov:00}
is actually the Nyquist theorem~\cite[Eq.~(78.3)]{LL9}
\begin{align}
(\mathcal{E}^2)_\omega=2\hbar\omega R(\omega)/\tanh(\hbar\omega/2T),
\end{align}
where $(\mathcal{E}^2)_\omega$ is the spectral density of the voltage
$\mathcal{E}=E_zc_0$ between CuO$_2$ planes with distance $c_0$,
$E_z$ is thermally fluctuating electric field between conducting layers,
$R(\omega)=c_0/a_0^2\sigma_z(\omega)$ is the resistance 
between two plaquettes with area $a_0^2$, and $\sigma_z(\omega)$
is the conductivity of the layered cuprate in the dielectric direction.
As it was recently proved, the general Callen-Welton fluctuation-dissipation theorem
can be considered as a consequence of the Nyquist theorem~\cite{Reggiani:19,FDT,Reggiani:20}.

The calculated in such a way averaged square of the fluctuating charge $Z^2=Q^2/q_e^2$
has to be substituted in the differential Eq.~(\ref{differential}) or transport
Eq.~(\ref{transport}) cross-section.
Additionally, for the area density of the ``impurities'' we have to substitute 
in the mean free path the density of the plaquettes $n_\mathrm{imp}=1/a_0^2$.
At these conditions the Drude formula Eq.~(\ref{Drude}) gives 
for 2D resistivity per square of CuO$_2$ plane
\be
4\pi\epsilon_0\varrho=\frac{m^2T}{8\hbar^3n_e^2d_0}
\ee
and the 2D dimensional conductivity $\sigma_\mathrm{Drude}/4\pi\epsilon_0$
has dimension velocity. In Gaussian system $4\pi\epsilon_0=1$, 
but all equations in the present paper are system invariant.
For a bulk material where separate bi-layers are at distance $c_0$,
the 3D resistivity parallel to the conducting planes $\rho_{ab}$
can be evaluated as
\be
4\pi\epsilon_0\rho_{ab}=\frac{m^2c_0}{8\hbar^3n_e^2d_0}T.
\ee
In short, the linear behavior of the resistivity reveals that in layered materials 
thermal fluctuations of electric field determine the density fluctuations.
Electrons scatter on the fluctuation of their own density
which in some sense is a self-consistent procedure.
A slightly different realization of the same idea is described in 
Ref.~\cite[Chap.~8]{MishonovPenev:11}.
Analogously, the wave scattering of the sunlight by the density fluctuations of the atmosphere determines the color of the sky;
who would be blind for the blue sky~\cite{Mishonov:00}?
In a maximal traditional interpretation,
resistivity of the layered high $T_c$ cuprates is simply Rayleigh scattering 
of Fermi quasiparticles on the electron density fluctuations 
in a layered metal.

However, our formula for the scattering rate $\Gamma=1/\tau$, Eq.~(\ref{Drude})
naturally explains an isotropic scattering which does not agree
with the spectroscopic data.
If we consider the energy in \Eqref{Drude} to be equal to the Fermi one $\eF$
the formula for the cross-section \Eqref{transport}
predicts negligible anisotropy if it is applied to the CuO$_2$ plane
while ARPES 
(Angle Resolved Photo-Emission Spectroscopy) data~\cite{Damascelli:03,Handbook,Shen:21}
reveals remarkable anisotropy of $\Gamma(\varphi)$
when we rotate on angle $\varphi$ around $(\pi,\,\pi)$-point 
i.e. the center of the hole pocket.

\textcolor{black}
{Analysis of the normal conductivity per square CuO$_2$ plane in some cuprates
gives that $\sigma/4\pi\varepsilon_0\gg v_\mathrm{Bohr}\equiv e^2/\hbar$.
This leads that in CuO$_2$ plane we have well-defined quasiparticles 
(in the sense of Landau) as it is confirmed by ARPES data in nodal direction.
That not only Fermi liquid but even Fermi gas approach do not reveals an
''enemy of working class'' in sense of Kivelson and Fradkin 
\cite[Chap.~15, page 583]{Handbook}.}

It is obvious that the Coulomb scattering is not the only mechanism 
for creation of the scattering rate $\Gamma$ and Ohmic resistivity.
The purpose of the present work is to take into account the 
\textit{s-d} exchange interaction which creates the pairing in the superconducting phase.

In the next chapter we recall the generic 4-band model for the  CuO$_2$ plane
and Shubin-Kondo-Zener exchange interaction applied to this ``standard model''.
\chapter{Basic electronic properties of C\lowercase{u}O$_2$ plane}
\label{Basic electronic properties of CuO$_2$ plane}

Ideas and notions of the Landau-Fermi liquid theory were widely used to analyze
the properties of normal high-$T_c$ cuprates~\cite{Plakida:10}.
See, for example, the early papers by Carrington et al.
\cite{Carrington:92}, Hlubina and Rice \cite{Hlubina:95},
Stojkovic and Pines \cite{Stojkovic:97},
and Ioffe and Millis \cite{Ioffe:98} and the recent review by Varma
\cite{Varma:20}.
The large anisotropy in the scattering rate 
along the Fermi surface of these materials was reported for first time by Shen and Schrieffer \cite{Shen:97}
and later by Valla \textit{et al.} \cite{Valla:00}.
The cornerstone of the Boltzmann equation analysis of carriers
kinetics is the
strong anisotropy of their charge lifetime $\tau_\mathbf{p}$
along the Fermi contour. 
The central concepts here are the ``hot spots'' where close to $(\pi,\,0)$
and $(0,\,\pi)$ regions of the Fermi contour the electron lifetime 
is unusually short and the
angle Resolved Photoemission Spectroscopy (ARPES)
spectral function is very broad~\cite{Shen:95,Randeria:97} 
suggesting strong scattering~\cite{Ioffe:98}.
For contemporary ARPES studies see Ref.~\cite{Shen:21}
and references therein.
Ioffe and Millis~\cite{Ioffe:98} however emphasized that the concept
of ``cold spots'' along the BZ where electron lifetime is significantly longer
and ARPES data reveal well defined quasiparticle peak,
suggesting relatively weak scattering,
which increases rapidly along the Fermi contour moving away 
from ``cold spots''.
Recent research on the physics of hot and cold spots can be found in
Refs.~\cite{He:11,Lee:16,Fink:19}.
This hot/cold spot phenomenology is applicable roughly speaking when cuprates demonstrate Fermi liquid behavior for optimally and overdoped regime.

\REM{
The angle Resolved Photoemission Spectroscopy (ARPES) technique has
been widely used to experimentally study cuprates
~\cite{Shen:95,Randeria:97,Feng:01,Takeuchi:01,Armitage:02,Feng:02,Damascelli:03,
Kaminski:05,Inosov:07,Vishik:10,He:11,Hank:17,Yu:20,Yu:21,Shen:21,Zonno:21}
and for the observed scattering rate anisotropy
the phenomenology of ``hot spots'' by
Hlubina and Rice~\cite{Hlubina:95} and
``cold spots'' by Ioffe and Millis~\cite{Ioffe:98} on the Fermi surface was introduced.
The regions with strong scattering and short lifetime are coined hot spots
and the cold spots are the regions with the longest lifetime or the weakest scattering.
This hot/cold spot phenomenology is applicable roughly speaking when cuprates demonstrate Fermi liquid behavior for optimally and overdoped regime.
}

For optimally doped and overdoped cuprates the Linear Combination of Atomic Orbitals (LCAO) approach for the electron bands~\cite{MishonovPenev:11} agrees with Local Density Approximation (LDA) band 
calculations~\cite{Andersen:95,Andersen:96}.
A remarkable property of the high-temperature superconducting hole doped cuprates
is the correlation between the critical temperature $T_c$ and the Cu$4s$ energy level
obtained within LDA band calculations~\cite{Pavarini:01}.


It took almost 50 years since the experimental discovery of
superconductivity till its theoretical explanation.
Nowadays history repeats itself with regard to high-temperature superconductivity;
it's been more than 35 years since its experimental realization
with no convincing theory in sight
in spite the fact that qualitative understanding has been already
achieved~\cite{Kivelson:15,Zhou:21}.
One of the first proposed mechanisms for cuprates was based on the
extended Hubbard model involving interation between
Cu3$d$ and O2$p$ states in the CuO$_2$ plane~\cite{Emery:87,Zhang:88}, 
however, it is still an open question whether the Hubbard model can quantitatively describe high temperature superconductivity~\cite{Arovas:22}.
For the time being, the $t$-$J$~\cite{Spalek:07} and
$t$-$J$-$U$~\cite{Spalek:17} models, both descendants of the Hubbard
model, are on the agenda of theoretical studies~\cite{Spalek:22}.

In this work, we explore the properties of the Shubin-Kondo-Zener \textit{s-d} exchange interaction~\cite{Shubin,Zener1,Kondo}
within LCAO Hamiltonian
in the CuO$_2$ plane of optimally and overdoped cuprates.
To this end, we take advantage of the observation that 
in Cu$4s$-3$d_{x^2-y^2}$ hybridization seems to be the crucial quantum chemical parameter controlling related electronic degrees of freedom~\cite{Roehler:00,Roehler:00a}.
Moreover, we describe the anisotropy of the 
lifetime of the normal charge carriers using
the \textit{s-d} exchange interaction, which naturally explains the gap 
anisotropy of the superconducting phase. The main advantage of this
our approach is the use of one and the same Hamiltonian to describe
both the normal and superconducting phase.

This work is organized as follows.
We review briefly the basic electronic properties of the CuO$_2$ plane
in Chap.~\ref{Basic electronic properties of CuO$_2$ plane} using
(i) the band structure in LCAO approximation,
(ii) the Shubin-Kondo-Zener \textit{s-d} exchange interaction,
(iii) BCS reduction of the exchange interaction and (iv) Pokrovsky
theory of anisotropic gap superconductors to evaluate $T_c$ of CuO$_2$
plane.
in Sec.~\ref{Fermi liquid reduction}, we perform
Fermi liquid reduction of the exchange \textit{s-d} Hamiltonian
and suggest a possible explanation of the 
phenomenology of ``hot'' and ``cold'' spots used to 
describe the normal properties of high-$T_c$ cuprates.
For an illustration of the Fermi liquid theory we 
analyze in Sec.~\ref{Zero sound for ferromagnetic sign of s-d exchange interaction}
the case study of a nonsuperconducting layered perovskite involving
a ferromagnetic $J_{sd}$.
For these perovskites, we unveil the propagation of zero sound.
In Sec.~\ref{Fermi liquid behavior of overdoped cuprates}
using the same Hamiltonian and notions, we analyze
Fermi liquid behavior of overdoped cuprates where the Ohmic resistance
is quadratic in temperature $\varrho_{_\Omega}=A_\Omega T^2$.
We calculate the coefficient $A_\Omega$ which is determined by \textit{s-d}
exchange interaction.
The explicit formula for the scattering rate 
in the Landau Fermi liquid theory for the Stoss-integral is the central result of
the present study.
The main qualitative conclusion of the present article is that 
the phenomenology of the normal properties can be derived 
from the \textit{s-d} Hamiltonian, see Chap.~\ref{Discussion and
conclusions}.

\REM{
\section{Basic notions of the elementary kinetics}
\label{Basic notions of the elementary kinetics}

\subsection{Transport cross-section of two dimensional coulomb scattering}
Let us consider scattering by two dimensional Coulomb potential in a text-book style
\be
U(r)=\frac{Ze^2}{r},\quad r=|\mathbf{r}|=\sqrt{x^2+y^2},\quad
e^2\equiv\frac{q_e^2}{4\pi\varepsilon_0}.
\ee
Our first step is to calculate the matrix elements between normalized plane waves
\begin{align}
& \psi_\mathrm{i}({\mathbf{r}})
=\frac{1}{\sqrt{S}}\mathrm{e}
^{\im \mathbf{P}_\mathrm{i}\cdot\mathbf{r}/\hbar}, \qquad
 \psi_\mathrm{f}({\mathbf{r}})
=\frac{1}{\sqrt{S}}\mathrm{e}
^{\im \mathbf{P}_\mathrm{f}\cdot\mathbf{r}/\hbar},\\
& P=P_\mathrm{i}=P_\mathrm{f},\qquad
\mathbf{P}_\mathrm{f}=\mathbf{P}_\mathrm{i}+\hbar\mathbf{K},\qquad
\hbar K=2P\sin(\theta/2), \nn
\end{align}
where $\theta$ is the angle between the initial $p_\mathrm{i}$ and
final $p_\mathrm{f}$ momentum.
For the distances $L_x$ and $L_y$ we suppose periodic boundary conditions 
and $S=L_xL_y$.
Using the well-known integral
\be
\int_0^{2\pi}
\frac{\md \varphi}{a+b\cos(\varphi)}
=\frac{2\pi}{\sqrt{a^2-b^2}},
\ee
after some regularization and analytical continuation
for the Fourier transform we obtain
\be
\left(\frac1{r}\right)_\mathbf{\!K}
=\int \frac{1}{r}\, \mathrm{e}^{-\im \mathbf{K}\cdot\mathbf{r}}\,
\md x\,\md y=\frac{2\pi}{K},
\ee
and for the matrix elements between the initial and final states we have
\be
U_\mathrm{f,i}=\int \psi_\mathrm{f}^*({\mathbf{r}})U(r)
\psi_\mathrm{i}({\mathbf{r}})\,\md x\md y
=\frac{2\pi Ze^2}{2(P/\hbar)\sin(\theta/2)S}.
\ee
Then for the density of final states per unit angle for free particles
$P_E \equiv \sqrt{2mE}$ we have
\begin{align}
 \rho_f (E) &
=\frac{1}{2\pi}\sum_{\mathbf{P}}
\delta(E-E_\mathbf{P}) \\
& =\frac{1}{2\pi}\frac{S}{(2\pi\hbar)^2}
\int\limits_0^\infty
\delta\left(\frac{P^2}{2m}-\frac{P_E^2}{2m}\right)\mathrm{d}(\pi P^2)
=\frac{mS}{(2\pi\hbar)^2}.\nn
\end{align}
And for the flux of the probability of incoming electron we have the product of the velocity
$v$ and the density of the probability $1/S$ of a plane wave
\be
j_\mathrm{i}=\frac{V_\mathrm{i}}{S}, \qquad V_\mathrm{i}=\frac{P}{m}.
\ee

According to the second Fermi golden rule for the cross-section 
with dimension length in 2D we derive
\be
\sigma(\theta)
=\frac{2\pi}{\hbar}\left\vert U_\mathrm{f,i}\right\vert^2\frac{\rho_f}{j_\mathrm{i}}
=\frac{\pi}{4\hbar} \frac{(Ze^2)^2}{VE \sin^2(\theta/2)}, \;\;\;
E=\frac{P^2}{2m}
\label{differential}
\ee
and using 
$\sin^2(\theta/2)=\frac12\left(1-\cos(\theta)\right)$ 
we easily calculate the transport section
\be
\sigma_\mathrm{tr}
=\int_0^\pi \sigma(\theta)\left(1-\cos\theta\right)\mathrm{d}\theta
=\frac{\pi^2}{2\hbar} \frac{(Ze^2)^2}{VE}.
\label{transport}
\ee
For the applicability of the Born approximation the effective charge $|Z|\ll1.$
In the next subsection we incorporate this cross-section in the formula for the
temperature dependence of the resistivity.

\subsection{Linear temperature dependence of the in-plane resistivity}
\label{Plane_capacitor_model}

The mean free path $l$, 
impurity concentration $n_\mathrm{imp}$ 
and transport section $\sigma_\mathrm{tr}$
are involved in the well-known relation
\be
ln_\mathrm{imp}\sigma_\mathrm{tr}=1
\ee
which determines the electrical conductivity in the Drude formula which we apply to the
2D case
\begin{align}
\label{Drude}
& \frac{1}{\varrho}=\sigma_\mathrm{Drude}=\frac{n_eq_e^2\tau}{m},
\qquad \tau=\frac{l}{V},\\
& \Gamma_\mathrm{\! C}\equiv\frac1{\tau}=n_\mathrm{imp}\sigma_\mathrm{tr}V
=\frac{\pi^2(Ze^2)^2\,n_\mathrm{imp}}{2\hbar E},
\nn
\end{align}
where $\Gamma_\mathrm{\! C}$ is the Columb scattering rate, 
$\tau$ is the mean free time, and $\varrho$ is the resistivity
of the 2D conductor with dimension $\Omega$ in SI units.
For a general introduction of kinetics of metals, see 
Refs.~\cite{LL10,AbrGorDzya,Abrikosov,LifAzbKag}.

High-$T_c$ cuprates are layered materials, but in order to evaluate the contribution
of the classical fluctuation of the electric field between conducting 2D layers
in Ref.~\cite{Mishonov:00} was analyzed a plane capacitor model for a (CuO$_2$)$_2$
bi-layer.
Imagine that a 2D plane is divided in small squares (plaquettes)
with a side equal to the Cu-Cu
distance, the in-plane lattice constant $a_0$
and the distance between the planes (double or single) is $d_0$.
The capacity of the considered small capacitor
\be
C=\varepsilon_0\frac{a_0^2}{d_0}.
\ee
For the square of the fluctuation charge $Q=Zq_e$ of this plaquette 
the equipartition theorem\cite{Herapath:21,Waterston:51,Waterston:92}
with temperature in energy units 
\be
\frac{\left<Q^2\right>_T}{2C}=\frac{T}{2}
\label{PlaneCapacitor}
\ee
gives
\be
(Zq_e)^2=\left<Q^2\right>=CT=\varepsilon_0\frac{a_0^2}{d_0}T,
\ee
where for brevity from now on we omit the brackets $\left<\,\right>_T$
 denoting thermal averaging.
The used here and in Ref.~\cite{Mishonov:00}
is actually the Nyquist theorem \cite{LL9}
\begin{align}
(\mathcal{E}^2)_\omega=2\frac{\hbar\omega }{\tanh\frac{\hbar\omega}{2T}}R(\omega),
\end{align}
where $(\mathcal{E}^2)_\omega$ is the spectral density of the voltage
$\mathcal{E}=E_zc_0$ between CuO$_2$ planes with distance $c_0$,
$E_z$ is thermally fluctuating electric field between conducting layers,
$$
\quad R(\omega)=\frac{c_0}{a_0^2\sigma_z(\omega)}
$$
is the resistance 
between two plaquettes with area $a_0^2$, and $\sigma_z(\omega)$
is the conductivity of the layered cuprate in the dielectric direction.
As it was recently proved, the general Callen-Welton fluctuation-dissipation theorem
can be considered as a consequence of Nyquist theorem.\cite{Reggiani:19,FDT,Reggiani:20}

The calculated in such a way averaged square of the fluctuating charge $Z^2=Q^2/q_e^2$
has to be substituted in the differential Eq.~(\ref{differential}) or transport
Eq.~(\ref{transport}) cross-section.
Additionally, for the area density of the ``impurities'' we have to substitute 
in the mean free path the density of the plaquettes $n_\mathrm{imp}=1/a_0^2$.
At these conditions the Drude formula Eq.~(\ref{Drude}) gives 
for 2D resistivity per square of CuO$_2$ plane
\be
4\pi\epsilon_0\varrho=\frac{m^2T}{8\hbar^3n_e^2d_0}
\ee
and the 2D dimensional conductivity $\sigma_\mathrm{Drude}/4\pi\epsilon_0$
has dimension velocity. In Gaussian system $4\pi\epsilon_0=1$, 
but all equations in the present paper are system invariant.
For a bulk material where separate bi-layers are at distance $c_0$,
the 3D resistivity parallel to the conducting planes $\rho_{ab}$
can be evaluated as
\be
4\pi\epsilon_0\rho_{ab}=\frac{m^2c_0}{8\hbar^3n_e^2d_0}T.
\ee
In short, the linear behavior of the resistivity reveals that in layered materials 
thermal fluctuations of electric field determine the density fluctuations.
Electrons scatter on the fluctuation of their own density
which in some sense is a self-consistent procedure.
A slightly different realization of the same idea is described in 
Ref.~\cite[Chap.~8]{MishonovPenev:11}.
Analogously, the wave scattering of the sunlight by the density fluctuations of the atmosphere determines the color of the sky;
who would be blind for the blue sky?~\cite{Mishonov:00}
In a maximal traditional interpretation,
resistivity of the layered high $T_c$ cuprates is simply Rayleigh scattering 
of Fermi quasiparticles on the electron density fluctuations 
in a layered metal.

However, our formula for the scattering rate $\Gamma=1/\tau$, Eq.~(\ref{Drude})
naturally explains an isotropic scattering which does not agree
with the spectroscopic data.
If we consider the energy in \Eqref{Drude} to be equal to the Fermi one $\eF$
the formula for the cross-section \Eqref{transport}
predicts negligible anisotropy if it is applied to the CuO$_2$ plane
while ARPES (Angle Resolved Photo-emission Spectroscopy) data\cite{Damascelli:03,Handbook,Shen:21}
reveals remarkable anisotropy of $\Gamma(\varphi)$
when we rotate on angle $\varphi$ around $(\pi,\,\pi)$-point 
i.e. the center of the hole pocket.

\textcolor{magenta}
{Analysis of the normal conductivity per square CuO$_2$ plane in some cuprates
gives that $\sigma/4\pi\varepsilon_0\gg v_\mathrm{Bohr}\equiv e^2/\hbar$.
This leads that in CuO$_2$ plane we have well-defined quasiparticles 
(in the sense of Landau) as it is confirmed by ARPES data in nodal direction.
That not only Fermi liquid but even Fermi gas approach do not reveals an
''enemy of working class'' in sense of Kivelson and Fradkin 
\cite[Chap.~15, page 583]{Handbook}.}

It is obvious that the Coulomb scattering is not the only mechanism 
for creation of the scattering rate $\Gamma$ and Ohmic resistivity.
The purpose of the present work is to take into account the 
\textit{s-d} exchange interaction which creates the pairing in the superconducting phase.

}

\section{Band structure in LCAO approximation}

A general review of electron band calculations in cuprates is given
by Pickett \cite{Pickett:89}, here we use and interpolation 
scheme of the band structure convenient for theoretical treatment 
of the exchange interaction, see e.g. Ref. \cite{MishonovPenev:11} for
more details.
LCAO method completely dominates in the intuitive picture in quantum chemistry and simple quantum calculations.
In LCAO, we have a Hilbert space spanned by the valence orbitals. 
Here we wish to point out that LCAO was 
used by Abrikosov \cite{Abrikosov:03} 
in an attempt to explain the metal-insulator phase transition in
CuO$_2$.
Applying LCAO to a CuO$_2$ plane, we have 
\begin{align}
\hat\psi_\mathrm{LCAO,\alpha}(\mathbf{r}) =
\sum_\mathbf{n}&\left[\hat{D}_{\mathbf{n},\alpha}
\psi_{\mathrm{Cu3d}_{x_2-y_2}}(\mathbf{r}-\mathbf{R}_\mathrm{Cu}-a_0\mathbf{n})
\right. \nn \\
&+\hat{S}_{\mathbf{n},\alpha}
\psi_{\mathrm{Cu}4s}(\mathbf{r}-\mathbf{R}_\mathrm{Cu}-a_0\mathbf{n})
\notag \\
&+\hat{X}_{\mathbf{n},\alpha}\psi_{\mathrm{O2p}_x}
(\mathbf{r}-\mathbf{R}_\mathrm{O_x}-a_0\mathbf{n}) \notag\\
&\left. +\hat{Y}_{\mathbf{n},\alpha}\psi_{\mathrm{O2p}_y}
(\mathbf{r}-\mathbf{R}_\mathrm{O_y}-a_0\mathbf{n})
\right],
\end{align}
where
$\mathbf{n}=(\tilde x, \tilde y)$ designates the elementary cell with integer
2D coordinates
$\tilde x,\,\tilde y=0,\pm1,\pm2,\pm3,\dots$
In the elementary cell with constant $a_0$
we have for the coordinates of the Cu ion
$\mathbf{R}_\mathrm{Cu}=\left(0,0\right)$,
and for the oxygen ions in $\tilde x$- and $\tilde y$-direction we have
$\mathbf{R}_{\mathrm{O},x}=\left(\frac12,0\right)a_0$
and 
$\mathbf{R}_{\mathrm{O},y}=\left(0,\frac12\right)a_0$.
We write the LCAO wave function in the second quantization
assuming that the atomic amplitudes
$\hat{D}_{\mathbf{n},\alpha}$, 
$\hat{S}_{\mathbf{n},\alpha}$, 
$\hat{X}_{\mathbf{n},\alpha}$, and
$\hat{Y}_{\mathbf{n},\alpha}$
in front of the atomic wave functions identified by $\psi$
are Fermi annihilation operators.
For illustration, we consider atomic functions of neighboring atoms as orthogonal.

In the generic 4 orbitals and 4 band model we have to take
single site energies $\epsilon_d$, $\epsilon_s$ and $\epsilon_p$
and the transfer integrals between neighboring atoms $t_{sp}$, $t_{pd}$ and $t_{pp}$.
Starting from the coordinate space $\mathbf{n}$,
we arrive at the momentum space symmetric Hamiltonian 
\be
H_\mathrm{LCAO}=
\begin{pmatrix}
\epsilon_d   & 0                    & t_{pd}s_x      &  -t_{pd}s_y  \\
0                & \epsilon_s       & t_{sp}s_x       & t_{sp}s_y     \\
t_{pd}s_x  & t_{sp}s_x       & \epsilon_\mathrm{p}      & -t_{pp}s_xs_y \\
-t_{pd}s_y &t_{sp}s_y        & -t_{pp}s_xs_y& \epsilon_\mathrm{p}
\end{pmatrix},
\label{H_LCAO}
\ee
where 
$$
s_x=2\sin\left(\frac12p_x\right), \quad
s_y=2\sin\left(\frac12p_y\right).
$$
Here a remark is due:
in electron band calculations the Coulomb repulsion is not neglected but only calculated
in a self-consistent way.
Roughly speaking, the Hubbard $U$ is incorporated in the single-site energies
and the experimental observation of the Fermi surface, by ARPES for example, 
is a small hint of the applicability of the self-consistent approach as
initial approximation.

The dimensionless quasi-momenta or phases $0\leq p_x,\,p_y \leq 2\pi$ 
belong to 2D Brillouin zone (BZ)
and for the eigenfunctions we have
\be
\Psi_{\mathbf{p}}=
\begin{pmatrix}
D_\mathbf{p}\\[0.2cm] S_\mathbf{p}\\[0.2cm] X_\mathbf{p}\\[0.2cm] Y_\mathbf{p}
\end{pmatrix}
=
\begin{pmatrix}
-\varepsilon_s\varepsilon_\mathrm{p}^2+4\varepsilon_pt_{sp}^2
	(s_x+s_y)-32t_{pp}\tau_{sp}^2s_xs_y\\[0.2cm]
	-4\varepsilon_\mathrm{p} t_{sp} t_{pd}(s_x-s_y) \\[0.2cm]
	-(\varepsilon_s\varepsilon_\mathrm{p}-8\tau_{sp}y)t_{pd}s_x\\[0.2cm]
(\varepsilon_s\varepsilon_\mathrm{p}-8\tau_{sp}s_x)t_{pd}s_y
\end{pmatrix},
\label{Psi_LCAO}
\ee
where
$\varepsilon_s=\epsilon-\epsilon_s$,
$\varepsilon_d=\epsilon-\epsilon_d$,
$\varepsilon_\mathrm{p}=\epsilon-\epsilon_\mathrm{p}$,
$\tau_{sp}^2=t_{sp}^2-\frac12\varepsilon_st_{pp}$.
The true quasi-momentum is given by $\mathbf{P}=\frac{\hbar}{a_0} \,
\mathbf{p}$; the dimensionless variables simplify the complicated notations
and provide formulas convenient for numerical analysis.
Moreover, the band wave functions $\Psi_{\mathbf{p}}$ have to be
normalized. 
Let us mention also that we make use of the assumption that atomic
wave functions of neighboring atoms as orthogonal thus neglecting
atomic overlapping.

In terms of energy, the secular equation of the band Hamiltonian
\be
\mathrm{det}(H_\mathrm{LCAO}-\epsilon\oone)
=\mathcal{A}xy+\mathcal{B}(x+y)+\mathcal{C}=0
\label{spectrum}
\ee
is a 4-degree polynomial having 4 zeros, say
$\varepsilon_{b,\mathbf{p}}$,
with band index
$b=1,2,3,4$.
For the coefficients in the secular equation \eqref{spectrum},
after some algebra, we obtain
\begin{subequations}
\label{ABC}
\begin{align}
\mathcal{A}(\epsilon) &= 16(4t_{pd}^2t_{sp}^2+2t_{sp}^2t_{pp}\varepsilon_d
-2t_{pd}^2t_{pp}\varepsilon_s
-t_{pp}^2\varepsilon_d\varepsilon_s),
\\
\mathcal{B}(\epsilon) &= -4\varepsilon_\mathrm{p}(t_{sp}^2\varepsilon_d+t_{pd}^2\varepsilon_s),
\\
\mathcal{C}(\epsilon) &=
	\varepsilon_d\varepsilon_s\varepsilon_\mathrm{p}^2.
\end{align}
\end{subequations}
Introducing the auxiliary parameters
\begin{align}
t=\frac{\mathcal{A}}8+\frac{\mathcal{B}}4,\qquad
t^\prime=\frac{\mathcal{A}}{16} ,\qquad
\eta=-\frac{\mathcal{A}}{4} -\mathcal{B}-\mathcal{C}
\end{align}
the secular equation \eqref{spectrum} providing the profile of the Constant Energy Curve (CEC) 
that can be written in the convenient formula
\be
\eta=-2t\,[\cos (p_x)+\cos (p_y)]+4t^\prime\cos (p_x)\cos (p_y).
\ee
This exact form with energy dependent coefficients  
inspired many theorists to approximate
LCAO CEC for $\epsilon_\mathbf{p}=\eF$ 
by expressions inherent to relatively simple tight
binding models on a square lattice.
However, this is related to the profile of CEC only at fixed energy
and cannot be used to describe the whole energy dependence of the conduction band 
or calculation of the Fermi velocity.
As a rough approximation for small transfer integrals,
one uses the approximation
\be
\mathcal{C}=\varepsilon_d\varepsilon_s\varepsilon_\mathrm{p}^2
\approx
(\epsilon_\mathbf{p}-\epsilon_d)(\epsilon_d-\epsilon_s)(\epsilon_d-\epsilon_p)^2
\ee
and thus $\eta$ can be considered as a linear function of the band energy $\epsilon_\mathbf{p}$. 

On the other hand, the shape of the hole pocket can be experimentally observed 
by analysing ARPES data.
Then, CEC curve passes through the points
$\tilde{\mathcal{D}} =
(p_d,\,p_d)$ and
$\tilde{\mathcal{C}} = (\pi,\,p_c)$,
where
\begin{subequations}
\be
x_d=\frac{-\mathcal{B}+\sqrt{\mathcal{B}^2-\mathcal{A}\mathcal{C}}}{\mathcal{A}}
=\sin^2\left(\frac{p_d}{2}\right), \qquad
x_c=-\frac{\mathcal{B}+\mathcal{C}}{\mathcal{A}+\mathcal{B}}=\sin^2\left(\frac{p_c}{2}\right).\nn
\ee
\end{subequations}
The parameters $x_c$ and $x_d$ can be used to fit CEC to the experimental data 
introducing
\begin{subequations}
\label{CEC_fit}
\begin{align}
& \mathcal{A}_f=2x_d-x_c-1, \\ 
& \mathcal{B}_f=x_c -x_d^2, \\ 
& \mathcal{C}_f=x_d^2(x_c+1)-2x_c x_d,
\end{align}
satisfying the equation
\begin{equation}
\mathcal{A}_f \, xy+\mathcal{B}_f \, (x+y)+\mathcal{C}_f =0,
\end{equation}
with
$$
\frac{\mathcal{A}_f}{\mathcal{B}_f}=\left.\frac{\mathcal{A}}{\mathcal{B}}\right|_{\varepsilon=\eF}.
$$
\end{subequations}
The fitting parameters $x_c$ and $x_d$ can be used 
to compare the result of electron band calculations 
and photo-emission data \cite{MishonovPenev:11}.
For our further analysis, we introduce convenient formulae in different representations
\begin{align}
\frac{t}{t'}
=2+4\dfrac{\mathcal{B}_f}{\mathcal{A}_f},
\label{t'/t}
\end{align}
i.e. the ratio $t^\prime/t$ can be calculated from electron band
structure obtained from ARPES data for the Fermi contour.
For example, According to ARPES data for
Bi$_2$Sr$_2$Cu$_1$O$_{6+\delta}$, Ref. \cite{Takeuchi:01}
obtains $p_d=0.82\,$rad and $p_c=0.129$~rad, which gives for this cuprate
$t^\prime/t=0.492$.
Furthermore we refer also to the dimensionless parameters \cite{Pavarini:01}
\begin{align}
r\equiv\frac1{2(1+s)},\qquad 
	s(\eF)\equiv
	\frac{(\epsilon_s-\eF)(\eF-\epsilon_\mathrm{p})}{(2t_{sp})^2}.
\label{reF}
\end{align}

The secular LCAO equation \eqref{spectrum} provides the opportunity to calculate CEC
in the BZ analytically 
\be
p_y=\pm\arcsin\sqrt{y},\qquad
0\le y=-\frac{\mathcal{B}x+\mathcal{C}}{\mathcal{A}x+\mathcal{B}}
\le1.
\ee
After performing diagonalization,
the band Hamiltonian of the free charge carriers takes the standard form
\begin{equation}
\hat{H}^{\prime (0)}
=\sum_{b,\mathbf{p},\alpha}(\varepsilon_{b,\mathbf{p}},
-\eF)\,
\hat{c}_{b,\mathbf{p},\alpha}^\dagger
\hat{c}_{b,\mathbf{p},\alpha}
\end{equation}
where $\hat{c}_{b,\mathbf{p},\alpha}^\dagger$ are the Fermi
creation operators for electron in band ``b'' with momentum $\mathbf{p}$
and spin projection $\alpha=\pm\tfrac12$.
The change of the Fermi operators \eqref{n_p} is related to diagonalization 
of the single particle Hamiltonian $\hat{H}^{\prime (0)}$.
After summation on the bands, momenta and spin projections,
we can return from momentum representation  
to the real space lattice wave function
\be
\hat\Psi_{\mathbf{n},\alpha}
=\begin{pmatrix}
\hat D_{\mathbf{n},\alpha}\\[0.2cm]
\hat S_{\mathbf{n},\alpha}\\[0.2cm]
\hat X_{\mathbf{n},\alpha}\\[0.2cm]
\hat Y_{\mathbf{n},\alpha}
\end{pmatrix}
=\frac1{\sqrt{N}}\sum_{b,\mathbf{p}}
\mathrm{e}^{\mathrm{i}\mathbf{p}\cdot\mathrm{n}}
\begin{pmatrix}
D_{b,\mathbf{p}}\\[0.2cm]
S_{b,\mathbf{p}}\\[0.2cm]
\mathrm{e}^{\mathrm{i}\varphi_x}X_{b,\mathbf{p}}\\[0.2cm]
\mathrm{e}^{\mathrm{i}\varphi_y}Y_{b,\mathbf{p}}
\end{pmatrix}
\hat{c}_{b,\mathbf{p},\alpha},
\label{n_p}
\ee
where the phases 
$
\varphi_x=p_x/2$, $\varphi_y=p_y/2$ are chosen so that the band Hamiltonian (\ref{H_LCAO})
and its eigenfunctions (\ref{Psi_LCAO}) are real.
Moreover, for the Fermi operators we have
$\{  \hat{c}_{b,\mathbf{p},\alpha},\, 
\hat{c}_{\mathrm{b^\prime},\mathbf{p}^\prime,\alpha^\prime}^\dagger \}=
\delta_{b\mathrm{b^\prime}}$,
$\delta_{\mathbf{p}\mathbf{p}^\prime}
\delta_{\alpha\alpha^\prime},
\{\hat D_{\mathbf{n},\alpha}, \hat D_{\mathbf{n^\prime},\alpha^\prime}^\dagger\}=
\{\hat S_{\mathbf{n},\alpha}, \hat S_{\mathbf{n^\prime},\alpha^\prime}^\dagger\}
=\{\hat X_{\mathbf{n},\alpha}, \hat X_{\mathbf{n^\prime},\alpha^\prime}^\dagger\}
=\{\hat Y_{\mathbf{n},\alpha}, \hat Y_{\mathbf{n^\prime},\alpha^\prime}^\dagger\}
=\delta_{\mathbf{n}\mathbf{n^\prime}}\delta_{\alpha,\alpha^\prime},
$ 
and the other anti-commutators are identically zero.
In \eqref{n_p} $N=N_xN_y$ designates the number of elementary cells subjected to periodic boundary conditions along $x$- and $y$-axes;
we imagine that the CuO$_2$ plane lies over a torus with $N_x$
and $N_y$ plaquettes.
The spectrum is calculated with the aid of the secular equation~(\ref{spectrum})
and the eigenvalues $\epsilon_{b}$,
thus, we can calculate the corresponding band wave function
$\Psi_{b,\mathbf{p}}=\Psi(\varepsilon_{b,\mathbf{p}})$
given in Eq.~(\ref{Psi_LCAO}) for every band and momentum.

\begin{figure}[ht!]
\centering
\includegraphics[scale=0.75]{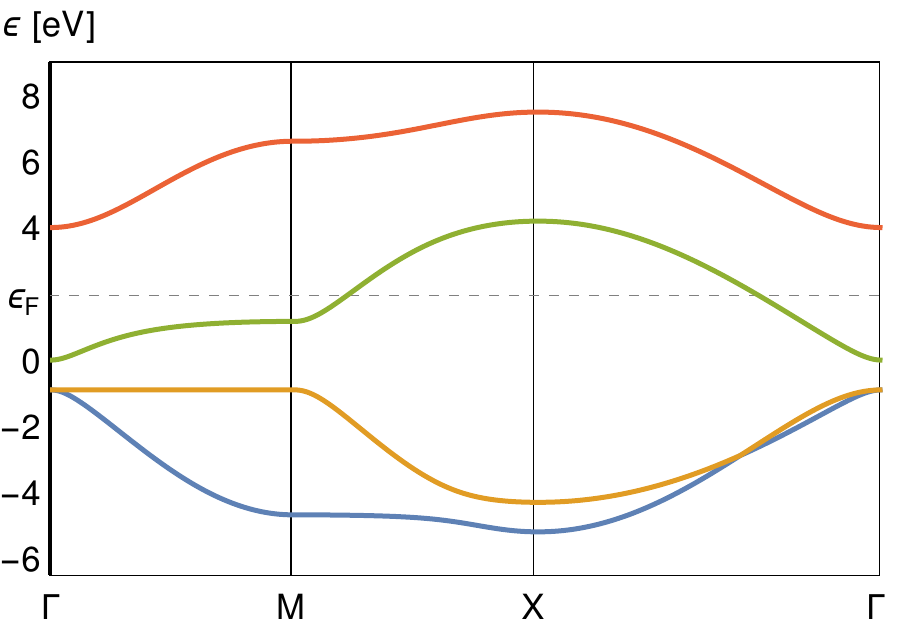}
\caption{Energy bands $\epsilon_{\mathbf{p},b}$ of LCAO Hamiltonian (\ref{H_LCAO})
with input parameters
$\epsilon_s=4$ eV, $\epsilon_p=-0.9$ eV, $\epsilon_d=0$ eV, $t_{sp}=2$
eV, $t_{pp}=0.2$ eV, $t_{pd}=1.5$ eV, $f_h=0.58$, $a_0=3.6$ \AA\  and
$T_c=90$ K, chosen to be close to the valus given in
Refs.~\cite{Andersen:95,Pavarini:01}.
The Fermi energy $\epsilon_{_\mathrm{F}}$ is given with dashed line.
The labeled points in the quasi-momentum space are: 
$\Gamma=(0,\,0)$,
$\mathrm{M}=(\pi,\,0)$,
$\mathrm{X}=(\pi,\,\pi)$.
The conduction Cu3$d_{x^2-y^2}$ band ($b=3$) coincides in $\Gamma$ point 
with the  Cu3$d_{x^2-y^2}$ atomic level $\epsilon_d=0$
which is chosen for the zero of the energy scale.
We have two completely filled oxygen bands $b=1,\,2$ 
($\epsilon_{_{\Gamma,1}}=\epsilon_{_{\Gamma,2}}=\epsilon_p$),
and one completely empty Cu$4s$ band $b=4$; $\epsilon_{_{\Gamma,4}}=\epsilon_s$.
}
\label{Fig:eps_bands}
\end{figure}

Our first approximation in this problem of physics of metals is to take into account 
only the conduction $d$-band of the CuO$_2$ plane and to omit further in the summation
the completely empty $s$-band or the completely filled
oxygen~$2p$-bands.
The band structure computed with the aid of the Newton method
\cite{MishonovPenev:11} is depicted in \Fref{Fig:eps_bands}.
The conduction \textit{d}-band, shown in green, is the 3$^\mathrm{rd}$
($b=3$) from the bottom and it is crossed by the dashed Fermi level.
Hereafter, we will omit the band index in all expressions supposing
that only the band $b=3$ contributes the most.

\begin{table}[ht]
\centering
\begin{tabular}{ >{\centering}p{0.8cm} >{\centering}p{0.8cm} >{\centering}p{0.8cm}
				      >{\centering}p{0.8cm} >{\centering}p{0.9cm} >{\centering}p{0.8cm} 
				    c >{\centering}p{0.9cm} c >{\centering}p{0.9cm} }	
		\hline \hline
		&  \\ [-1em]
		$\epsilon_s$  & $\epsilon_p$  & $\epsilon_d$  &
$t_{sp}$  & $t_{pp}$~\cite{Mishonov:96} & $t_{pd}$  & $f_h$	& $a_0$ & $T_c$ \\ 
			&  \\ [-1em] 
4.0	& -0.9 & 0.0 &	2.0	& 0.2 & 1.5 & 0.58 &3.6~\r{A}& 90~K \\
\hline \hline
\end{tabular}	\caption{
Single site energies $\epsilon$ and hopping amplitudes $t$ 
and Fermi energy $\eF$ according to Eq.~(\ref{Fermi_energy}) of the
LCAO Hamiltonian Eq.~(\ref{H_LCAO}) in eV. 
The parameters values are chosen close to the ones given in
Refs.~\cite{Andersen:95,Pavarini:01}.
}
\label{tbl:in_energy}
\end{table}

For the sake of simplicity, we can start with the Cu$3d_{x^2-y^2}$
level, $\epsilon_\mathbf{p}^{[0]}=\epsilon_d$, and apply several Newton iterations
\be
\epsilon_\mathbf{p}^{[i+1]}=\epsilon_\mathbf{p}^{[i]}
-\left.\frac{\mathcal{A}xy+\mathcal{B}(x+y)+\mathcal{C}}
        {\mathcal{A}^\prime xy+\mathcal{B}^\prime(x+y)+\mathcal{C}^\prime}\right\vert
        _{\epsilon=\epsilon_\mathbf{p}^{[i]}}.
\ee
Starting from the $\Gamma$ point ($\epsilon(0,0)=\epsilon_d$),
we calculate the conduction band energy at some neighboring point in the momentum grid.
The calculated in this way electron band structure 
is drawn in \Fref{Fig:eps_bands},
cf.~Ref.~\cite{MishonovPenev:11}
where different parameters for the calculation were used.
In such a way we can tabulate the energy $\epsilon_\mathbf{p}$
and further necessary $\chi_\mathbf{p}\equiv S_\mathbf{p}D_\mathbf{p}$
in a rectangular grid
\begin{align}
& p_x=\Delta p_x\, i_x,\qquad i_x=0,\,\dots,\, N_x, \qquad\Delta p_x=\frac{2\pi}{N_x}, \nn \\
& \quad p_y=\Delta p_y\, i_y, \qquad i_y=0,\,\dots,\, N_y,\qquad \Delta p_y=\frac{2\pi}{N_y}, \nn\\
& N_x=2K_x\gg1,\qquad N_y=2K_y\gg1,\nn\\
& \epsilon_{_\Gamma}=\epsilon_\mathrm{bottom}
=\epsilon_{0,0}=\epsilon(0,0)=\epsilon_d = 0,\nn\\
& \epsilon_\mathrm{_M}=\epsilon_\mathrm{Van\,Hove}
=\epsilon_{0,\pi}=\epsilon_{\pi,0}
=\epsilon(K_x,0)=\epsilon(0,K_y),\nn\\
& \epsilon_\mathrm{_X}=\epsilon_\mathrm{top}=\epsilon_{\pi,\pi}=\epsilon(K_x,K_y).
\end{align}
Further we can use those tables for interpolation in arbitrary point of the momentum space 
$\mathbf{q}$ in a rectangular grid, for example
\begin{align}
& q_x=\Delta p_x\, i, \qquad i=0,\,\dots,\, M_x, \qquad \Delta q_x=\frac{2\pi}{M_x} \nn \\
& q_y=\Delta q_y\, j, \qquad j=0,\,\dots,\, M_y, \qquad \Delta q_y=\frac{2\pi}{M_y}\nn\\
&M_x=2L_x\gg N_x,\quad M_y=2L_y\gg N_y. \nn
\end{align}
And further
\begin{align}
&
i_x=\mathrm{Int}\left (\frac{q_x}{\Delta p_x} \right ),\qquad
c_x=\frac{q_x}{\Delta p_x}-i_x\in (0,\,1),\nn\\
&
i_y=\mathrm{Int}\left (\frac{q_y}{\Delta p_y}\right ),\qquad
c_y=\frac{q_y}{\Delta p_y}-i_y\in (0,\,1),\nn\\
\epsilon_\mathbf{q} \approx &
   (1-c_x)(1-c_y)\,\epsilon(i_x,i_y)
+c_x(1-c_y)\,\epsilon(i_x+1,i_y)\nn\\
&
+(1-c_x)c_y\,\epsilon(i_x,i_y+)
+c_x c_y\,\epsilon(i_x,i_y),
\label{bi_linear_interpolation}
\end{align}
and analogous bi-linear approximation for the hybridization function

\begin{figure}[ht]
\centering
\includegraphics[scale=0.8]{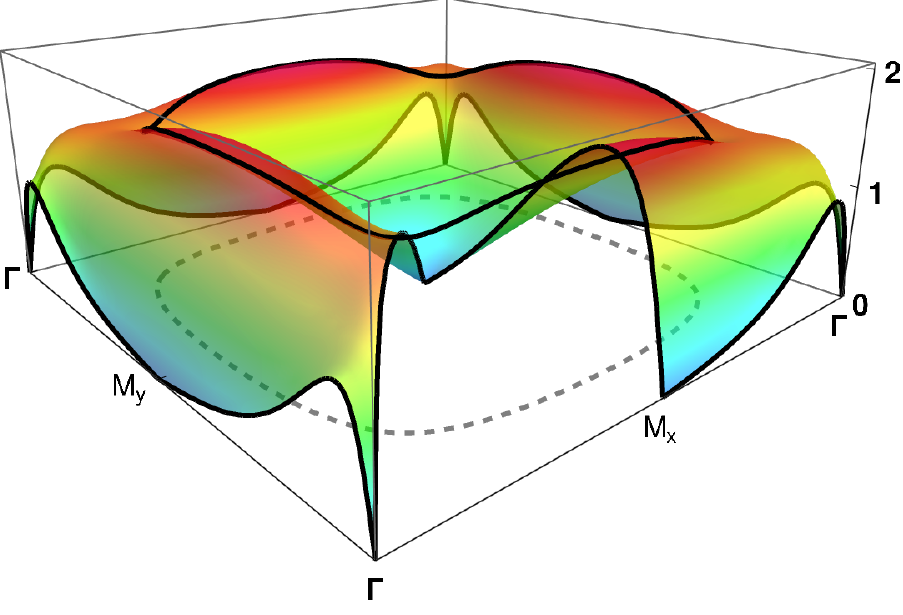}
\caption{The velocity $v_\mathbf{p}$ 
of the conduction band 
as a function of quasi-momentum $p_x,\,p_y\in (0,2\pi)$
with dimension energy and given in eV.
The variable $V=(a_0/\hbar)v$ has dimension m/s.
In the special points
$\Gamma=(0,\,0)$,
$\mathrm{M}=(\pi,\,0)$,
$\mathrm{X}=(\pi,\,\pi)$
the band velocity 
is zero. 
}
\label{Fig:Vzone}
\end{figure}

The electron band velocity
$$
v(\mathbf{p})=\left|\frac{\partial\epsilon_\mathbf{p}}{\partial\mathbf{p}}\right|
$$
of the conduction band is given in \Fref{Fig:Vzone}.

\begin{figure}[ht!]
\centering
\includegraphics[scale=0.8]{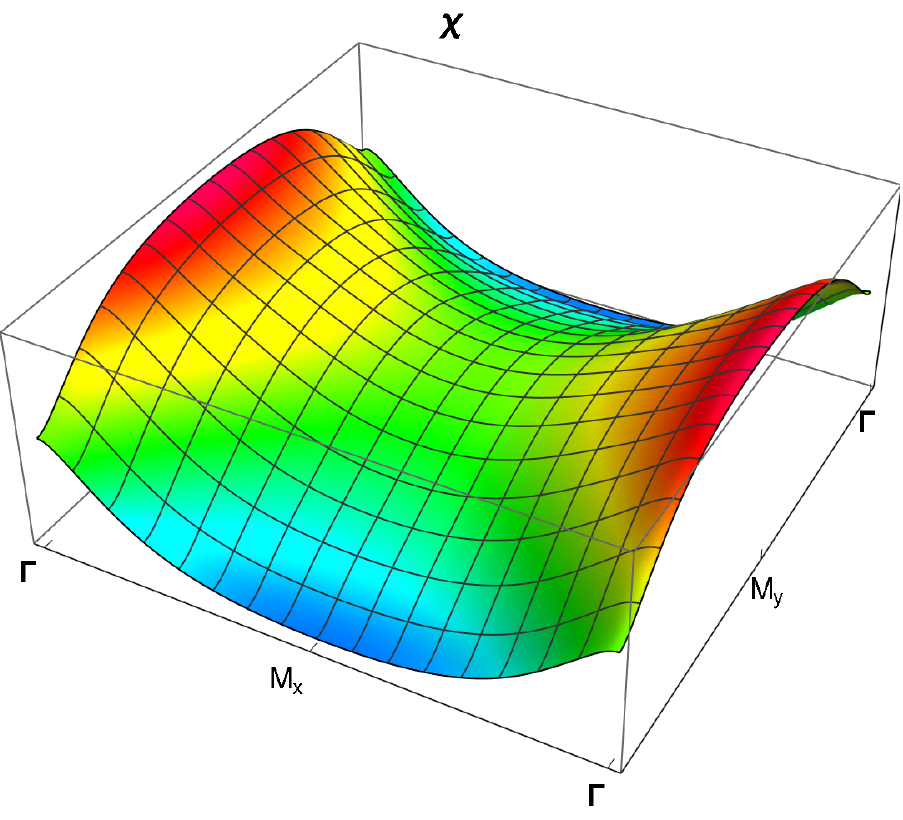}
\caption{
The hybridization function 
as function of quasi-momentum $\mathbf{p}$.
This hybridization describes the amplitude electron from conduction 
Cu3$d_{x^2-y^2}$ band to be simultaneously Cu$4s$ electron.
}
\label{Fig:chi}
\end{figure}

The behavior of the bilinear approximation for the hybridization function
\begin{align}
\chi_\mathbf{p} & \equiv S_\mathbf{p}D_\mathbf{p} 
\label{chi_analytical}
\end{align}
which is an important ingredient in our further consideration
is shown in Fig. \ref{Fig:chi}.
Recall that the real dimensional
quasimomentum is $\mathbf{P}=(\hbar/a_0) \, \mathbf{p}$.
Moreover, we have to emphasize that
the Coulomb interaction between the electrons is taken into account
in a self-consistent way and we consider that the LCAO method is only
an interpolation scheme of the local density band structure calculations. 
The inter-atomic transfer integrals and single site energies 
are the parameters of the named interpolation scheme.

From the canonical equation \eqref{spectrum} for the spectrum,
we easily derive the explicit equation
for CEC
\begin{equation}
p_y(p_x;\epsilon)=\pm2\arcsin(\sqrt{y}).
\end{equation}
\REM{
its derivative
\be
\tan^2(\alpha)\equiv
\left(\frac{\md p_y}{\md p_x}\right)^{2}
=\frac{(1-x)x}{(1-y)y}\left(\frac{\mathcal{A}y+\mathcal{B}}{\mathcal{A}x+\mathcal{B}}\right)^{\!2},
\ee
and the cosine of the same angle $\alpha$
\begin{equation}
\frac1{\cos(\alpha)}
=
\frac{\md p_l}{|\md p_x|}
=
\sqrt{1+\frac{(1-x)x}{(1-y)y}\left(\frac{\mathcal{A}y+\mathcal{B}}{\mathcal{A}x+\mathcal{B}}\right)^{\!2}},
\end{equation}
with
$$
\md p_l\equiv \sqrt{(\md p_x)^2+(\md p_y)^2}.
$$
}

The Fermi energy $\epsilon_{_\mathrm{F}}$
is determined by the hole filling factor,
i.e. the relative area of the hole pocket $S_p$, 
and the area of the Brillouin zone $(2\pi)^2$, i.e.
\begin{align}
f_h&=\overline{\theta(\varepsilon_\mathbf{p}-\eF)}=\frac{S_p}{(2\pi)^2},
\label{Fermi_energy}
\end{align} 
where the over-line stands for BZ averaging
\be
\overline{ F(\mathbf{p})}
\equiv \int_{0}^{2\pi}\int_{0}^{2\pi}F(p_x,p_y)\frac{\md p_x\md p_y}{(2\pi)^2}.
\ee

In our brief review of the results of the electron properties of the CuO$_2$ plane
it is also instructive to introduce the averaging on the Fermi surface
$\varepsilon_\mathbf{p}=\eF$;
the Fermi contour in the 2D case
\begin{equation*}
\langle f(\mathbf{p})\rangle
=\frac{\overline{f(\mathbf{p})\delta(\varepsilon_\mathbf{p}-\eF)}}
{\overline{\delta(\varepsilon_\mathbf{p}-\eF)}}.
\end{equation*}


Averaging over the Fermi contour we introduce
\be
\chi_\mathrm{av}
\equiv\exp\left\{\frac{\langle\chi_\mathbf{p}^2\ln|\chi_\mathbf{p}|\rangle}{\langle\chi_\mathbf{p}^2\rangle}\right\}
\ee
and change of the normalization of the hybridization amplitude
$\tilde\chi\equiv\chi/\chi_\mathrm{av}$
whose averaging on the Fermi contour gives
\be
\langle \tilde\chi_\mathbf{p}^2 \ln(\tilde\chi_\mathbf{p}^2)\rangle=0.
\ee
The renormalized gap anisotropy is maximal in modulus amplitude
in the pairing X-M direction
\be
\tilde\chi_\mathrm{max}=|\tilde\chi(p_x=p_c,p_y=\pi)|,\qquad \epsilon=\eF.
\ee
Within these notations we introduce 
the effective mass of the charge carriers at the center of the hole pocket
\begin{align}
\frac1{m_\mathrm{top}}=-\frac{1}{E_0}
\left.\frac{\partial^2\epsilon_\mathbf{p}}{\partial p_x^2}\right\vert_{(\pi,\pi)},
\end{align}
where
\begin{align}
E_0\equiv\frac{\hbar^2}{m_ea_0^2}. \notag
\end{align}
Using the mass of the free electron, the introduced effective mass is dimensionless
and $E_0$ is an energy parameter characterizing the CuO$_2$ plane.

Analogously we introduce the effective cyclotron mass $m_c$
which for almost cylindrical in 3D Fermi surfaces is parameterized by the
density of states per plaquette.
According to the Shockley formula \cite{LL9}, we have 
\begin{align}
& m_c= \frac{1}{2\pi m_e }\frac{\md \mathcal{S}_{_P}}{\md \eF}
=2\pi E_0\rho_F,\qquad
\mathcal{S}_{_P}\equiv\frac{\hbar^2}{a_0^2}\,\frac{f_h}{(2\pi)^2},
\end{align}
where $m_e$ is the mass of the free electron, 
$\mathcal{S}_{_P}$ is the area of the hole pocket in the quasi-momentum space 
$\mathbf{P}$, and $m_c$ is again a dimensionless parameter.

Let's assume that \hlc[cyan!20]{in some space homogeneous high
frequency vector-potential slightly changes all}
momenta of the electrons with an evanescent $\mathbf{Q}$.
Therefore, we have $\mathbf{P}\rightarrow\mathbf{P}+\mathbf{Q}$
the total change of the electron energy $\Delta E$ (per plaquette) is 
parameterized by the reciprocal tensor of the effective optical mass,
say $\mathfrak{m}_\mathrm{opt}$, 
\begin{align}
\Delta E &= 2\sum_\mathbf{p} [\epsilon(\mathbf{p+q})-\epsilon(\mathbf{p})] \,
\theta(\epsilon(\mathbf{p})-\eF))\label{Delta_E} \nn \\
	&=\mathbf{Q}\cdot \frac{N_e}{2m_e} \mathfrak{m}_\mathrm{opt}^{-1} \cdot\mathbf{Q},
\qquad \mathbf{q}\equiv \frac{a_0}{\hbar}\mathbf{Q}, 
\end{align}
where 
$$
N_e=2\sum_\mathbf{p}\theta[\epsilon(\mathbf{p})-\eF)]
$$
is the total number of holes per plaquette
and the factor 2 in front of momentum summation takes into account spin summation.
In the brackets in \Eqref{Delta_E} we recognize the second
derivative, which
in 2D space using the Gauss theorem for quadratic symmetry
gives for the dimensionless optical mass
\be
\frac1{m_\mathrm{opt}}
=\frac{\rho_F}{2E_0f_h}\langle v^2\rangle.
\ee
As a numerical test, if $\eF$ is slightly below $\epsilon_\mathrm{top}$,
all masses are equal.

The optical mass of the hole pocket at $T=0$ defined as
\be
\frac1{m_\mathrm{opt}}=
\left\langle\dfrac{\partial^2\epsilon_\mathbf{p}}{\partial P_x^2}\right\rangle
\label{m_opt}
\ee
is an important ingredient of the Kubo single band sum rule \cite{Kubo:57}
for the frequency dependent real part of the conductivity
$\sigma_{xx}(\omega)$ we have
\cite{Norman:02,Norman:03,Guy:05}
\be
\frac{q_e^2n_h}{2m_\mathrm{opt}}
=\int_{-\infty}^{\infty}\sigma_{xx}(\omega)\frac{\mathrm{d}\omega}{2\pi}.
\label{Kubo_sum}
\ee

Let $c_0^{-1}$ be the density of CuO$_2$ planes in the $c$-direction,
then the volume density of the holes reads
\be
n_h=\frac{2f_h}{c_0a_0^2}.
\ee
For $T\ll T_c$ all charge carriers are superconducting, $n_s=n_h$,
and for the in-plane penetration depth we obtain
\begin{align}&
\frac{1}{\lambda_{ab}^2(0)}
=\frac{q_e^2}{\varepsilon_0c^2}\frac{a_0^2}{\hbar^2}\,\langle v^2\rangle
\nu_F,\quad
\nu_F\equiv\frac{\rho_F}{a_0^2c_0}.
\end{align}
To derive a general formula for finite temperatures 
$\lambda(0)\rightarrow \lambda(T)$  
we have to regularize the Fermi surface averaging as follows \cite{MishonovPenev:11}
\begin{equation}
\langle v^2\rangle\rightarrow \left\langle
	v^2r_d\left(\frac{\Delta_\mathbf{p}}{2T}\right)\right\rangle,
\end{equation}
where 
\begin{align}
& r_d(y)\equiv \left(\frac y\pi\right)^2\sum_{n=0}^\infty
\left[\left(\frac y\pi\right)^2
+\left(n+\frac12\right)^{\!\!2}
\,\,\right]^{-3/2}.
\end{align}
Asymptotically we have
$$
r_d(y)\approx 7\zeta(3)\left(\frac y\pi\right)^2 \
\ \mathrm{for}\ \ y \ll 1,\qquad r_d(\infty)=1.
$$
In such a way we reveal how the results of the LCAO~\textit{s-d} approximation 
are incorporated into the standard theory of anisotropic gap BCS superconductors.
The general formula for the tensor of the reciprocal squares of penetration depths
reads \cite{Kogan:02} 
\begin{align}
\left(\frac{1}{\lambda^{2}(T)}\right)_{i,j}=
\frac{q_e^2}{\varepsilon_0c^2}\,2\nu_F
\langle V_i V_j r_d\rangle,
\end{align}
where $\varepsilon_0^{-1}$ is an \hlc[cyan!20]{eccentric manner to write $4\pi$ in the good old system
and $i,\,j=1,\,2,\,3$.
For clean superconductors and low temperatures $T\ll T_c$,
$2m_\mathrm{opt}$ is the effective mass of Cooper pairs,
say $m_\mathrm{opt}$ is the mass of the superfluid
charge carrier (per particle).}
This important for the physics of CuO$_2$ superconductors parameter
is experimentally accessible by electrostatic charge modulation of thin
superconducting films~\cite{Fiory:91}.
For such significant \hlc[cyan!20]{energy reduction} unexplained broad maximum of
mid infrared  (MIR) absorption of CuO$_2$ plane
finds natural interpretation as
direct inter-band transition between the conduction Cu$3d_{x^2-y^2}$
band and the completely empty Cu$4s$ band.
For more details on optical conductivity and spatial inhomogeneity of cuprate superconductors
see e.g. the review Ref.~\cite{Orenstein:07}.

For comparison of the results for the optical mass and penetration depth 
here we also give the conductivity $\sigma_{i,j}$ tensor in $\tau_\mathbf{p}$
approximation \cite{LifAzbKag,LL10} 
\be
\sigma_{i,j}=2q_e^2\nu_F
\langle V_i V_j \tau_\mathbf{p}\rangle.
\ee

Here we wish to emphasize the significant discrepancy between optical mass
of the conduction CuO$_2$ plane according to Table~\ref{tbl:out_energy} and
Ref.~\cite{Fiory:91}.
We do not exclude that all energy scales of the electron bands have to be re-examined.
Another weak point of all electron band calculations is the very high position of the 
Cu4\textit{s} level. 
We consider $\epsilon_s\simeq 4\,$eV to be unusually high as
the energy difference between the ground state level of the Cu atom 
$3d^{10}4s^1$ and the first excited level $3d^{9}4s^2$ is 
(after multiplet fine structure averaging) $\Delta E\approx 1.5\,\mathrm{eV}$
is much smaller than all values of $\epsilon_s$, which describe
the energy difference between Cu$3d$ and Cu$4s$ levels;
for atomic data see Ref.~\cite{NIST:Cu}.
This difference is unlikely to be ascribed to influence of oxygen on the ligands.

\hlc[cyan!20]{After this long introduction of notions and notations, 
we calculate the matrix elements of the exchange interaction
in the next section.}

\section{Influence of strong \textit{s-d} correlation on Cu site}
A reliable theory of CuO$_2$ plane must incorporate strong electron correlations.
Two fermion terms describe self-consistent single particle motion.
Strong correlations are fast processes which in the effective low-frequency Hamiltonians
give four-fermion terms. 
Heitler-London 2-electron correlations in two atom molecules are perhaps the most famous example.
Two electrons are newer in one at the same atom and in the second-quantization language
we write down the 4-fermion Hamiltonian of the valence bound.
However, magneto-chemistry, the physics of magnetism and perhaps the exchange mediated 
superconductivity is based on the proximity of $4s$ and $3d$ levels.
There are no interesting magnetic properties for light elements before the group of iron.
Shubin-Kondo-Zener \textit{s-d} exchange interaction (or \textit{c-l} exchange in the general case)
is actually the most usual \textit{s-d} exchange is described practically in all textbooks on condensed matter physics and physics of magnetism.
It was introduced in the physics long time before the BCS theory.
We write it in the lattice representation
\be
\hat H_{sd}=-J_{sd}\sum_{\mathrm{n},\alpha,\beta}
\hat S^\dagger_{\mathbf{n}\beta} \hat D^\dagger_{\mathbf{n}\alpha}
\hat S_{\mathbf{n}\alpha} \hat D_{\mathbf{n}\beta};
\label{s-d_exchange_interaction}
\ee
one Cu$4s$ electron with spin $\alpha$ is annihilated in the lattice cell $\mathbf{n}$
and resurrected with the same spin in the Cu$3d_{x^2-y^2}$ orbital.
Simultaneously, one Cu$3d_{x^2-y^2}$ electron with spin $\beta$
jumps without spin flip in the Cu$4s$ orbital.
There is no charge transfer for this exchange process which we sum on 
all elementary cells $\mathbf{n}$.

The substitution here of the representation by momentum space operators 
Eq.~(\ref{n_p}) using the explicit eigenfunctions Eq.~(\ref{Psi_LCAO}),
the exchange Hamiltonian for the conduction band 
\begin{align}&
\hat H_{sd}=-\frac{J_{sd}}{N} \! \!
\sum_{\substack{
\mathbf{p}^\prime+\mathbf{q}^\prime=\mathbf{p}+\mathbf{q}\\
\alpha, \, \beta}} \! \!
S_{\mathbf{q}^\prime} D_{\mathbf{p}^\prime}
\hat c^\dagger_{\mathbf{q}^\prime\beta} \hat c^\dagger_{\mathbf{p}^\prime\alpha}
\hat c_{\mathbf{p}\alpha}\hat c_{\mathbf{q}\beta}S_\mathbf{p}D_\mathbf{q}
\label{H_sd_1}
\end{align}
Let us remind that the band index $b=3$ is omitted
and we keep further only the conduction \textit{d}-band.
Next we perform a BCS reduction of this exchange Hamiltonian
and after the success of the analysis of the description of the superconducting properties 
we later carry out a Fermi liquid reduction.

\section{BCS reduction}

If we wish to obtain a homogeneous in space order parameter with zero momentum
in Hamiltonian (\ref{H_sd_1}), we have to perform the BCS reduction
\be
\mathbf{p}^\prime+\mathbf{q}^\prime=\mathbf{p}+\mathbf{q}=0, \quad \beta=-\alpha.
\ee
In other words, we create singlet Cooper pairs:
creation of an electron with momentum $\mathbf{p}$ and spin $\alpha$
with simultaneous creation of another electron with momentum 
$-\mathbf{p}$ and opposite spin projection $\beta=-\alpha$,
i.e. in the sum in the r.h.s of Eq.~(\ref{H_sd_1}) we have to take into account
only the terms with $\mathbf{q}= -\mathbf{p}$ and $\beta=-\alpha$.
Analogously for creation without spin flip we have to take only terms with 
$\mathbf{q}^\prime = -\mathbf{p}^\prime$.
Formally this initial reduction can be represented by insertion of $\delta$-functions
in the summand in \Eqref{H_sd_1}, i.e.
\begin{align*}
	\hat c^\dagger_{\mathbf{q}^\prime\beta} \hat c^\dagger_{\mathbf{p}^\prime\alpha}
\hat c_{\mathbf{p}\alpha}\hat c_{\mathbf{q}\beta}
&
\rightarrow
\delta_{\mathbf{q}^\prime+\mathbf{p}^\prime,0}\,
\delta_{\mathbf{q}+\mathbf{p},0} \,  
\delta_{\beta,\overline\alpha}\,     
\hat c^\dagger_{\mathbf{q}^\prime\beta} \hat c^\dagger_{\mathbf{p}^\prime\alpha}
\hat c_{\mathbf{p}\alpha}\hat c_{\mathbf{q}\beta}\nn\\
&
\rightarrow
\delta_{\mathbf{q}^\prime,-\mathbf{p}^\prime}\,
\delta_{\mathbf{q},-\mathbf{p}} \,  
\delta_{\beta,\overline\alpha}\, 
\left(
\delta_{\alpha,+}\,\hat B_\mathbf{p^\prime} \hat B_\mathbf{p}
+\delta_{\alpha,-}\,\hat B_{-\mathbf{p}^\prime} \hat B_{-\mathbf{p}}
\right), 
\label{BCS_reduction} \\
\end{align*}
where
$$
\hat B_\mathbf{p} \equiv\hat c_{-\mathbf{p},-}\hat c_{\mathbf{p}+}.\nn
$$
Whence, the reduced BCS Hamiltonian can be written as
\begin{align}
\hat H_\mathrm{BCS}=\frac{1}{N}\sum_{\mathbf{p},\mathbf{p}^\prime}
\hat B^\dagger_\mathbf{p}f(\mathbf{p},\mathbf{p}^\prime)\hat B_{\mathbf{p}^\prime},
\end{align}
with
\begin{align}
f(\mathbf{p},\mathbf{p}^\prime)\equiv-2J_{sd}\chi_\mathbf{p}\chi_{\mathbf{p}^\prime}.
\label{separable_1}
\end{align}
The multiplier 2 stems from the summation on $\alpha$.
We partially follow the notations from 9-th volume of Landau Lifshitz course of theoretical physics
\cite{LL9} in order to emphasize that the only difference 
is the $\chi$ factors in the reduced Hamiltonian.

Using the Bogolyubov transformation
\begin{align}
	\hat c_{\mathbf{p}+}&=u_\mathbf{p}\hat b_{\mathbf{p}+}+v_\mathbf{p}\hat b_{-\mathbf{p},-}^\dagger,
\qquad u_{-\mathbf{p}}=u_\mathbf{p}, \nn \\
	\hat c_{\mathbf{p}-}&=u_\mathbf{p}\hat b_{\mathbf{p}-}-v_\mathbf{p}\hat b_{-\mathbf{p},+}^\dagger,
\qquad v_{-\mathbf{p}}=v_\mathbf{p}, \nn
\end{align}
with $u_\mathbf{p}$ and $v_\mathbf{p}$ the parameters the Bogolyubov 
rotation $u_\mathbf{p}^2+v_\mathbf{p}^2=1$ and the new Fermi operators $\hat b_\mathbf{p}$.
The BCS self-consistent approximation yields
\begin{align*}
\langle B^\dagger_\mathbf{p}\hat B_{\mathbf{p}^\prime}\rangle
\approx \langle B^\dagger_\mathbf{p}\rangle\langle \hat B_{\mathbf{p}^\prime}\rangle,
\end{align*}
thus using
\begin{align*}
n_{\mathbf{p}+}=n_{\mathbf{p}-}\equiv n_\mathbf{p}
=\langle\hat b_{\mathbf{p}-}^\dagger\hat b_{\mathbf{p}-}\rangle
=\langle\hat b_{\mathbf{p}+}^\dagger\hat b_{\mathbf{p}+}\rangle,
\end{align*}
we get
\begin{align}
\langle B^\dagger_\mathbf{p}\rangle=\langle B_\mathbf{p}\rangle
=u_\mathbf{p}v_\mathbf{p}\,(1-n_{\mathbf{p}+}-n_{\mathbf{p}-})
. \nn
\end{align}
Finally, for the averaged interaction energy we have the standard functional
$E_\mathrm{BCS}(\{u_\mathbf{p}\},\{n_\mathbf{p}\})=\langle\hat H_\mathrm{BCS}\rangle$.
The non-interacting part of the Hamiltonian $\langle\hat{H}^{\prime (0)}\rangle$
has the same form as in Ref.~\cite{LL9}.

Minimization of the variational energy first with respect of $u_\mathbf{p}$
and taking into account that 
\be
n_\mathbf{p}=\frac{1}{\exp(E_\mathbf{p}/T)+1},
\ee
the Fermi distribution gives the standard equation for the superconducting gap
$\Delta_\mathbf{p}$ in cuprates, i.e.
\begin{align}
\label{BCS_gap_equation}
2J_{sd}\, \overline{\frac{\chi_\mathbf{p}^2}{2E_\mathbf{p}}
\tanh\left(\frac{E_\mathbf{p}}{2T}\right)} =1,
\end{align}
where
\begin{align}
E_\mathbf{p}=\sqrt{\eta_\mathbf{p}^2+\Delta_\mathbf{p}^2},\quad
\eta_\mathbf{p}=\epsilon_\mathbf{p}-\eF,
\quad \Delta_\mathbf{p}=\Xi(T)\,\chi_\mathbf{p}. \nn
\end{align}
The occurrence of the BCS spectrum for cuprates was
analyzed in the review \cite{Campuzano:18}.
In the next section we recall the main results of the Pokrovsky 
theory \cite{Pokr:61,PokrRiv:63} for the thermodynamics of 
anisotropic gap superconductors.

\section{Pokrovsky theory of anisotropic gap superconductors}

The \textit{s-d} exchange interaction is localized in a single transition ion in elementary cell
which automatically gives separable kernel of the BSC gap equation
\be
V_{\mathbf{q},\mathbf{p}}\equiv f(\mathbf{q},\mathbf{p})
=-2J_{sd}\chi_\mathbf{q}\chi_\mathbf{p}.
\label{separable_2} 
\ee
At the Fermi surface we have
\begin{align}\label{EigenValue}
\langle V_{\mathbf{q},\mathbf{p}}\chi_\mathbf{p}\rangle_\mathbf{p}
=-V_0\chi_\mathbf{q}, 
\end{align}
where $V_0=2J_{sd}\langle \chi^2\rangle$ is the eigenvalue of the interaction kernel.
In the general case the BCS gap equation reads
\begin{align}
\Delta_\mathbf{q}=\overline{V_{\mathbf{q},\mathbf{p}}
\frac{\Delta_\mathbf{p}}{E_\mathbf{p}}
\tanh\left(\frac{E_\mathbf{p}}{2T}\right).
}
\end{align}
Yet the separability of the kernel reduces the equation above to the
problem described by \Eqref{BCS_gap_equation}.

The general consideration by Pokrovsky reveals that in the BCS weak 
coupling limit we have to solve the corresponding eigenvalue problem 
and to use the maximal in modulus eigenvalue $V_0$.
The LCAO \textit{s-d} approximation simply gives us a text-book example of the
Pokrovsky theory for the anisotropic gap superconductors.

Inspired by Euler and Mascheroni definition for the popular constant
we introduce a new notion, namely the Euler-Mascheroni energy of the gap
anisotropy 
\begin{align}
E_\mathrm{C}\equiv\lim_{\epsilon\rightarrow 0}
\left [ \epsilon\exp\left\{
	\frac{\overline{\theta(|\eta_\mathbf{p}|-\epsilon)\chi_\mathbf{p}^2/|\eta_\mathbf{p}|}
	}{2\langle\chi^2\rangle\rho_F}
\right\} \right ].
\label{Euler-Mascheroni_energy}
\end{align}
Introduced in such a way the Euler-Mascheroni energy 
is a convenient notion to describe the BCS properties 
of superconductors with 4-fermion (two particle) interaction.
Within the so introduced notations we use
the well-known BCS formulas for the critical temperature
\begin{align}
T_c=\frac{2\gamma}{\pi}\,E_\mathrm{C}\, \exp\left(-\frac1\lambda\right), 
\label{TcBCS}
\end{align}
where the BCS coupling parameter
\begin{align}
\lambda\equiv V_0\rho_F=2J_{sd}\langle \chi^2\rangle\rho_F, \quad
V_0 \equiv 2 J_{sd}  \langle \chi^2\rangle,
\label{lambdaBCS}
\end{align}
the order parameter at zero temperature
\begin{align}
&\tilde \Xi(0)=2E_\mathrm{C}\,\exp\left(-\frac1\lambda\right),\quad
\frac{2\tilde\Xi(0)}{T_c}=\frac{2\pi}{\gamma}\approx 3.53,
\end{align}
and the superconducting gap
\begin{align}
\Delta_\mathbf{p}(T)=\tilde\Xi(T)\tilde\chi_\mathbf{p}=\Xi(T)\chi_\mathbf{p}.
\end{align}
that is factorizable function of the temperature and momentum.

In the original works by Pokrovsky \cite{Pokr:61,PokrRiv:63}
the factorizable order parameter and separable approximation of the pairing kernel 
are results of the weak coupling BCS approximation,
but for our Hamiltonian it is an immanent property.
Then for the maximal gap at zero temperature and the jump of the
\hlc[cyan!20]{heat capacity}
at the critical temperature the result by Pokrovsky reads \cite{MishonovPenev:11}
\begin{align}
\frac{2\Delta_\mathrm{max}}{T_c}
=\frac{2\pi}{\gamma}\frac{|\chi|_\mathrm{max}}{\chi_\mathrm{av}},
\qquad
\frac{\Delta C}{C_n(T_c)}=\frac{12}{7\zeta(3)}
\frac{\langle\chi_\mathbf{p}^2\rangle^2}{\langle\chi_\mathbf{p}^4\rangle}.
\end{align}
Perhaps the most important ingredient for the thermodynamics 
is the Pokrovsky equation for the temperature dependence of the order parameter 
\begin{align}
\ln\frac{\tilde\Xi(0)}{\tilde \Xi(T)}
=2\langle  \tilde\chi_\mathbf{p}^2 I(\tilde\Xi(T)\tilde\chi_\mathbf{p}/T) \rangle,
\label{Pokrovsky_anisotropy}
\end{align}
with
\begin{align*}
I(u)\equiv\int\limits_{0}^\infty\frac{\md x}{\sqrt{u^2+x^2}[\exp(\sqrt{u^2+x^2})+1]}.
\end{align*}
\hlc[cyan!20]{The brackets $\langle\dots\rangle_{_\mathrm{F}}$} here denote averaging
on the Fermi contour.
Note that the only difference between the isotropic BCS model is given by the $\chi$-factors.

The result for the temperature dependence of the superconducting gap 
Eq.~(\ref{Pokrovsky_anisotropy})
for anisotropic superconductors is derived by Pokrovsky~\cite{Pokr:61,PokrRiv:63} in the early BCS epoch.
The CuO$_2$ plane gives a simple analytical example of the gap anisotropy
which for qualitative purposes can be approximated by
\be
\chi_\mathbf{p}\simeq 
\frac{t_{sp} t_{pd}}{(\epsilon_s-\epsilon_d)(\epsilon_d-\epsilon_\mathrm{p})}
\cos(2\theta),
\qquad \eF \approx \epsilon_d,
\ee
where $\theta$ is the angle along the Fermi contour.
The sequence of the perturbation theory denominators reveals
that in first order the $d$-electron becomes $p$-one and 
only in the second order perturbation theory $p$-electron obtain 
$s$-amplitude necessary to be involved in the \textit{s-d}
interaction.
In short we follow the 3$d$-to-4$s$-by-2$p$ highway to superconductivity
\cite{MishIndPen:02} where the \textit{s-d} interaction was proposed as pairing

Finally,as a test example of the used approach
we rewrite $\chi_\mathbf{p}$ for the phonon model
\begin{align}
\chi_\mathbf{p}=\theta(E_C-|\eta_\mathbf{p}|),
\quad
E_\mathrm{C}=\hbar\omega_\mathrm{D},
\quad
\omega_\mathrm{D}=\sqrt{\frac{K}{M}}
\end{align}
for which the Euler-Mascheroni $E_c$ energy is nothing but the Debye
energy $\hbar\omega_\mathrm{D}$.
This statement reveals the applicability criterion of the BCS theory
$\exp(-1/\lambda)\ll1$ which gives the condition $T_c/E_\mathrm{C}\ll1$
which is perfectly satisfied according to the results presented in Table~\ref{tbl:out_energy}.
For the exchange mediated superconductors the reader may consult Ref. \cite{Manske:04}.

Now we can continue with technical details for application
of the Pokrovsky theory for anisotropic superconductors 
for LCAO \textit{s-d} approximation applied to the CuO$_2$ plane.

\section{Calculation of $T_c$ of CuO$_2$ plane}
Our first task is to calculate the exchange amplitude $J_{sd}$
supposing that for a 90~K superconductor LCAO electron band parameters 
are determined by the fit to band calculations, for example.
According to \Eqref{BCS_gap_equation}
the reciprocal exchange integral can be expressed by momentum integration
\begin{align}&
\frac1{J_{sd}}=I_\mathrm{sum}\equiv
\overline{\frac{\chi_\mathbf{p}^2}{\eta_\mathrm{p}}
\tanh\left(\frac{\eta_\mathrm{p}}{2T_c}\right)}.
\label{T_c_Equation}
\end{align}

\REM{
The behaviour of the $\tanh$ multiplier of the integrand is depicted in \Fref{Fig:tanh}
\begin{figure}[ht]
\centering
\includegraphics[scale=0.8]{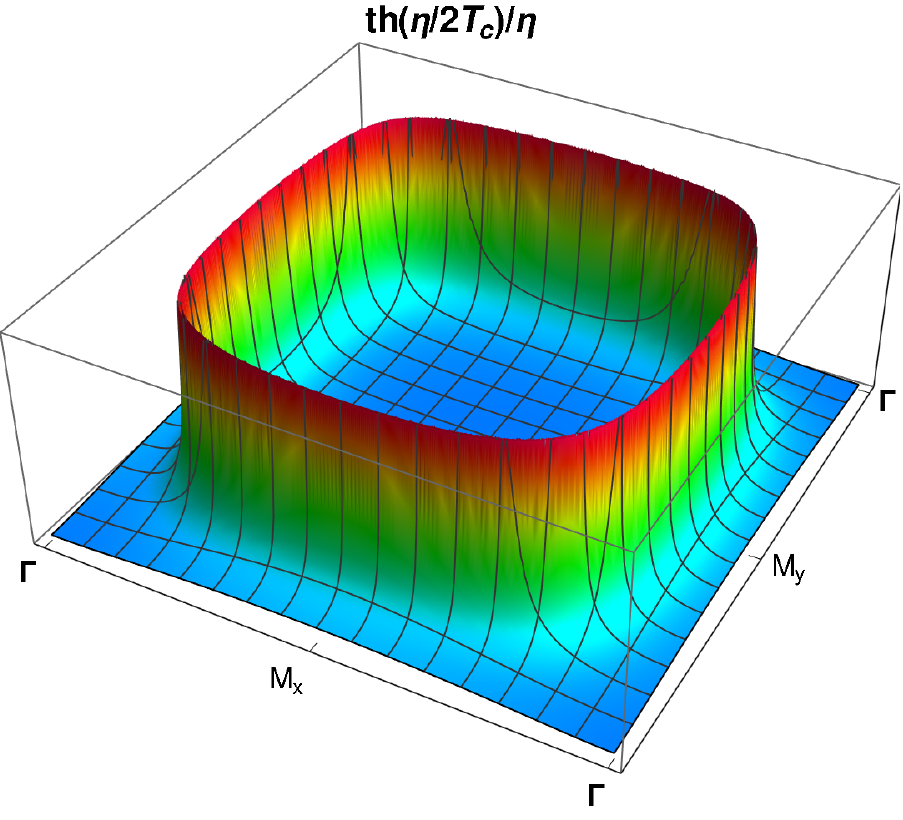}
\caption{
The multiplier $\tanh(\eta/2T_c)/\eta$
from the BCS equation for the critical temperature
\Eqref{T_c_Equation}
as function of quasi-momentum $(p_x,\,p_y)$.
This function has a sharp maximum $1/2T_c$ along the Fermi contour
while far from the Fermi contour is small.
}
\label{Fig:tanh}
\end{figure}}

\begin{figure}[ht]
\centering
\includegraphics[scale=0.8]{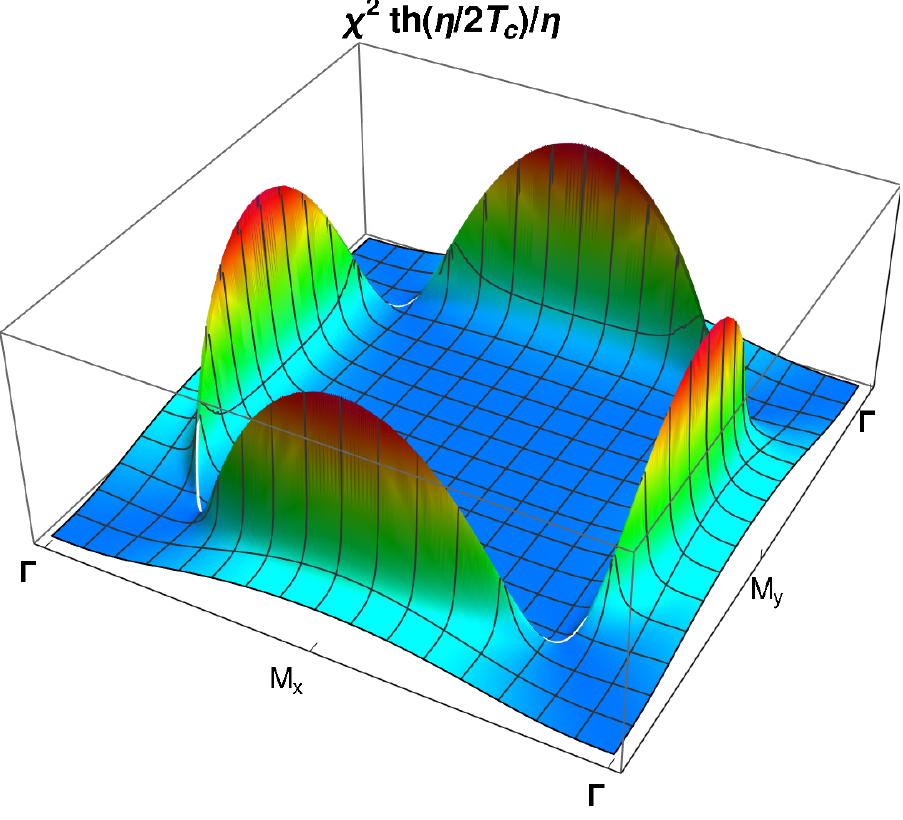}
\caption{The integrand of the equation \Eqref{T_c_Equation} 
for the critical temperature drawn at 300~K.
}
\label{Fig:Integrand}
\end{figure}

The behavior of the integrand is depicted in \Fref{Fig:Integrand}
at higher temperature $T=300$~K in order for a sharp function to be visible.
To evaluate the integral in Eq. \eqref{T_c_Equation}, we split the
integration interval into three sections.
The narrow first domain corresponds to
$\epsilon_a \gg T_c$
(with $\epsilon_a =5~T_c$), where
the density of states can be taken as constant.
Simultaneously, we suppose that 
$\epsilon_a \ll \epsilon_b \sim (\eF-\epsilon_\mathrm{Van\,Hove})/2$,
i.e. the energy parameter is much smaller than the typical band 
energies, for example, the distance between the Fermi level and the energy of the Van Hove
$\epsilon_\mathrm{Van\,Hove}\equiv\epsilon_\mathrm{_M}\epsilon_\mathrm{0,\,\pi}$.
The second energy parameter $\epsilon_b$ ensures that topology will not be changed 
in the intermediate energy interval.
In short, the integral in \Eqref{T_c_Equation}
can be represented by a sum of three integrals, i.e,
$I_\mathrm{sum}=I_a+I_{ab}+I_b$.
\begin{table}[ht]
\centering
\caption{Output parameters of our numerical calculation, the extra numbers are only for a numerical test.
The new quantities are the values of the \textit{s-d} exchange amplitude $J_{sd}$ and the effective masses derived from the parameters of electron band 
calculations~\cite{Pavarini:01}. 
The value of $\tilde{\chi}_\mathrm{max}=1.167$ is within  10\% accuracy 
 of its theoretical value for a pure \textit{d}-wave gap with isotropic Fermi velocity $2/\sqrt{\e}=1.213$.}
	\label{tbl:out_energy}
\begin{tabular}{ r p{0.5cm} r p{0.5cm} r }	
		\hline \hline
		&  \\ [-1em]
		$E_\mathrm{C}$ = 1.928~eV  & & $\lambda$ = 0.177  & &
		$m_\mathrm{top}$ =  0.839 \\
		&  \\ [-1em] 
		$\eF$ = 1.851~eV & & $\tilde{\chi}_\mathrm{max}$ = 1.167 &
		  & $m_c$ = 	0.931 \\
		&  \\ [-1em] 
  	     $\epsilon_{\mathrm{_M}}$ = 1.167~eV  &  & $\langle \chi^2 \rangle$ = 0.065&   & $m_\mathrm{opt}$ = 0.890 \\
     		&  \\ [-1em] 
  	     $\epsilon_{\mathrm{_X}}$ = 4.193~eV   &  &
  	    $\langle \chi^2 \rangle^2/\langle \chi^4 \rangle$ = 0.737& & $r$ = 0.365~eV \\
     		&  \\ [-1em] 
    	     $E_0$ = 0.528~eV  & & $\rho_F = 0.281~\mathrm{eV}^{-1}$	& &  $2/\sqrt{\e}$ = 1.213\\
    	     $J_{sd}$ = 7.230~eV & & $2 \Delta_\mathrm{max}/T_c$ = 4.116 & &$V_0$ = 0.940~eV \\
		&  \\ [-1em] 
\hline \hline
\end{tabular}
\end{table}

For the first integral assuming that density of states is almost equal,
using the well known integral limit
\be
\lim_{M\rightarrow\infty}\left(\int_0^M\frac{\tanh x}{x}\md x-\ln M\right)
=\ln\left(\frac{4\gamma}{\pi}\right)
\ee
we obtain after energy integration
\be
I_a=\overline{\theta(\epsilon_a-|\eta_\mathrm{p}|)
\,\frac{\chi_\mathbf{p}^2}{\eta_\mathrm{p}}
\tanh\left(\frac{\eta_\mathrm{p}}{2T_c}\right)} 
\approx2\langle\chi^2\rangle\rho_F\ln\left(\frac{2\gamma\epsilon_a}{\pi T_c}\right).
\ee
For the rest of the momentum space when $\epsilon_a\gg 2T_c$ we use the
$\tanh(\epsilon_a/2T_c)\approx1$ approximation
and for the second integral we obtain
\begin{align}
I_{ab}&=\overline{\theta(\epsilon_a<|\eta_\mathrm{p}|<\epsilon_b)
\,\frac{\chi_\mathbf{p}^2}{\eta_\mathrm{p}}
\tanh\left(\frac{\eta_\mathrm{p}}{2T_c}\right)} \nn \\
&\approx \int_{\epsilon_a}^{\epsilon_b}
\left[
\left.\langle\chi^2\rangle\rho\right\vert_{(\eF-\eta)}
+\left.\langle\chi^2\rangle\rho\right\vert_{(\eF+\eta)}
\right]
\frac{\md\eta}{\eta} 
\approx 2\ln\left(\frac{\epsilon_b}{\epsilon_a}\right)\langle\chi^2\rangle\rho_F,
\end{align}
where for the last approximation we suppose constant density of states.
If $\epsilon_b=\epsilon_a$ this second integral is annulled.
The third integral 
\begin{align}
I_b= 
\overline{\theta(|\eta_\mathrm{p}|-\epsilon_b)\,\frac{\chi_\mathbf{p}^2}{|\eta_\mathbf{p}|}}
\end{align}
is simply an energy integration far from the Fermi energy.
Supposing that $\epsilon_a$ is almost zero i.e. much smaller than the band parameters,
we recognize the Euler-Mascheroni energy \Eqref{Euler-Mascheroni_energy}
\begin{align}
&\ln\epsilon_a+\overline{\theta(|\eta_\mathrm{p}|-\epsilon_a)
\,\frac{\chi_\mathbf{p}^2}{\eta_\mathrm{p}}}
(2\langle\chi^2\rangle\rho_F)^{-1}
\approx \ln E_\mathrm{C}.
\end{align}
The numerical integration here can be performed by a Riemann sum of the
bi-linear approximation of the integrand functions.
As a result the summary integral can be expressed as 
\be
I_\mathrm{sum}\approx 2\langle\chi^2\rangle\rho_F
\left[\ln\left(\frac{2\gamma}{\pi T_c}\right)+\ln E_\mathrm{C}\right] \nn
\ee
and finally after substitution in \Eqref{T_c_Equation} we 
arrive at \Eqref{TcBCS}.

The most important ingredient of the BCS formula for the critical temperature $T_c$
is the BCS coupling constant $\lambda$ defined by \Eqref{lambdaBCS}.
On the other hand,
Pavarini \textit{et al.}~\cite{Pavarini:01} observed a remarkable correlation
between their range parameter $r(\eF)$ defined by \Eqref{reF}
and the critical temperature $T_c$. 
What is hidden in this band trend correlation?
Emphasizing the importance of this empirical correlation 
Patrick Lee pointed out that it is not a simple task for the theory~
\cite[\textit{t-J} Model and Gauge Theory Description of Underdoped Cuprates]{Handbook}.
For the application of  $t$-$J$ model 
in the physics 
of high-$T_c$ cuprates see also the reviews by Spalek~\cite{Spalek:07},
P.~Lee~\cite{Lee:06} and Livelong~\cite{Kivelson:03}.

\begin{figure}[ht]
\centering
\includegraphics[scale=0.5]{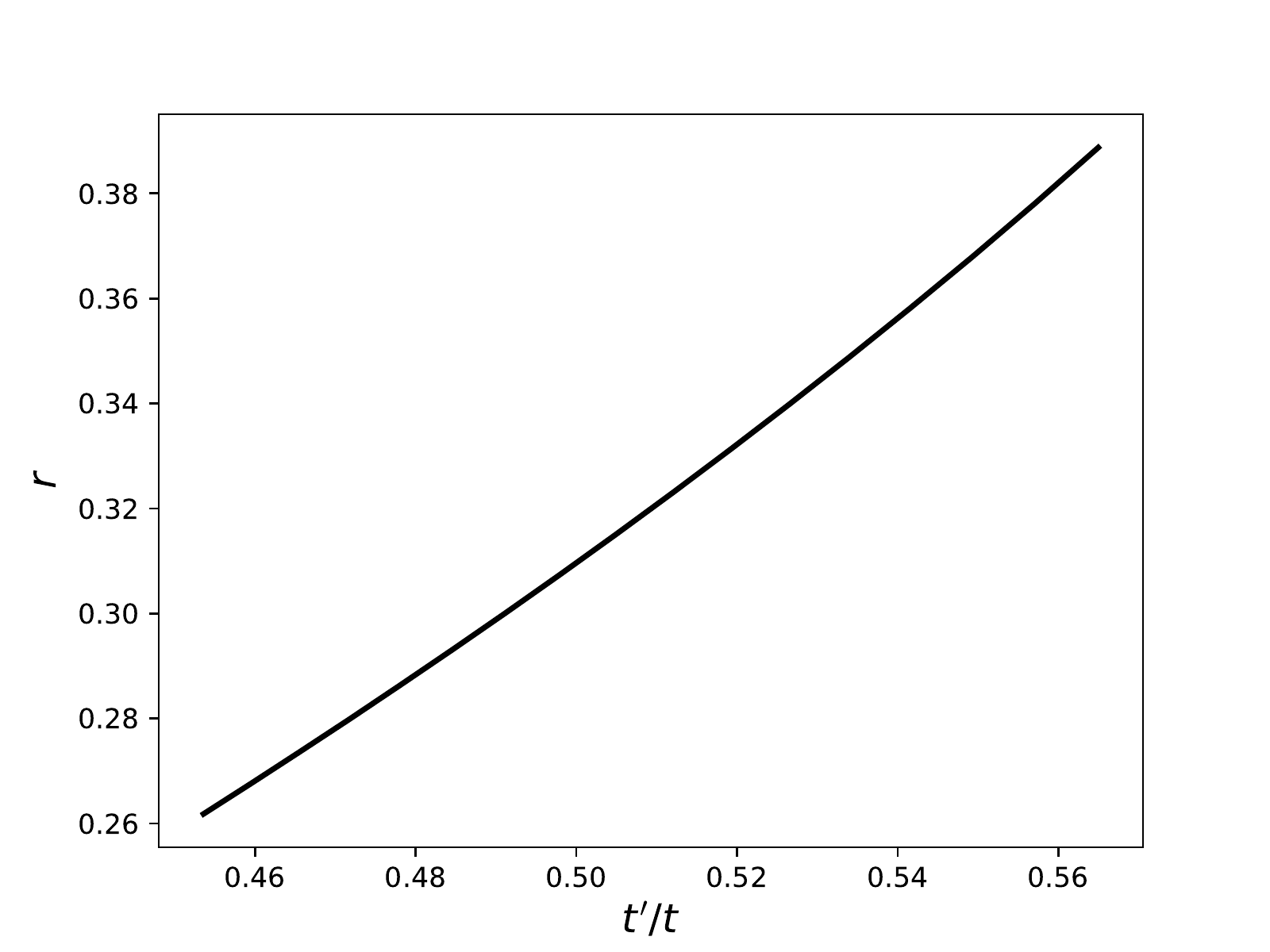}
\caption{Almost linear dependence between range parameter 
$r(\eF)$ \Eqref{reF} and the ratio $t^\prime/t$ \Eqref{t'/t}.
These parameters are introduced in Ref.~\cite{Pavarini:01}.
}
\label{Fig:tptr}
\end{figure}

The range parameter $r(\eF)$ defined by \Eqref{reF} 
is also almost linear function of the ratio $t^\prime/t$ \Eqref{t'/t}
as shown in \Fref{Fig:tptr}.

\begin{figure}[ht]
\centering
\includegraphics[scale=0.5]{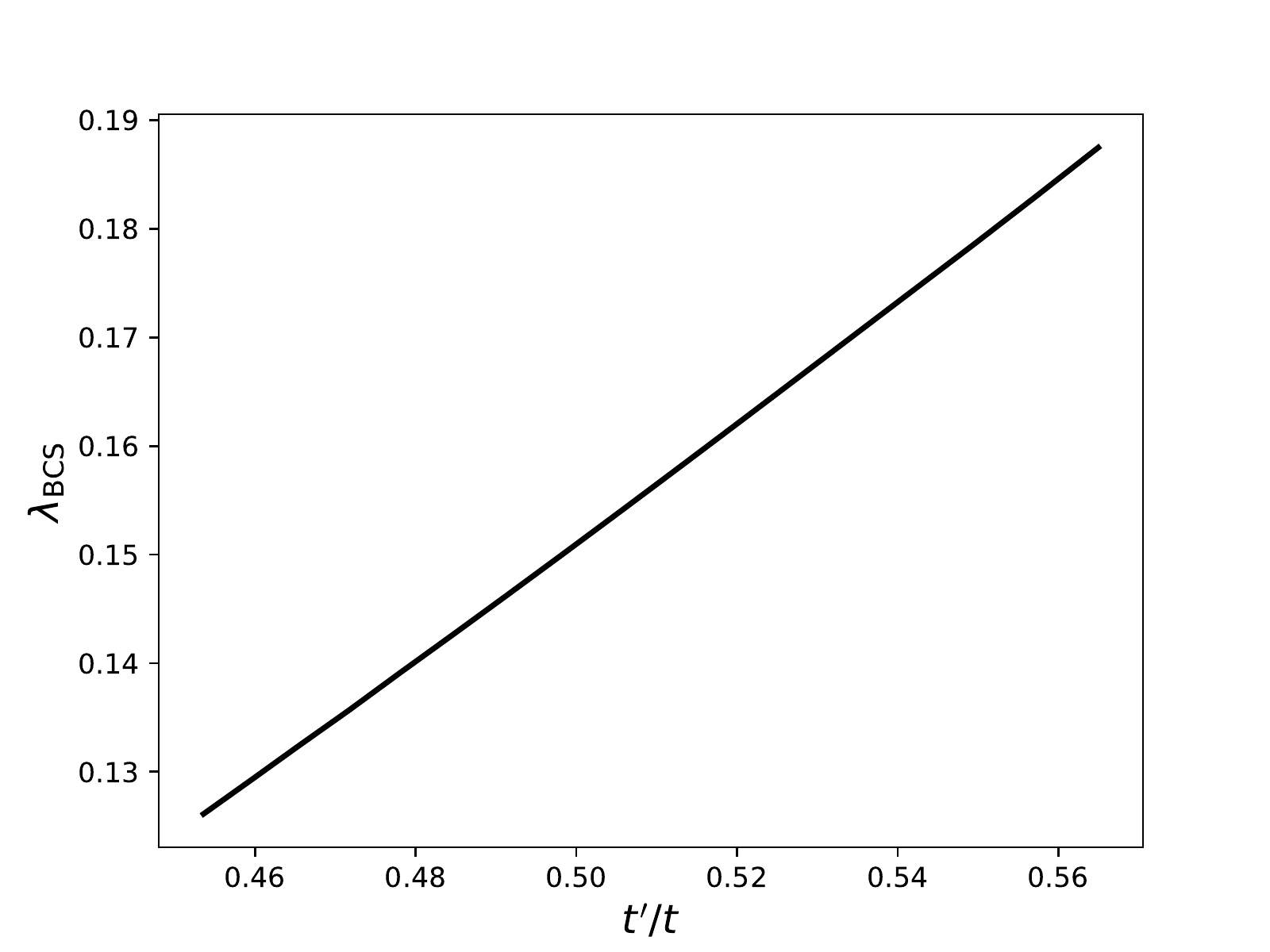}
\caption{This non-interesting straight line (within the accuracy of
the numerical calculation) represents the relation between the
BCS coulpling constant $\lambda$ defined in \Eqref{lambdaBCS}
admitting common $J_{sd}$ given in Table~\ref{tbl:out_energy}
for all cuprates 
and the ratio of the tight binding parameters $t^\prime/t$ 
calculated in \Eqref{t'/t} where \Eqref{ABC} is substituted.
It is well-known according to \Eqref{TcBCS} that $\lambda$
has the main influence on the critical temperature $T_c$.
The complicated integral representing $\langle\chi^2\rangle\rho_F$ 
with analytical expression \Eqref{chi_analytical} substituted
in Fermi contour averaging
gives little hope for an analytical solution.
In such a way we have only the graphical solution that
BCS coupling constant $\lambda$ is in good approximation
linear function of the $t^\prime/t$ parameter determining the shape of the 
Fermi surface; for both variables we derived complicated explicit expressions 
\Eqref{t'/t} and \Eqref{TcBCS} exact in the used LCAO approximation
for electronic the band structure.
}
\label{Fig:tpt-lambda}
\end{figure}

\begin{figure}[ht]
\centering
\includegraphics[scale=0.5]{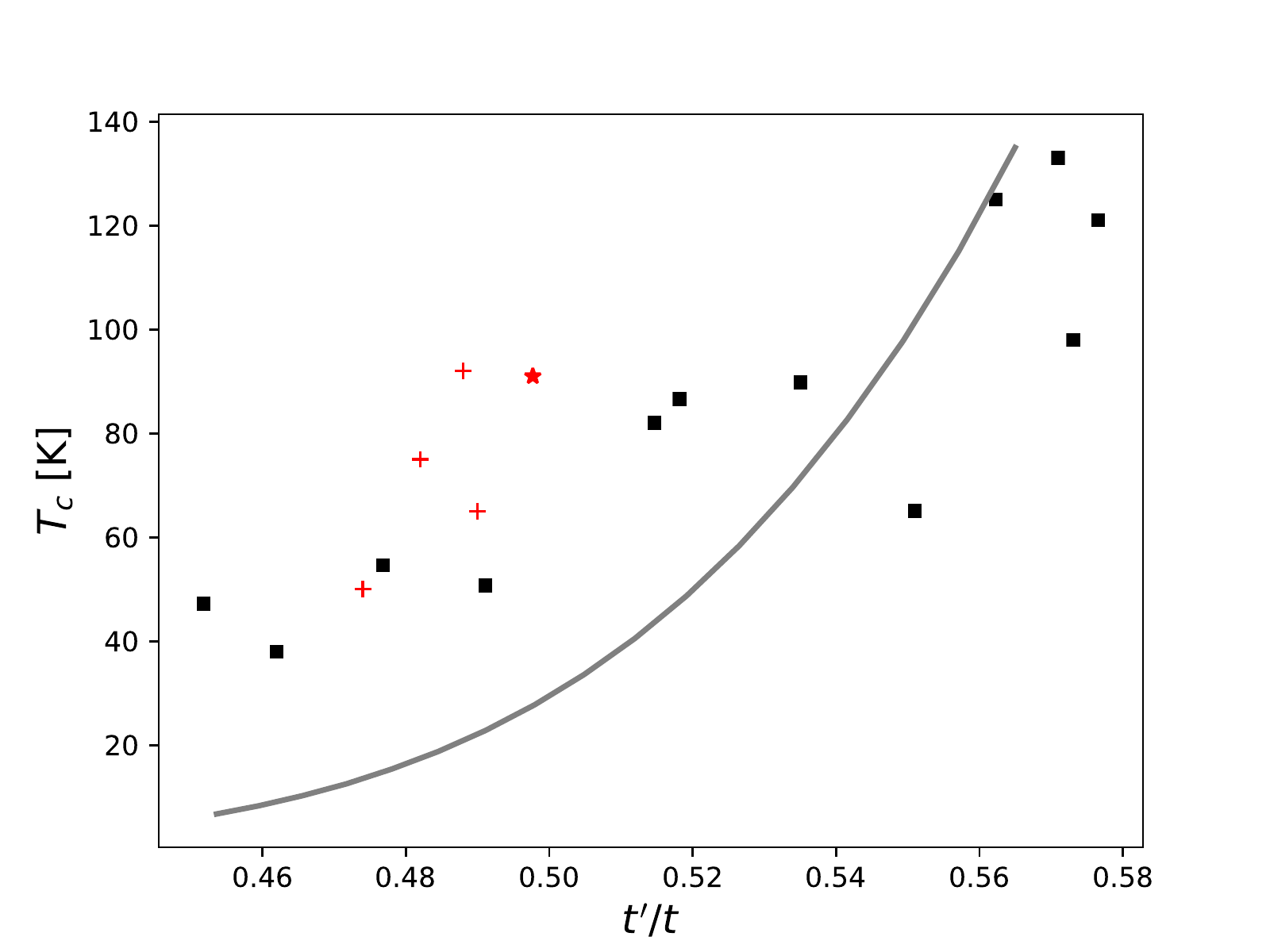}
\caption{Pavarini \textit{et al.}~\cite{Pavarini:01} ($_\blacksquare$)
correlation between the critical temperature $T_c$
and $t^\prime/t$ 
which is almost linear function of 
their range parameter $r(\eF)$ 
drawn in Fig.~\ref{Fig:tptr},
The $t^\prime/t$  parameter 
itself is exactly a linear function
of the BCS coupling parameter $\lambda$
supposing constant $J_{sd}$ in Fig.~\ref{Fig:tpt-lambda}.
According to our traditional BCS interpretation (solid line)
this band-structure trend describes $T_c$-$\lambda$ correlation
for the \textit{s-d} exchange amplitude $J_{sd}$
approximately equal for all cuprates. 
With ({\color{red}$\mathbf{\star}$}) we have included ARPES data by Zonno et al.~\cite[Fig.~5c]{Zonno:21} and
Ref.~\cite{Kaminski:05} for which $p_c=0.589\,$rad, and 
$p_d=1.155\,$rad
and with ({\color{red}\textbf{+}}) ARPES data from Vishik et al.~\cite{Vishik:10} for underdoped Bi-2212.
}
\label{Fig:rTc}
\end{figure}

On the other hand, the  dimensionless BCS coupling constant 
defined in the present article by \Eqref{lambdaBCS}
is exactly linear function of the $t^\prime/t$ \Eqref{t'/t} ratio
depicted in \Fref{Fig:tpt-lambda}.
In such a way the Pavarini \textit{et al.}~\cite{Pavarini:01} $T_c$-$r$
correlation we redraw in \Fref{Fig:rTc},
reveals correlation between the critical temperature $T_c$
and the BCS coupling constant $\lambda$ according the well-known
BCS formula \Eqref{TcBCS}.
In short, Pavarini \textit{et al.}~\cite{Pavarini:01}
empirically discovered the BCS correlation between the coupling constant and
the critical temperature.
We express our respect of this indirect confirmation of the BCS theory obtained by
observation of correlations between the shape of the Fermi contours 
and critical temperatures of hole doped cuprates.
This is a result of a huge volume of electron band calculations.
In the next section we try qualitatively to interpret this result.

\section{Why the exchange amplitude $J_{sd}$ can have antiferromagnetic sign}
In our work the exchange amplitude $J_{sd}$ is just a parameter of the theory
which can be determined to fit to one experiment and then to be used to predict
the results of many others.
Actually the \textit{s-d} interaction was introduced by Schubin and Wonsovsky and later by Zener long time before some \textit{ab initio} calculations to give even 
a small chance for reliable calculation.
However, the Kondo effect gave the proof that in many cases $J_{sd}$
can have antiferromagnetic sign.
Let suppose for a while that we have Coulomb repulsion 
only in one Cu ion.
In this case we have a single impurity Anderson model~\cite{Anders:12}.
Let us try to adapt a well-known from the textbooks formula 
by White and Geballe~\cite[Eq.~(7.17)]{WhiteGeballe:79} 
for CuO$_2$ plane.
Starting from the notations from these books
we try to trace an analogy making the replacement
\begin{align}
J&=2\left\vert  V_{0\mathbf{k}_{_\mathrm F}}\right\vert^2\frac{U}{E_0(E_0+U)}\\&
\rightarrow - J_{sd}\simeq 2
\left\vert  t_{pd} \right\vert^2\frac{U_{dd}}
{(\epsilon_d -\eF)[(\epsilon_d -\eF)+U_{dd}]}.\nn
\end{align}
Even for infinite Hubbard repulsion in Cu3$d_{x^2-y2}$
we have finite antiferromagnetic sign for the Kondo interaction.
Then we should admit that for an array or lattice of transition Cu ions the result could be qualitatively the same.
The theory of Kondo arrays and lattices is an intensive topic of contemporary 
researches~\cite{Anders:18,Anders:20,Anders:21a,Anders:21b}
and we are just waiting a CuO$_2$ oriented study to be performed.
There is perhaps no need of more apology.
As in mathematical physics many theorists consider negative $U$ Hubbard model,
we following Kondo do have antiferromagnetic $J_{sd}$.
If the phenomenology is successful, \textit{ab initio} consideration will
not come with substantial delay.

Up to now we can only conclude that LCAO projection of electron degree of freedom
works successfully and we put 
the calculation of $J_{sd}$
in the agenda of condensed matter physics.
First considerations of the $J_{sd}$ calculations from first principles will be only qualitative,
the development of the atomic physics is an excellent example for such a scenario.
The only one reliably calculated single electron hopping is between 
\be
t=2\, \frac{\hbar^2}{m_ea_\mathrm{_B}^2}\,R\,\exp(-R-1),\qquad
R=\frac{r}{a_\mathrm{_B}}
\ee
between 2 protons in the H$_2^+$ ion~\cite{LL3},
where $r$ is the distance between protons and $a_\mathrm{_B}$ is the Bohr radius,
and $m_e$ is the mass of the free electron.
This is the well-known result by Landau~\cite[Sec.61]{LL3} and Herring~\cite{Herring:62}
The second one is the two-electron exchange in the Hydrogen quasi-molecule
considered by Herring and Flicker~\cite{Herring:64}
\be
J=1.64\,\frac{\hbar^2}{m_ea_\mathrm{_B}^2}\,R^{5/2}\exp(-R).
\ee
Before these two old problems to become a part of the solid state physics,
it is premature to calculate quantitatively exchange parameters
but qualitative consideration is one indispensable first step.

\section{Short consideration of the unique properties $\mathrm{CuO}_2$ plane}

Close to the winter solstice or two moons later \textit{Homo sapiens}
exchange season greetings.
But in the spring there is another, even bigger occasion for season greetings related to
triple coincidence which we are going to consider qualitatively.

The conduction band in the cuprate plane CuO$_2$ can be considered as the energy
of atomic Cu$3d_{x^2-y^2}$ level 
smeared by the transition amplitudes between neighboring ions.
In this sense we can say that the single conduction band is a Cu$3d$ band.
However, the pairing exchange interaction is between 
Cu$3d$ and Cu$4s$ states in every Cu ion.
But what is necessary for the band electron function with momentum $\mathbf{p}$ to have 
significant Cu$4s$ component $S_\mathbf{p}$ according to Eq.~(\ref{Psi_LCAO})?
The qualitative analysis is transparent in the model case if all inter-ionic transfer amplitudes are much smaller than the differences between the atomic levels. 
In this case the Fermi energy of the almost half filled Cu$3d$ band 
is approximately equal to the atomic level 
$\eF\approx\epsilon_d$ and $D_\mathbf{p}\approx1$.
In the same approximation
\be
S_\mathbf{p}\approx-\frac{t_{sp}t_{dp}}{(\epsilon_s-\epsilon_d)(\epsilon_d-\epsilon_p)}
\left(s_x^2-s_y^2\right). \nn
\ee
Taking into account 
\be
s_x^2-s_y^2=-2\left(\cos p_x-\cos p_y\right), \nn
\ee
we obtain
\be
S_\mathbf{p}\approx\frac{2t_{sp}t_{dp}}{(\epsilon_s-\epsilon_d)(\epsilon_d-\epsilon_p)}
\left(\cos p_x-\cos p_y\right).
\label{perturbative}
\ee
As the hopping between planes is going through the big radius Cu$4s$ orbital,
the inter-layer hopping is proportional to $S_\mathbf{p}^2$, i.e.
\be
t_\perp(\mathbf{p})=t_{ss}S_\mathbf{p}^2=\frac{t_0}4 (\cos p_x-\cos p_y)^2,
\ee
with $t_0\approx150\,$meV for Bi$_2$Sr$_2$CaCu$_2$O$_8$~\cite{Ioffe:98}.
This behavior of inter-layer hopping and corresponding $S_\mathbf{p}$ amplitude
is in agreement with many band-structure 
calculations~\cite{Andersen:95,Andersen:96}
and further LCAO analysis~\cite{Mishonov:00}.

Now following the perturbative formula \Eqref{perturbative},
we can better understand the 
causes of the high-$T_c$ in cuprates.
The perturbative formula has transfer amplitudes in the numerator and energy denominators.
For example, O$2p$ amplitudes $X_\mathbf{p}$ and $Y_\mathbf{p}$ have multiplier 
$t_{pd}/(\epsilon_d-\epsilon_p)$
describing hopping between Cu$4d$ and O$2p$ with corresponding energy denominator.
Continuing from $X_\mathbf{p}$ to Cu$4s$ we obtain an additional factor 
$t_{sp}/(\epsilon_s-\epsilon_d)$.
Finally for the Cu$4s$ amplitude we have dimensionless energy factor
\be
\mathcal{Q}=\frac{t_{sp}t_{dp}}{(\epsilon_s-\epsilon_d)(\epsilon_d-\epsilon_\mathrm{p})}\ll 1, \nn
\ee
which together with the angular dependence participates in the BCS gap equation \Eqref{BCS_gap_equation}.
As 
\be\chi_\mathbf{p}\simeq\mathcal{Q} \, (\cos p_x-\cos p_y) \nn \ee
in order to have maximal $T_c$,
the dimensionless BCS-coupling constant (here we omit the $\langle\chi^2\rangle$
factor in the exact definition for $\lambda$)
\be G_0\equiv J_{sd}\rho_F \mathcal{Q}^2 \ll 1 \nn \ee
has to be as big as possible.
Typically $\rho_F J_{sd}\lesssim 1$ but simultaneously $\mathcal{Q}\lesssim 1$
and as a result the product of those three factors are small enough in order 
weak coupling BCS theory to be in its habitat of applicability.
On the other hand, the exchange integral $J_{sd}$ is much bigger than the 
Debye frequency and it is not necessary to take into account
Eliashberg type corrections for the ratio $2\Delta(0)/T_c$, for example.
In this sense the CuO$_2$ plane is closer to the original BCS weak coupling theory than
strong coupling conventional superconductors like Sn and Pb.

Perhaps for the CuO$_2$ plane we have the closest triple coincidence 
of the 3 levels of the transition metal
and the chalcogenide $\epsilon_p<\epsilon_d<\epsilon_s$.
Like after the spring equinox we wait for the full moon and next weekend
in order to have a Great holiday -- happy Easter to the CuO$_2$ plane:
from $3d$ to $4s$ by $2p$ the highway of high-$T_c$ superconductivity~\cite{MishIndPen:02}.

It is remarkable that the correlation between the band parameters
\begin{equation}
s(\eF)=\frac{(\epsilon_s-\eF)(\eF-\epsilon_\mathrm{p})}{2t_{sp}^2},
\quad r=\frac{1}{2(1+s)},
\end{equation}
and maximal critical temperature $T_{c,\mathrm{max}}$ at optimal doping
was observed by Pavarini \textit{et al.}~\cite{Pavarini:01}
analyzing the band structure of many hole doped cuprates.
This band structure trend is a strong hint that cuprate superconductivity 
is the modern face of the ancestral two-electron exchange~\cite{MishIndPen:03,BJP:11}.

The band theory has proven to be successful in deriving parameters for an effective Hamiltonian, 
and in capable hands can explain the trends in various members of the cuprate family. 
Nevertheless, this is only the starting point for achieving a deeper understanding of a strongly correlated problem, and the game is by no means over.

Before we begin the description of the normal state properties 
we had to recall and complete the superconducting ones.
It is challenging to try to use one and the same Hamiltonian to explain 
simultaneously normal and superconducting properties of the high-$T_c$ cuprates
which is the main purpose of the present work.
Next we analyze the \textit{s-d} exchange Hamiltonian in the spirit of Fermi
liquid theory.
\chapter{Fermi liquids}

\section{Fermi liquid reduction and inter-layers electric field fluctuations}
\label{Fermi liquid reduction}

\REM{
Ideas and notions of the Landau-Fermi liquid were widely used to analyze
normal properties of high-$T_c$ cuprates.
See, for example, papers by Carrington et al.,\cite{Carrington:92}
Hlubina and Rice,\cite{Hlubina:95}
Stojkovic and Pines,\cite{Stojkovic:97} 
and Ioffe and Millis.\cite{Ioffe:98}
We have to mention that large anisotropy in the scattering rate 
along the Fermi surface was reported for first time by Shen and Schrieffer\cite{Shen:97}
and Valla \textit{et al.}\cite{Valla:00}
The central detail of the Boltzmann equation analysis is the 
strong anisotropy of the charge carriers lifetime $\tau_\mathbf{p}$
along the Fermi contour. 
The central concepts are the ``hot spots'' where close to $(\pi,\,0)$
and $(0,\,\pi)$ regions of the Fermi contour the electron lifetime 
is unusually short and ARPES
spectral function is very broad\cite{Shen:95,Randeria:97} 
suggesting strong scattering.\cite{Ioffe:98}
For contemporary ARPES studies see also Ref.~\cite{Shen:21}
and references therein.
Ioffe and Millis~\cite{Ioffe:98} however accented on the concept
of ``cold spots'' along the BZ where electron lifetime is significantly longer
and ARPES data reveal well defined quasiparticle peak,
suggesting relatively weak scattering
which increase rapidly as one moves along the Fermi contour away 
from cold spots.
Recent research on hot and cold spots can be found in
Refs.~\cite{He:11,Lee:16,Fink:19} for instance.
}

Let us consider what is necessary to be supposed in order the
``cold spot'' concept to be derived sequentially from the 
\textit{s-d} Shubin-Kondo-Zener Hamiltonian 
\Eqref{H_sd_1}
which we write again in the momentum representation
\be
\hat H_{sd}=-\frac{J_{sd}}{N} \!\!
\sum_{\substack{
\mathbf{p}^\prime+\mathbf{q}^\prime=\mathbf{p}+\mathbf{q}\\
\alpha,\, \beta}} \!\!
S_{\mathbf{q}^\prime} D_{\mathbf{p}^\prime}
\hat c^\dagger_{\mathbf{q}^\prime\beta} \hat c^\dagger_{\mathbf{p}^\prime\alpha}
\hat c_{\mathbf{p}\alpha}\hat c_{\mathbf{q}\beta}S_\mathbf{p}D_\mathbf{q}.
\label{H_sd_2}
\ee
Now we perform Landau-Fermi liquid reduction taking from the sum above
only the terms with
$\mathbf{p}^\prime=\mathbf{p}$ and
$\mathbf{q}^\prime=\mathbf{q}$, and
introducing standard operators for the electron numbers
$\hat n_{\mathbf{p},\alpha}=\hat c^\dagger_{\mathbf{p\alpha}}\hat c_{\mathbf{p\alpha}}$
in the conduction Cu$3d_{x^2-y^2}$ band of cuprates.
For comparison with \Eqref{BCS_reduction}
now we have to insert different $\delta$-function multipliers, formally
\begin{align}
\label{FL_reduction}
\hat c^\dagger_{\mathbf{q}^\prime\beta} \hat c^\dagger_{\mathbf{p}^\prime\alpha}
\hat c_{\mathbf{p}\alpha}\hat c_{\mathbf{q}\beta}
&
\rightarrow
\delta_{\mathbf{q}^\prime,\mathbf{q}}\,
\delta_{\mathbf{p}^\prime,\mathbf{p}}\,  
\hat c^\dagger_{\mathbf{q}^\prime\beta} \hat c^\dagger_{\mathbf{p}^\prime\alpha}
\hat c_{\mathbf{p}\alpha}\hat c_{\mathbf{q}\beta}\\
&
=\delta_{\mathbf{q}^\prime,\mathbf{q}}\,
\delta_{\mathbf{p}^\prime,\mathbf{p}}\,  
\left(\hat{n}_{\mathbf{p},\alpha}
\hat{n}_{\mathbf{q},\beta} 
+\delta_{\mathbf{p},\mathbf{q}}\,\delta_{\alpha,\beta}\,\hat{n}_{\mathbf{p},\alpha}
\right).
\nn
\end{align}
The last term with $\delta_{\mathbf{p},\mathbf{q}}$ is irrelevant for the interaction 
and we omit it in the further considerations.
In such a way we obtain a separable Fermi liquid Hamiltonian
\be
\hat H_{_\mathrm{FL}}=\frac{1}{2\mathcal{N}}\sum_{\mathbf{p},\mathbf{q},\,\alpha,\beta}
\hat n_{\mathbf{p},\alpha}f(\mathbf{p},\mathbf{q})\hat n_{\mathbf{q},\beta}
\label{H_FL}
\ee
for which we are going to use the self-consistent approximation
\be
\langle\hat n_{\mathbf{p}\alpha}\hat n_{\mathbf{q}\beta}\rangle
\approx \langle\hat n_{\mathbf{p}\alpha}\rangle \langle\hat n_{\mathbf{q}\beta}\rangle
\ee
and when necessary apply thermal averaging and spin summation
$n_\mathbf{p}=\sum_\alpha\langle\hat n_{\mathbf{p},\alpha}\rangle$.

We wish to emphasize that in \Eqref{H_FL} we again arrives at the same 
separable kernel $f(\mathbf{p},\mathbf{q})=-2J_{sd}\chi_\mathbf{p}\chi_\mathbf{q}$
as for the BCS reduction and gap anisotropy \Eqref{separable_1} and \Eqref{separable_2}.

According to the Landau idea Eq.~(2.2) and Eq.~(39.20) of Ref.~\cite{LL9}
the influenced by the interaction electron band spectrum we express
by the functional derivative
\be
\varepsilon(\mathbf{p},\mathbf{r})
=\epsilon_\mathbf{p}
+\frac{\partial \hat H_{_\mathrm{FL}}}{\partial\hat n_{\mathbf{p},\alpha}}
\rightarrow\epsilon_\mathbf{p}+\frac1{N}\sum_{\mathbf{q},\beta}f(\mathbf{p},\mathbf{q}) 
\hat n_{\mathbf{q},\beta}(\mathbf{r}).
\label{WKB_Hamiltonian}
\ee

In the spirit of BCS averaged variational energy we can use Fermi liquid averaged
energy
\be
E(\{n_\mathrm{p}\})=\langle\hat H_{_\mathrm{FL}}\rangle \nn
\ee
and single particle spectrum
\be
\varepsilon(\mathbf{p},\mathbf{r})
=\epsilon_\mathbf{p}
+\frac{(-2J_{sd})}{N}\chi_\mathbf{p}\sum_\mathbf{q}\chi_\mathbf{q} 
n_{\mathbf{q}}(\mathbf{r},t)
\label{Fermi_liquid_spectrum}
\ee
in which space argument $\mathbf{r}$ can be introduced only in the quasi-classical WKB approximation.
Considering here the second $\mathbf{r}$-dependent term as perturbative scattering potential the perturbative scattering amplitude is $\propto \chi_\mathbf{p}$, 
and for the scattering rate we have
\be 1/\tau_\mathbf{p}\propto |\chi_\mathbf{p}|^2 \label{scattering_rate}.\ee
This way, after a long chain of approximations, we arrive at the conclusion
that strong anisotropy of the lifetime is a consequence of the \textit{s-d}
interaction and the specific $d$-type symmetry $S_\mathbf{p}$ of the empty $4s$ band.

In a qualitative consideration we can extend the WKB concept in order to analyze
even short wavelength thermal fluctuations of the electron density.
For the dispersion of this random variable, the $\chi$ factor is of order of one
and can be omitted in the qualitative considerations.
Summation on the momentum $\mathbf{p}$ gives simply the local 
fluctuations of the electron density $\delta n(\mathbf{r})$ around
space point $\mathbf{r}=a_0\mathbf{n}$ or CuO$_2$ plaquette $\mathbf{n}$.
The local thermal fluctuations of the electron density  $\delta n(\mathbf{n})$ are
related to the thermally excited random charge $Q$ in the plane
capacitor model \cite{Mishonov:00}
\be
\frac1{N}\sum_\mathbf{q}\chi_\mathbf{q} \, \delta n_{\mathbf{q}}(\mathbf{r},t)
\simeq \frac{Q}{e}\propto T.
\ee

Here we repeat the qualitative arguments related to the physics of the linear resistivity.
Layered cuprates are metals in the $ab$-plane CuO$_2$  but in the perpendicular 
$c$-direction in the normal phase there is no coherent electron transport.
Along this ``dielectric'' $c$-direction or $z$-direction, indispensably 
there are thermal fluctuations of the electric field $E_z$ electrostatically connected 
to the 2D charge density of single or doubled CuO$_2$ planes.
In such a way local thermal fluctuations of the electron density $Q$
substituted in the WKB formula Eq.~(\ref{Fermi_liquid_spectrum})
give a random potential
$U(\mathbf{r})=\varepsilon(\mathbf{p},\mathbf{r})$
on which charge carriers scatter.
The scattering rate $1/\tau_\mathbf{p}$ in the WKB approximation
in Born approximation is proportional to the 
matrix elements of the random potential
$1/\tau_\mathbf{p}\propto |U_\mathrm{p}|^2\propto \chi_\mathbf{p}^2$.
In such a way our qualitative model consideration leads
that the scattering rate is proportional 
to the square of the \textit{s-d} hybridization amplitude and temperature.
Calculating in the Born approximation
the scattering amplitude we have in
Eq.~(\ref{Fermi_liquid_spectrum}) $\chi_\mathbf{p}$,
giving for the scattering rate $\propto \chi_\mathbf{p}^2$
and explaining Ioffe and Millis~\cite{Ioffe:98} ``cold spots'' 
simply as zeros of the $\chi_\mathbf{p}$ factor 
in the separable interaction kernel general for BCS pairing and FL approach.
In \Eqref{Fermi_liquid_spectrum} $\chi_\mathbf{p}$ is momentum dependent
while the sum depends on the space vector $\mathbf{r}$
and thermally activated number of quasi-particles 
$\langle n_{\mathbf{q}}(\mathbf{r},t)\rangle \propto T$ 
are proportional to the temperature
and this is simple consequence of the classical fluctuations of the electric field 
perpendicular to the planes of the
layered conductor.

\begin{figure}[ht]
\centering
\includegraphics[scale=1.0]{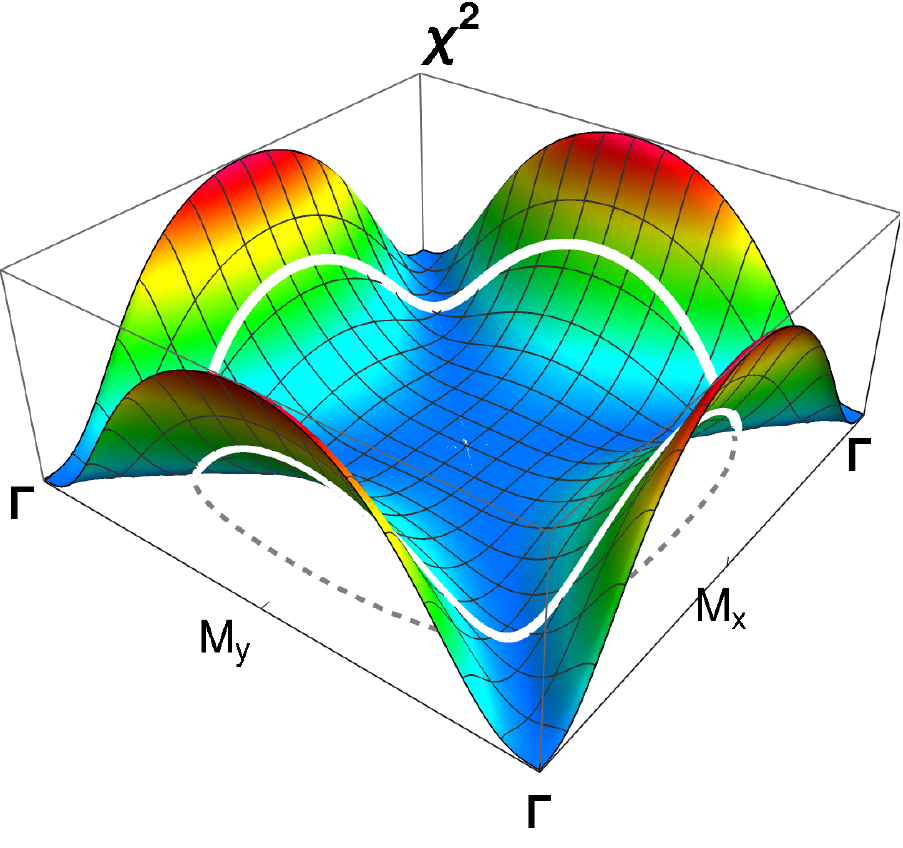}
\caption{The hybridization probability 
$\chi_\mathbf{p}^2=S_\mathbf{p}^2D_\mathbf{p}^2$
which participates in the BCS gap equation 
\Eqref{BCS_gap_equation} 
and scattering rate of the normal charge carriers 
by exchange interaction \Eqref{Gamma_p}.
The heights correspond hot spots
while navigation channels in the deep blue sea
correspond to cold spots in $(0,\,0)$-$(\pi,\,\pi)$ direction.
The Fermi contour (dashed line) projected on this $\chi_\mathbf{p}^2$
surface and this is curve is important detail of scattering rate theory;
after [ Fig.~2.3.c] \cite{MishonovPenev:11}.
}
\label{Fig:chi2}
\end{figure}

On the Fermi contour a hole pocket around $(\pi,\,\pi)$ point
has shape of a rounded square but conserving topology (in the spherical cow approximation)
can be approximated by a circle.
Making Fourier analysis in acceptable approximation $d$-wave gap anisotropy function
can be approximated by $d$-wave with $l=2$ giving
$\chi_\mathbf{p}\propto \cos(2\theta)$.
In this model approximation for the separable kernel, we obtain
exactly the angular dependence of
Ioffe and Millis~\cite{Ioffe:98}
\begin{align}
f(\mathbf{p},\mathbf{q})=I \cos(2\theta) \cos(2\theta^\prime),
\label{separable_kernel}
\end{align}
where
\begin{align}
I\simeq(-J_{sd})\left(\frac{t_{sp} t_{pd}}{(\epsilon_s-\epsilon_d)(\epsilon_d-\epsilon_\mathrm{p})}\right)^{\!2},
\end{align}
where the authors take into account angles from the BZ diagonal 
$\tilde\theta=\theta -\frac{\pi}4$ which converts 
$\cos(2\theta)=\sin(2\tilde\theta)$.
In such a way the electron scattering rate $\Gamma_\mathbf{p}=1/\tau_\mathbf{p}$
proportional to the imaginary part of the energy according second Fermi golden rule
take the ``cold spot'' angular form speculated by Ioffe and Millis in Ref.~\cite{Ioffe:98}
\begin{align}
-\mathrm{Im} (\epsilon_\mathbf{p})
\propto\Gamma_\mathbf{p}
&=\frac{\Gamma_0}4\sin^2(2\tilde\theta)+\frac1{\tau_0}
\approx \Gamma_0\tilde\theta^2+\frac1{\tau_0},
\label{Gamma_p}
\end{align}
where $\Gamma_0=k_1T+k_2T^2$.
The exact scattering rate $\Gamma_\mathbf{p}\propto\chi_\mathbf{p}^2$ is
represented in Fig.~\ref{Fig:chi2}.
The small constant $1/\tau_0=\Gamma_\mathrm{\! C}$
describes the scattering rates in cold spot direction which could have Coulomb 
scattering origin described in the beginning of the present work, 
i.e. $\tau_0\equiv \tau_\mathrm{cold}$.
The coefficient $k_1$ describes classical fluctuations of the electric field perpendicular to the 
CuO$_2$ plane when $k_2$ is negligible.
If however, for overdoped cuprates we have significant conductivity in the $c$-direction
and small fluctuations of the electric field,
we have the condition of applicability of the most conventional Landau-Fermi liquid theory
with $k_1=0$ and $k_2$ calculated according to 4-fermion $s$-$d$ Hamiltonian
using the general scheme described in Sec.~76 ``Absorption of sound in Fermi liquid''
of the textbook by Lifshitz and Pitaevskii (X-th volume of the Landau-Lifshitz course) 
\cite{LL10}.
In other words having strongly anisotropic scattering rate we have
to average the momentum dependent relaxation time $\tau_\mathbf{p}$
along the Fermi contour and with this intuitively clear notion
we can use the Drude formula again. Thus using the relation
\begin{align}
\frac1{\tau(\theta)}&=\frac{1}{\tau_\mathrm{hot}}\cos^2(2\theta)+\frac1{\tau_\mathrm{cold}}.
\end{align}
The Drude relaxation time is given by
\begin{align}
	\label{Drude}
\tau_\mathrm{Drude}=\langle \tau(\theta)\rangle
\approx
\sqrt{\tau_\mathrm{cold}\tau_\mathrm{hot}}\gg\tau_\mathrm{hot}\equiv\frac4{\Gamma_0},
\end{align}
and the scattering rate yields
\begin{align}
\sigma_{ab}=
q_e^2n_e\tau_\mathrm{Drude}/m_c. 
\end{align}
Inequality \eqref{Drude} reveals that pure Coulomb scattering considered
in Ref.~\cite{Mishonov:00} is only the first step in a correct direction.
We use the Fermi liquid approach and Fermi liquid notions
but for superconductors with anti-ferromagnetic exchange amplitude 
$J_{sd}$ the zero sound is only a thermally activated dissipative mode,
as velocity of a Brownian particle.
Real zero sound is however possible to be observed in layered perovskites 
with ferromagnetic $J_{sd}$.

In order to trace a path to the derivation of hot and cold spots along the Fermi contour we perform a qualitative analysis in the spirit of the Migdal~\cite[``Qualitative methods in quantum mechanics'']{Migdal}
or de~Gennes[`` Simple views on condensed matter physics'']~\cite{de_Gennes}.
The natural explanations gives a hint that we are on a correct path and it is worthwhile 
to apply the methods of statistical physics
giving the possibility to analyze every kinetic problem.

\begin{figure}[ht]
\centering
\includegraphics[scale=1.5]{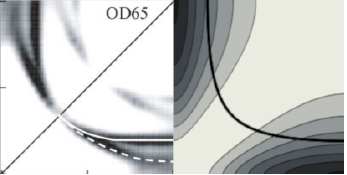}
\caption{Comparison of: 
(Left) ARPES data for spectral intensity for overdoped 
Bi$_2$Sr$_2$CaCu$_2$O$_{8+\delta}$ (Bi2212)
for $T_c=65\,$K OD65 from
Ref.~\cite{Feng:02}; the figure is taken from the arXiv version.
(Right) Scattering rate calculated in the framework of the 
\textit{s-d} exchange according to \Eqref{scattering_rate} depicted in \Fref{Fig:chi2}.
Continuous line in this theoretical calculation 
denotes the Fermi contour according to \Eqref{CEC_fit}.}
\label{Fig:Comp}
\end{figure}

Another hint for the correctness of our research is the qualitative agreement between our scattering rate calculation from \Fref{Fig:chi2}
and the published ARPES data\cite{Feng:02} shown in \Fref{Fig:Comp}.
We have reached this coherence in kinetics using the one and same Hamiltonian which describes the pairing and the $T_c$--Cu4$s$ energy correlation.
Broadening of the qualitative agreement of viable descriptions of variety of phenomena 
is an essential initial step towards the creation of the detailed theory.

But if we are on a correct path, we have to obtain more than we invest.
At least one new phenomenon has to be predicted 
if we have a general picture for superconducting pairing 
and anisotropic scattering rate in the normal phase.
The pendentive of the Landau-Fermi liquid theory is the prediction of zero sound
which is a property of a Fermi gas with repulsion.
The superconductivity is created by the attraction of the electrons and in this case
the zero sound is only a dissipation mode which can be only thermally activated.
In this section we have only touched to the normal state transport properties of the
high-$T_c$ superconductors, for an introduction in the problem see the excellent reviews \cite{Handbook,Hussey}.
If we are on a correct track the exchange interaction scattering can be taken in 
state-in-the-art way together with electromagnetic fluctuations in infinite media
which are well-described in the textbooks~\cite{AbrGorDzya,LL9}

In the next section we consider whether nevertheless it is possible to observe 
zero sound in layered transition metal perovskites. 

\section{Zero sound for ferromagnetic sign of $s$-$d$ exchange interaction}
\label{Zero sound for ferromagnetic sign of s-d exchange interaction}

The cuprates are high-$T_c$ superconductors because $J_{sd}$ has antiferromagnetic sign
and we have almost triple coincidence of the transition metal levels 3$d$ and 4$s$ and oxygen 2$p$.
But what will happen if 
in some perovskite the \textit{s-d} exchange integral has ferromagnetic sign with positive 
$(-J_{sd}>0)$?
This leads to a repulsion between electrons which prevents superconducting condensation
and opens the possibility for propagation of zero sound.
This topic is slightly lateral to our present study, 
but the developed system of notion and notations can be 
useful to study propagation of zero sound in similar compounds 
of layered perovskytes with transition metal ions.

Following the textbook by Lifshitz and Pitaevskii 
(IX volume of the Landau and Lifshitz course)~\cite{LL9}
we introduce notations and recall some basic notions of Landau-Fermi liquid theory.

The zero sound can be described as a collective degree of freedom related to local 
deformation of the Fermi surface 
considering in momentum space local change of the Fermi energy
$\eF\rightarrow\eF+\nu_\mathbf{p}$.
We repeat that quasi-momentum is represented by the dimensionless phases
$\mathbf{p}$
in the BZ,
and around the center of the hole pocket of CuO$_2$ plane
we can introduce polar coordinates
$\mathbf{p}=p(\cos\theta,\,\sin\theta)$.
In WKB wavelengths approximation we can consider
distribution of quasi-electrons per fixed spin projection
in  the phase space $n(\mathbf{p},\mathrm{r},t)$
by small linear deviation $\delta n(\mathbf{p},\mathrm{r},t)$ from 
equilibrium Fermi step $\theta(\eF-\epsilon_\mathbf{p})$
described by the Heavyside $\theta$-function.
Differentiating $\theta(\eF+\nu_\mathbf{p}-\epsilon_\mathbf{p})$
we obtain
\begin{align}&
n(\mathbf{p},\mathrm{r},t)=n_\mathbf{p}^{(0)}+\delta n(\mathbf{p},\mathrm{r},t),
\quad n_\mathbf{p}^{(0)}=\theta(\eF-\epsilon_\mathbf{p})\\&
\delta n(\mathbf{p},\mathrm{r},t)
=\delta(\eF-\epsilon_\mathbf{p})\nu_\mathbf{p}\exp(\im(\mathbf{K}\cdot\mathrm{r}-\omega t)),
\\&
n=\theta(\eF-\epsilon_\mathbf{p})
+\delta(\eF-\epsilon_\mathbf{p})\nu_\mathbf{p}\exp(\im(\mathbf{K}\cdot\mathrm{r}-\omega t)),
\end{align}
where plane wave amplitude $\nu_\mathbf{p}\exp(\im(\mathbf{K}\cdot\mathrm{r}-\omega t))$
with wave-vector $\mathbf{K}$ and frequency $\omega$ can be inserted in quasi-classical approximation
$a_0K\ll1$ and $\hbar\omega\ll\eF$.

The evolution of $n(\mathbf{p},\mathrm{r},t)$ quasi-particle distribution
we analyze in the initial collision approximation with zero substantial derivative 
in the phase space
\begin{align}
&0=\md_tn=\partial_tn+\partial_\mathbf{_P}n\cdot\dot{\mathbf{P}}+\partial_\mathbf{r}n\cdot\dot{\mathbf{r}},
\label{Boltzmann_kinetic-equation}
\end{align}
where we apply standard time and space derivatives
\begin{align}
&\partial_\mathbf{r}\delta n=\im\mathbf{K}\,\delta n,\quad
\mathbf{K}=K(\cos\beta,\,\sin\beta),\quad
\partial_{t}\delta n=-\im\omega\,\delta n,\nn\\
&\dot{\mathbf{r}}=\mathbf{V}_\mathbf{p},\quad
\mathbf{V}_\mathbf{p}=\partial_{_\mathbf{P}}\epsilon_\mathbf{p}
=\frac{a_0}{\hbar}\mathbf{v}_\mathbf{p},\quad
v_{\mathrm{_F},\mathbf{p}}=\left.v(\mathbf{p})\right\vert_{\varepsilon_\mathbf{p}=\eF},\nn
\end{align}
where $\epsilon_\mathbf{p}$ and $\mathbf{v}_\mathbf{p}$
have dimension energy, 
$\mathbf{V}_\mathbf{p}$ has dimension velocity,
$\mathbf{r}$ distance,
$\mathbf{P}$ momentum,
and $\mathbf{k}\equiv a_0\mathbf{K}$ is the dimensionless wave-vector.
The force acting on quasi-particles we calculate as space derivative 
of the Fermi liquid single particle Hamiltonian
Eq.~(\ref{WKB_Hamiltonian}) which gives
\be
\dot{\mathbf{P}}=\mathbf{F}=-\partial_\mathbf{r}\varepsilon(\mathbf{p},\mathbf{r})
=-\im\mathbf{K}
\int\limits_{\mathrm{BZ}}\!\! f(\mathbf{p},\mathbf{p}^\prime)\delta n_{\mathbf{p}^\prime}
\frac{\md p_x^\prime\md p_y^\prime}{(2\pi)^2}.
\ee
See also the well-known textbook by Nozieres~\cite{Nozieres}.
The plasma waves effects are negligible only for charge neutral oscillations
with zero amplitude oscillations of 2D charge density 
$\rho_\mathrm{el}(\mathbf{r},t)$
and current
\begin{align}&
\rho_\mathrm{el}(\mathbf{r},t)=\frac{e}{Na_0^2}\sum_\mathbf{p}\delta n(\mathbf{p},\mathbf{r},t),\\&
\mathbf{j}(\mathbf{r},t)=\int\limits_{\mathrm{BZ}}\!\!
e\mathbf{V}_\mathbf{p}\delta n(\mathbf{p},\mathrm{r},t)
\frac{\md p_x\md p_y}{(2\pi a_0)^2}.
\end{align}
In other words we can forget the electric force $e\mathbf{E}$
if we use only solutions of the kinetic equation with
$\langle\nu_\mathbf{p}\rangle_{\mathrm{_F}}=0$
and
$\langle \mathbf{k}\cdot\mathbf{v}_\mathbf{p}\,\nu_\mathbf{p}\rangle_\mathrm{_{F}}=0$.
The last condition in polar coordinates gives
$\langle\cos(\beta-\theta)\nu_\mathbf{p}\rangle_{\mathrm{_F}}=0$.

\begin{figure}[ht]
\centering
\includegraphics[scale=0.7]{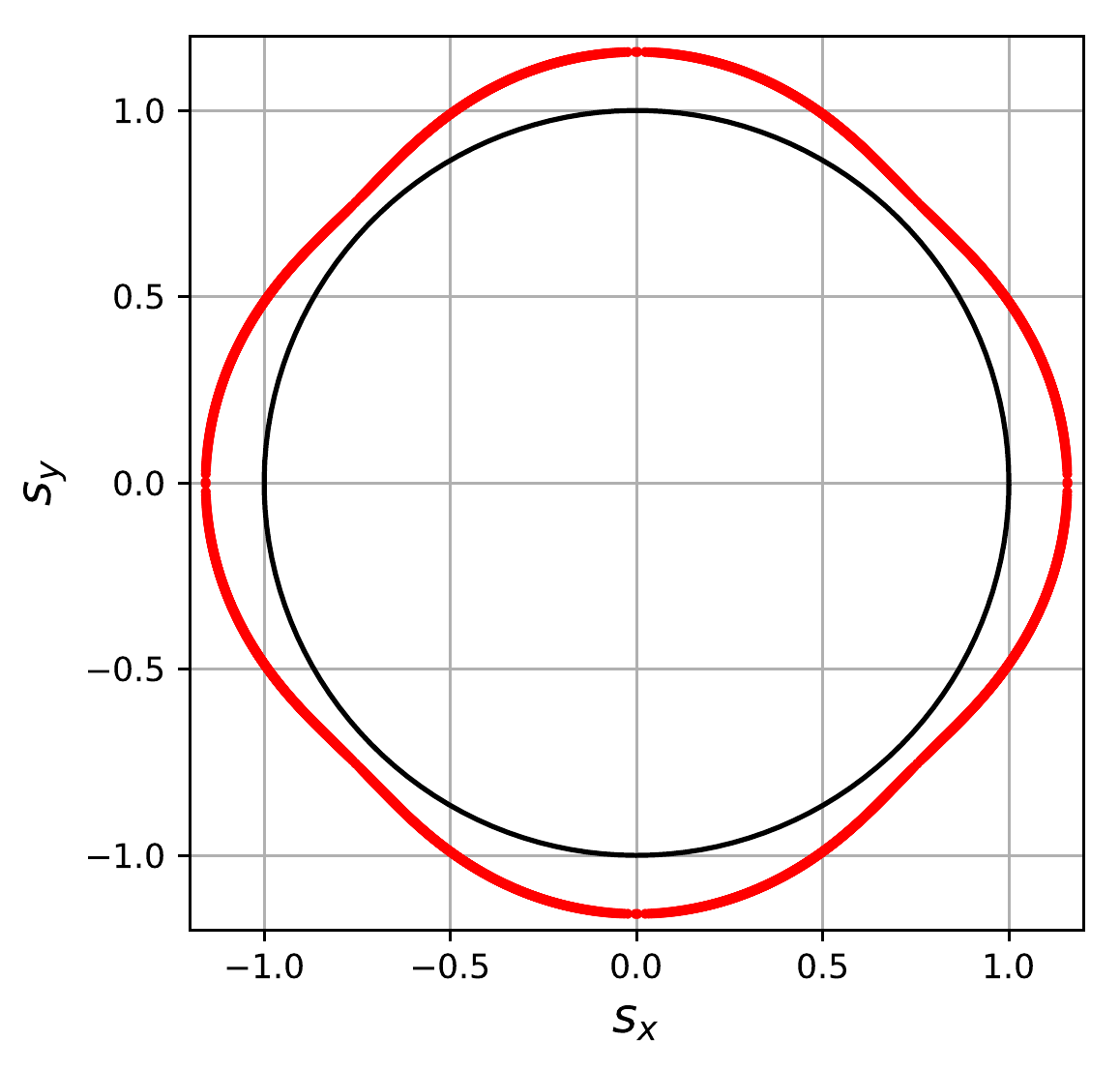}
\caption{
Two dimensional velocity space in units $v_\mathrm{_F}$. 
The unit circle is filled by electrons.
The zero sound phase velocity 
$\tilde{\mathbf{s}}=(\tilde s_x,\,\tilde s_y)=\tilde s(\cos\beta,\,\sin\beta)$ has 
several percent anisotropy 
with maxima along the pairing directions
and minima along the cold spots diagonals 
and zeros of the interaction function $\chi$.
No surfing electrons in all directions $\tilde{s}=\omega/kv_\mathrm{_F}>1$.
This is a model calculation with negligible plasmon effects which could be
acceptable approximation  when oscillations in neighboring planes are in anti-phase.
}
\label{Fig:Zero_Sound}
\end{figure}

\begin{figure}[ht]
\centering
\includegraphics[scale=0.6]{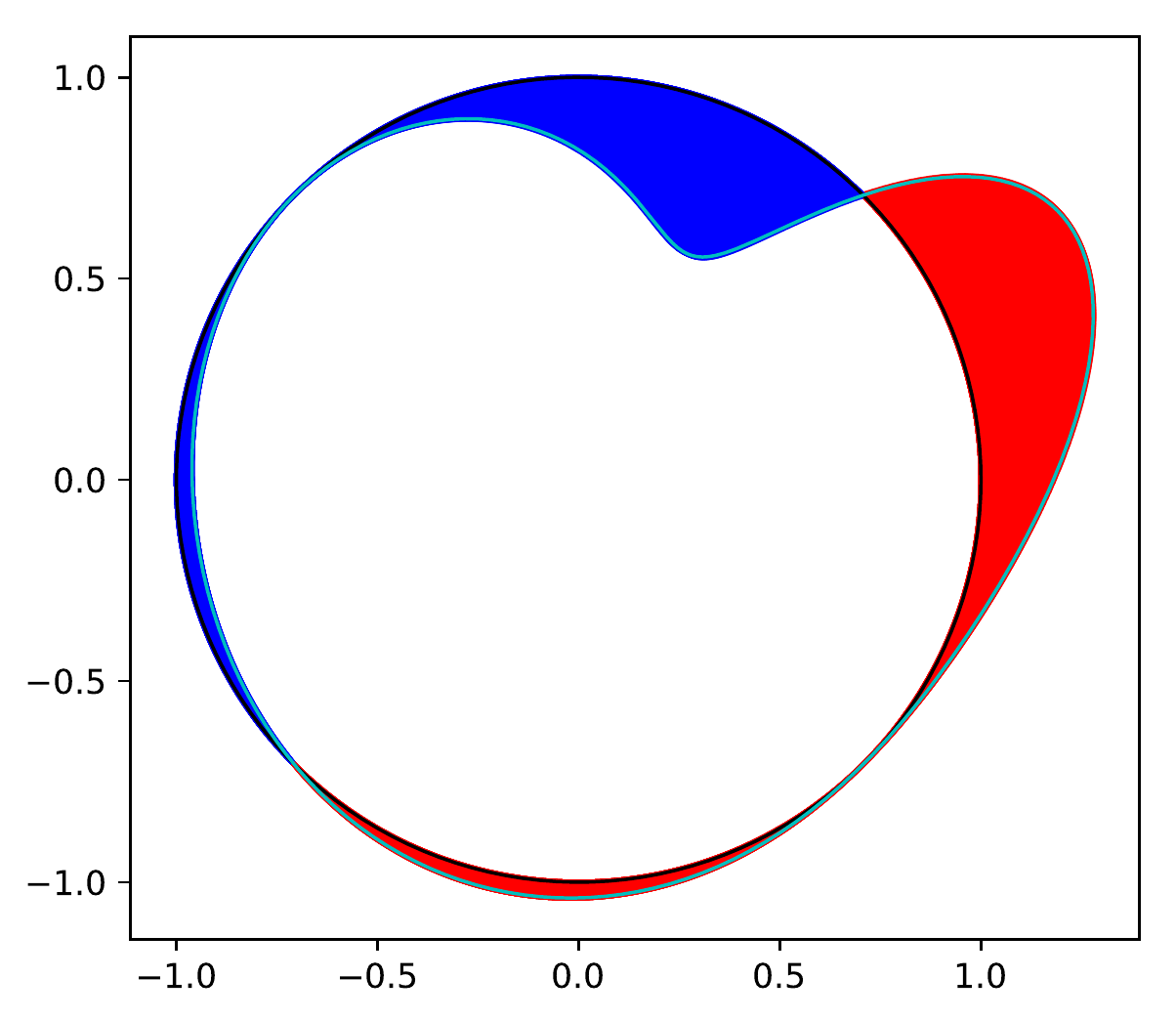}
\caption{Deformation of the Fermi contour in two dimensional 
momentum space $\mathbf{p}$ for zero sound 
propagating along cold spots diagonal $\beta=\pi/4$ 
in layered peroskites.
For this special case according to \Eqref{nu_sd_mod}
electric charge and current oscillations are zero.
For some appropriate wave-amplitude $a$ are given two contours
$\varepsilon_\mathbf{p}=\eF$ and 
$\varepsilon_\mathbf{p}=\eF+\nu_\mathbf{p}$.
For wave-vector in general direction it is necessary to take into account
Coulomb interaction and plasmon effects as it is done in works on kinetic theories for the 
electrodynamic response of Fermi liquids and anisotropic metals 
\cite{Valentinis:21,Valentinis:22, Baker:22}.
}
\label{Fig:sd_45}
\end{figure}

After substitution of the described details in the Boltzmann kinetic equation
Eq.~(\ref{Boltzmann_kinetic-equation}) we obtain the dispersion relation
\be
\left[\omega-\mathbf{K}\cdot\mathbf{V}_\mathrm{F}(\mathbf{p})\right]\nu_\mathbf{\mathbf{p}}
=\frac{\mathbf{K}\cdot\mathbf{V}_\mathrm{F}(\mathbf{p})}{(2\pi)^2}
\oint\limits_\mathrm{FC}f(\mathbf{p},\mathbf{p}^\prime)\,
\nu_{\mathbf{p}^\prime}\, 
\frac{\md p_l^\prime}{v_\mathrm{_F}(p_l^\prime)}
\ee
giving $\omega(\mathbf{K})$ dependence
\cite{AbrGorDzya,LL9}.
The separable kernel Eq.~(\ref{separable_kernel})
with positive $I$ and ferromagnetic
sign of the exchange integral $(-J_{sd})>0$
trivializes the calculation of the above integral.
For model evaluation here we ignore the relatively
weak Fermi velocity anisotropy and use parabolic dispersion
$\epsilon\approx E_0\,p^2/2m_\mathrm{eff}$;
\textit{i.e.} in spherical cow approximation.
Following the standard substitutions,
we easily obtain for the deformation of the Fermi circle
with amplitude $a$
\be
\nu(\theta;\beta)=a\frac{\cos(\theta-\beta)}{\tilde s-\cos(\theta-\beta)}\cos(2\theta)
\label{nu_sd_mod}
\ee
and the dispersion relation for the zero sound 
takes the form 
\be
F_0\left<\frac{\tilde\chi^2(\theta-\beta)}{\tilde s-\cos(\theta)}\right>_{\!\mathrm F}=1,
\quad \tilde s=\frac{\omega/K}{v_\mathrm{_F}},
\quad F_0=\rho_F I,
\ee
similar to the well-known results~\cite{Abrikosov,LL9}.
In linear approximation the zero-sound amplitude \textit{a} is arbitrary. 
The solution of the elementary integrals for the circular Fermi surface 
and $d$-type interaction Eq.~(\ref{separable_kernel}) is
\begin{align}
-\frac12+\frac{\tilde s}{2\varsigma}
\left\{1+4\tilde s\varsigma\left[1-2s\left(\tilde s-\varsigma\right)\right]\right\}
\cos(4\beta)=\frac1{F_0}
\end{align}
where $\varsigma\equiv\sqrt{\tilde s^2-1}.$
The solution for the dimensionless zero sound velocity $\tilde s$ 
as a function of the angle along the Fermi circle 
is depicted in Fig.~\ref{Fig:Zero_Sound}.
However, this illustration has only conditional sense
because of charge and current neutrality conditions
\begin{align}
&\int_0^{2\pi}\nu(\theta,\pi/4)\,\md\theta=0,\nn \\
&\int_0^{2\pi}\cos(\theta-\pi/4)\nu(\theta,\pi/4)\,\md\theta=0 \nn
\end{align}
give the restrictions
$\beta=\frac{\pi}4$ and $\cos\beta=-1$,
which means
that low frequency zero sound oscillations can propagate only
along the BZ diagonals of the layered 
transition metal oxides with basic elementary cell TO$_2$.
The deformation $\nu_\mathbf{p}$ of the FC for such charge neutral
oscillations is shown in Fig.~\ref{Fig:sd_45}.

One can speculate how strict the charge neutrality conditions 
are close to the ``cold spot'' diagonals.
Theoretically Coulomb interaction can be easily taken into account,
moreover we consider zero sound at the wave-vector $K_x=\pi/c_0$
when neighboring transition metal planes TO$_2$ have charge and current oscillations 
with opposite sign in $c$-direction so that zero sound oscillations 
are charge neutral only if averaged in small volumes.
However, these conditions are not universal and require 
consideration of the properties for every compound separately.

Here we wish to point out some contemporary studies on similar topics:
zero sound in two dimensions~\cite{Khoo:19}, 
shear zero sound~\cite{Marel:21} and zero sound for $p$-type interaction~\cite{Ding:19}.
We suppose  that except liquid $^3$He
layered structures with large ferromagnetic exchange interaction could become
interesting systems for implementation of old idea by
Landau~\cite{Landau_theory,Landau_zero_sound,Landau_3}.
Returning to superconducting cuprates with aniferromagnetic sign of $J_{sd}$ we have to point out that in the normal phase the zero sound modes 
are dissipative modes.
The thermal excitation of those modes creates only fluctuations of the electron density.
Finally the scattering of normal charge carriers by these density fluctuations 
creates the Ohmic resistivity.
In the next section we continue with general considerations of the non-resolved problems.

\section{Fermi liquid behavior of overdoped cuprates}
\label{Fermi liquid behavior of overdoped cuprates}

Our qualitative consideration of the electron scattering rate 
\Eqref{scattering_rate} can be precised in we consider 
state-of-the-art consideration of electromagnetic fluctuations in a layered
metal of CuO$_2$ plane following \cite[Chap.~VIII]{LL9} followed 
by consideration of Stoss-integral in plasmas \cite[Chap.~IV]{LL10}
with incorporated exchange interaction.
However linear behaviour of the resistivity demonstrated at least qualitatively
that even in gaseous approximation electrons are scattered by electron density 
fluctuations.
We suppose however that these fluctuations of density and perpendicular
to the CuO$_2$ layers electric field $E_z$ are gradually frozen with increasing of doping.

As it was emphasized by P.~Lee \cite{Lee:06} with increasing of the doping 
\textit{sanity gradually returns}. 
In this section we consider the case of strong inter-layer hybridization when the plasma frequency is higher than the temperature and the electric field $E_z$ fluctuations 
and related with them 2-dimensional density fluctuations are frozen. 
In this case we have usual Baber~\cite{Baber:37} and Landau-Pomeranchuk \cite{Landau:36} scattering qualitatively described in the monograph by
Mott~\cite[Chap~2, Sec~6, Eq~(18)]{Mott:90}.
For the calculation of the Stoss-integral of the scattering in cuprates 
we  apply  the second Fermi golden rule
for the transition rate~\cite[Chap.~7, Sec.~46, Eq.~(39)]{Dirac},
\cite[Eq.~(23.13)]{Fermi}, \cite[Eq.~(42.5)]{LL3} applied to the \textit{s-d}
exchange interaction \Eqref{s-d_exchange_interaction}
or rather \Eqref{H_sd_1}
where only conduction band $b=3$ is taken into account
\begin{align}
\mathrm{St}[n](\mathbf{p})
=\frac{2\pi}{\hbar}
\sum
\overline{
\left| \langle \mathrm{f}\vert \hat{H}_{sd}\vert\mathrm{i}\rangle \right|^2}
\delta(E_\mathrm{f}-E_\mathrm{i}-\hbar\omega).
\label{Fermi2}
\end{align}
In the initial state we have 2 charge carriers with momenta 
$\mathbf{p}^\prime$ and $\mathbf{p}_1^\prime$
which after scattering have momenta $\mathbf{p}$ and $\mathbf{p}_1$,
cf. Ref.~\cite[Eq.~(38)]{LL10}
\begin{align}
\vert\mathrm{i}\rangle= (\mathbf{p}^\prime,\alpha;\; \mathbf{p}_1^\prime,\beta),
\qquad
\vert\mathrm{f}\rangle= (\mathbf{p},\alpha;\; \mathbf{p}_1,\beta).
\end{align}
The Stoss-integral is a functional of the momentum distribution $[n]$
and a function of the fixed momentum $\mathbf{p}$,
i.e. we should not sum over the index $\mathbf{p}$.
Additionally as we precise later, $\mathbf{p}_1^\prime$ is fixed from the 
momentum conservation included in the \textit{s-d} interaction in momentum representation.

In our model calculation we omit umklapp processes 
which can be easily taken into account when necessary.
There is no summation over momentum $\mathbf{p}$ in \Eqref{Fermi2},
the summation is  over
$\mathbf{p}_1$, $\mathbf{p}^\prime$, and $\mathbf{p}_1^\prime$.
In this paper we analyze static Ohmic resistivity with zero frequency $\omega=0$.
The initial and final energies are just the band energies of the charge carriers scattered by exchange interaction
\be
E_\mathrm{i}=\varepsilon_{\mathbf{p}^\prime}+\varepsilon_{\mathbf{p}_1^\prime},
\qquad
E_\mathrm{f}=\varepsilon_{\mathbf{p}}+\varepsilon_{\mathbf{p}_1}.
\ee
The over-line in the Fermi golden rule \Eqref{Fermi2} denotes the square average of the 
modulus of the matrix element.
For the  electron number operators we have
\be
\hat{n}_{\mathbf{p},\alpha}
= \hat{c}_{\mathbf{p},\alpha}^\dagger\hat{c}_{\mathbf{p},\alpha},
\qquad
1-\hat{n}_{\mathbf{p},\alpha}
= \hat{c}_{\mathbf{p},\alpha}\hat{c}_{\mathbf{p},\alpha}^\dagger.
\ee
This averaging we perform with matrix element of the conduction band
Hamiltonian \Eqref{H_sd_1}.
In the averaged squares 
denoted by over-line in \Eqref{Fermi2}
of the modulus we now recognize the averaged momentum distribution 
of electrons
\be
\overline{
\vert\hat{c}_{\mathbf{p},\alpha}\vert^2}=n_{\mathbf{p},\alpha}\in (0,\,1),
\qquad
\overline{
\vert\hat{c}_{\mathbf{p},\alpha}^\dagger\vert^2}=1-n_{\mathbf{p},\alpha}.
\ee
More precisely, in this averaging we take into account creation and annihilation operators 
with coinciding momenta
\be
\overline{\hat{c}_{\mathbf{q},\beta}^\dagger\hat{c}_{\mathbf{p},\alpha}}
=\delta_{\mathbf{q}\mathbf{p}}\delta_{\alpha\beta}
n_{\mathbf{p},\alpha}.
\ee
In  \Eqref{H_sd_1} we suppose that all momenta are different and in this case
commutation relation between Fermi operators are irrelevant.

For  electron in equilibrium we have the Fermi distribution
\be
n_{\mathbf{p},\alpha}^{(0)}= \frac1{\exp((\varepsilon_\mathbf{p}-\eF)/T)+1}
\ee
with zero temperature limit
\be
n_{\mathbf{p},\alpha}^{(0)}=\theta(\eF-\varepsilon_\mathbf{p}),
\qquad\mbox{at}\quad T=0.
\ee

The substitution of the exchange Hamiltonian in Fermi~II golden rule \Eqref{Fermi2}
gives
\begin{align}
\mathrm{St}(\mathbf{p})
=& \frac{2\pi}{\hbar}\frac{J_{sd}^2}{N^2}
\sum_{\mathbf{p}_1,\,\mathbf{p}^\prime,\,\mathbf{p}_1^\prime}
\delta(\varepsilon+\varepsilon_1-\varepsilon^\prime-\varepsilon_1^\prime)\,
\delta_{\mathbf{p}+\mathbf{p}_1,\mathbf{p}^\prime+\mathbf{p}_1^\prime}
\nn\\&
\times [ (1-n)(1-n_1)n^\prime n_1^\prime
           -(1-n^\prime) (1-n_1^\prime) n n_1]\nn\\&
 \times [(SD_1S^{\prime} D_1^{\prime})^2+(DS_1D^\prime S_1^\prime)^2],       
\label{St}
\end{align}
where for brevity momentum and spin indices are omitted
\begin{align}
n&=n_{\mathbf{p},\alpha},\quad n_1=n_{\mathbf{p}_1,\alpha},\quad
n^\prime=n_{\mathbf{p}^\prime,\alpha},\quad
n_1^\prime=n_{\mathbf{p}_1^\prime,\alpha},\nn\\
\varepsilon &=\varepsilon_{\mathbf{p}},\qquad \varepsilon_1=\varepsilon_{\mathbf{p}_1},\qquad
\varepsilon^\prime=\varepsilon_{\mathbf{p}^\prime},\qquad \varepsilon_1^\prime=\varepsilon_{\mathbf{p}_1^\prime},\nn\\
S &=S_{\mathbf{p}},\quad \;\, S_1=S_{\mathbf{p}_1},\quad \;\,
S^\prime=S_{\mathbf{p}^\prime},\quad \;\,
S_1^\prime=S_{\mathbf{p}_1^\prime},\nn\\
D &=D_{\mathbf{p}},\quad D_1=D_{\mathbf{p}_1},\quad
D^\prime=D_{\mathbf{p}^\prime},\quad 
D_1^\prime=D_{\mathbf{p}_1^\prime}.
\label{notations}
\end{align}
The Kronecker symbol coming from the Hamiltonian \Eqref{H_sd_1}
removes the summation on $\mathbf{p}_1^\prime$
giving the condition
$\mathbf{p}_1^\prime=\mathbf{p}+\mathbf{p}_1-\mathbf{p}^\prime$.
The energy conservation is embedded in \Eqref{Fermi2}
while the momentum conservation is embedded in  \Eqref{H_sd_1}.
The first (income) term in brackets $[\dots]$ arises from creation
$\hat{c}_{\mathbf{p},\alpha}^\dagger$ operator
while in the second (outgo) term with minus sign corresponds 
to the process in which initial $\vert\mathrm{i}\rangle$ and
final states $\vert\mathrm{f}\rangle$ are exchanged.
According to our system of notations $\mathbf{p}$ are dimensionless momenta 
with components in $(0,\,2\pi)$ interval for which we have
\be
\frac1{N}\sum_\mathbf{q}\dots
=\int_0^{2\pi}\int_0^{2\pi}\frac{\md q_x\, \md q_y}{(2\pi)^2}\dots\;.
\ee
In such a way the Stoss-integral is in agreement with standard 
Landau theory for Fermi liquids \cite[Eq.~(74.5)]{LL10} applied to our two dimensional CuO$_2$ plane with \textit{s-d} exchange interaction
\begin{align}
\mathrm{St}[n](\mathbf{p})
=&\int_{\mathbf{p}_1,\,\mathbf{p}^\prime}
w(\mathbf{p},\mathbf{p}_1;\mathbf{p}^\prime,\mathbf{p}_1^\prime)\,
\delta(\varepsilon+\varepsilon_1-\varepsilon^\prime-\varepsilon_1^\prime)\nn\\&
\times \left[ (1-n)\left(1-n_1\right)n^\prime n_1^\prime
           -(1-n^\prime) \left(1-n_1^\prime\right) n n_1\right]\nn\\&
 \times  \frac{\md^D p_1\, 
                     \md^D p^\prime\,
                   }{[(2\pi)^D]^2} , 
                   \qquad \qquad\mbox{with}\quad
                   \mathbf{p}_1^\prime=\mathbf{p}+\mathbf{p}_1-\mathbf{p}^\prime,  
\label{St1}
\end{align}
where for the scattering rate with dimension of frequency we obtain
\begin{align}
w(\mathbf{p},\mathbf{p}_1;\mathbf{p}^\prime,\mathbf{p}_1^\prime)
=\frac{2\pi}{\hbar}J_{sd}^2
[(SD_1S^{\prime} D_1^{\prime})^2+(DS_1D^\prime S_1^\prime)^2],
\label{centrum} 
\end{align}
and notations in the integrant are introduced in \Eqref{notations}.
The collision frequencies $w$ have dimension energy times frequency.
It is instructive to compare the so derived Stoss-integral for classical gases~\cite[Eq.~(3.8)]{LL10}, electron-phonon interaction \cite[Eq.~(79.3)]{LL10},
plasmas \cite[Eq.~(46.7)]{LL10}
impurity scattering of phonons \cite [Eq.~(70.1)]{LL10},
electron impurity scattering  \cite [Eq.~(78.14)]{LL10},  \textit{etc}.
This form for the Stoss-integral exactly coincides
with the standard Landau one \cite[Eq.~(74.5)]{LL10} for the Fermi liquids
applied in our case for the two dimensional space $D=2$,
$p_x,\,p_y\in (0,\, 2\pi)$
for which~\cite[Eq.~(74.30)]{LL10} and in our case 
\be
\int\dots\delta(\varepsilon_\mathbf{p}-\eF)\, \md^2 p
=\oint_{\varepsilon_\mathbf{p}=\eF}
\dots \frac{\md p_l}{v_\mathrm{_F} (\mathbf{p})} .
\ee
In such a way we arrive at the standard Landau Stoss-integral for Fermi liquids
and for the function
\be
w(\mathbf{p},\mathbf{p}_1;\mathbf{p}^\prime,\mathbf{p}_1^\prime)
=w(\mathbf{p}^\prime,\mathbf{p}_1^\prime;\mathbf{p},\mathbf{p}_1)
=w(\mathbf{p}_1,\mathbf{p};\mathbf{p}^\prime,\mathbf{p}_1^\prime).
\ee
we derive the explicit expression in the framework of LCAO method applied to
the standard \textit{s-d} interaction.

\chapter{Solution of the kinetic equation and derivation of the conductivity}

According to the original Landau consideration the explicit form of the transition rate $w$
is irrelevant for the final conclusion that at low temperatures
in case of negligible impurity scattering the electron-electron interaction
leads to $\varrho_\Omega=A_\Omega T^2$ and the \textit{sanity} returns by the
freezing of $E_z$ and the related two dimensional density 
fluctuations \cite[Sec.~1, Eqs.~(1.11-12)]{LL9}.
Scattering rate determined by Drude fit of conductivity
can demonstrate significant temperature dependence
\cite[Bonn and Hardy, Fig.~4.24]{Handbook,Harris:06}.
For a nice review of normal state properties of high-$T_c$ cuprates
see \cite[Hussey, Chap.~10]{Handbook}.
For a qualitative description of mean free path
$l=V_\mathrm{F}\tau_{_\mathrm{Drude}}\propto T^{-2}$
see \cite[Sec.~1, Eqs.~(1.11-12)]{LL9}
and \cite[Eq.~(75.1)]{LL10}.

Now following the original we perform detailed calculation
in order to solve the kinetic equation 
\begin{align}
&
\md_t n(t,\mathbf{p}) = \mathrm{St}[n](\mathbf{p}),\qquad
\md_t=\frac{\partial}{\partial t} +
e\mathbf{E}\cdot\frac{\partial}{\partial\mathbf{P}}
\label{kinetic_equation_0}
\end{align}
in the case of small homogeneous in-plane electric field $\mathbf{E}=E\mathbf{e}_x$.

The momentum distribution $n$ we represent as small correction to the equilibrium distribution $n^{(0)}$
\begin{align}&
n(t,\mathbf{p})=n^{(0)}(\mathbf{p})+\delta n(t,\mathbf{p})= \overline{n}+\Delta,\nn\\&
n=n(t,\mathbf{p}),\qquad \delta n(t,\mathbf{p})\equiv\Delta(t,\mathbf{p}), 
\nn\\&
\overline{n}\equiv n^{(0)}(\mathbf{p})
=n^{(0)}(\varepsilon_\mathbf{p})
=\frac1{\exp\left((\varepsilon_\mathbf{p}-\eF)/T\right)+1}.
\label{Fermi_distribution}
\end{align}
For brevity we omit the arguments and keep only the superscripts and indices denoting the complete notation for $\Delta$, $\Delta_1$, $\Delta^\prime$, $\Delta_1^\prime$.

The term in brackets in \Eqref{St1} we rewrite as
\begin{align}
\mathcal{Z}\equiv&
(1-n)(1-n_1)(1-n^\prime)
(1-n_1^\prime) \label{Z}\\&
\times
\left[ 
\frac{n^\prime n_1^\prime}{(1-n^\prime)(1-n_1^\prime)}  
-\frac{n n_1}{(1-n)(1-n_1)}\right].\nn
\end{align}

Taking into account 
the conservation of energy
\be \varepsilon+\varepsilon_1=\varepsilon^\prime+\varepsilon_1^\prime\ee
and that in equilibrium 
\be
\frac{n^{(0)}_\mathbf{p}}{1-n^{(0)}_\mathbf{p}}
=\frac{\overline{n}}{1-\overline{n}}
=\exp\left(-\frac{\varepsilon_\mathbf{p}-\eF}{T}\right)
\label{n/(1-n)}
\ee
in \Eqref{St1}
we have
\begin{align}
\overline{\mathcal{Z}}=&(1-\overline n) (1-\overline n_1) 
(1-\overline n^\prime) (1-\overline n_1^\prime)\nn\\&
\times
\left[\exp((\varepsilon+\varepsilon_1)/T)
-\exp((\varepsilon^\prime+\varepsilon_1^\prime)/T)\right]=0,
\end{align}
i.e. in equilibrium the Stoss-integral is zero.
The current in equilibrium is of course also zero because 
$\varepsilon_\mathbf{p}=\varepsilon_{-\mathbf{p}}$
and
$\overline n_\mathbf{p}=\overline n_{-\mathbf{p}}$.

In order to calculate the two dimensional current density
\be
\mathbf{j}=\frac{e}{a_0^2N}\sum_\mathbf{p}\mathbf{V}_\mathbf{p}n_\mathbf{p}
=\sigma \mathbf{E}
\label{current}
\ee
at an evanescent electric field $\mathbf{E}=E\mathbf{e}_x\rightarrow 0$
we have to calculate the small deviation of the momentum distribution.
In 2D the current $\mathbf{j}$ has dimension A/m.

Following Enskog 1917 cf.~\cite[Eq.~(6.1) and Eq.~(74.15)]{LL10} we represent
\begin{align}
n\approx\overline n+\Delta=& n^{(0)}(\varepsilon(\mathbf{p}-\mathbf{q}_\mathbf{p}))
=n^{(0)}-\mathbf{q}_\mathbf{p} \cdot \frac{\partial n^{(0)}}{\partial \mathbf{p}}\nn\\
\approx& n^{(0)}(\varepsilon_\mathbf{p}-\varphi_\mathbf{p})
=n^{(0)}-\varphi_\mathbf{p}\frac{\partial n^{(0)}}{\partial \varepsilon}.
\end{align}

As $n^{(0)}=n^{(0)}(\varepsilon(\mathbf{p}))$
\be
\frac{\partial n^{(0)}}{\partial \mathbf{p}}
=\frac{\partial n^{(0)}}{\partial \varepsilon}
\,\mathbf{v}_\mathbf{p},
\qquad \mathbf{v}_\mathbf{p}=\frac{\partial \varepsilon}{\partial \mathbf p}
\ee
we have
\be
\varphi_\mathbf{p}=\mathbf{v}_\mathbf{p}\cdot \mathbf{q}_\mathbf{p},\quad
\Delta=-\frac{\partial \overline n}{\partial \varepsilon}\,\varphi_\mathbf{p}
=-\mathbf{q}_\mathbf{p}\cdot\frac{\partial \overline n}{\partial \mathbf{p}}
\ee
For the energy derivative we have
\be
-\frac{\md\overline n}{\md\varepsilon}
=\frac{(1-\overline n)\,\overline n}{T}
\rightarrow \delta(\varepsilon-\eF),\qquad\mbox{at}\;\;T\rightarrow 0,
\label{energy_derivative}
\ee
see \cite[Eq.~(74.16)]{LL10}.
Analogously we have
\be
-\frac{\partial n^{(0)}}{\partial \mathbf{p}}
=\frac{(1-\overline n)\,\overline n}{T}\,\mathbf{v}_\mathbf{p}
\rightarrow \delta(\varepsilon-\eF)\,\mathbf{v}_\mathbf{p}.
\ee

In order to understand the meaning of the energy $\varphi$ or momentum shift $\mathbf{Q}$ we analyse the kinetic equation in $\tau$ approximation.
This is actually independent mode approximation \cite[Eq.~(13.3)]{MermAsh}
\be
\mathrm{St}=-\frac{n(\mathbf{P},t)-\overline n(\mathbf{P})}{\tau},\quad
n(\mathbf{P},t)\approx\overline n(\mathbf{P}-\mathbf{Q_P}(t)).
\ee
Supposing small deviations from equilibrium
\be
\Delta_\mathbf{P}\approx- \mathbf{Q_P}\cdot\frac{\partial \overline n}{\partial \mathbf P},
\qquad
\mathrm{St}(\mathbf P)\approx  \frac{\mathbf{Q_P}}{\tau} \cdot \frac{\partial \overline n}{\partial \mathbf P}.
\ee
For the time derivative we have
\be
\frac{\partial n}{\partial t}
=-\frac{\partial \overline n}{\partial \mathbf P}\cdot\frac{\partial \mathbf{Q_P}}{\partial t}.
\ee
And for the term with electric field in \Eqref{kinetic_equation_0} we use the gradient from the non-perturbed
distribution
\be
\mathbf E(t)\cdot \frac{\partial \overline n}{\partial \mathbf P}.
\ee
We substitute all those terms in \Eqref{kinetic_equation_0} and omit
the common multiplier $\frac{\partial \overline n}{\partial \mathbf P}$
we arrive at the Newton equation for a fictitious particle moving in a viscous fluid
\be
\frac{\md \mathbf{Q}}{\md t}= e\mathbf{E}-\mathbf{Q}/\tau.
\ee
In the model case of parabolic dispersion $\varepsilon=P^2/2m$ with
$\mathbf{V_P}=\mathbf P/m$ and 
constant electric field $\mathbf{E}$ the whole Fermi surface, as
a rigid object, is shifted in velocity space at distance $\Delta \mathbf V= e\mathbf E\tau/m.$
And for the current we obtain the Drude formula
\be
\mathbf{j}=\sigma \mathbf E,\qquad \sigma=e^2 n_D\tau/m,
\ee
where $n_D$ is the electron volume density in $D$ dimensional space.

The motivation of the Enskog ansatz is that $\mathcal{Q}$ and $\varphi$
are smooth functions of energy and momentum and sharp dependence is concentrated 
in the energy and momentum derivatives.
\be
\Delta=\delta n\approx-\varphi\,\frac{\md\overline n}{\md\varepsilon}
=\frac{(1-\overline n)\,\overline n}{T}\,\varphi.
\ee
Using these variables we have
\be
\frac{n_\mathbf{p}}{1-n_\mathbf{p}}
=\exp(-(\varepsilon_\mathbf{p}-\varphi_\mathbf{p}-\eF)/T)
\ee
and for its variation 
\be
\delta \frac{n_\mathbf{p}}{1-n_\mathbf{p}}
=\frac{\varphi_\mathbf{p}}{T}\,\exp(-(\varepsilon_\mathbf{p}-\eF)/T)
=\frac{\overline n}{1-\overline n}\frac{\varphi_\mathbf{p}}{T}.
\ee
Taking into account that term in brackets $[\dots]$ in \Eqref{Z}
is in equilibrium zero, for calculation of variation $\delta \overline{\mathcal{Z}}$
we have to calculate only his variations.
In such a way using additionally \Eqref{n/(1-n)} we obtain
\begin{align}\nn
\delta \mathcal{Z}=&
(1-n)(1-n_1)(1-n^\prime)
(1-n_1^\prime) \label{Delta Z}\\&
\times\frac1{T}
\left[ \mathrm{e}^{-(\varepsilon^\prime-\eF)/T}
(\varphi^\prime+\varphi_1^\prime)
\mathrm{e}^{-(\varepsilon_1^\prime-\eF)/T}\right.\nn \\&\left.
\qquad\;
-\mathrm{e}^{-(\varepsilon-\eF)/T}
(\varphi+\varphi_1)
\mathrm{e}^{-(\varepsilon_1-\eF)/T}
\right].
\end{align}
For this variation on the energy surface 
$\varepsilon^\prime+\varepsilon_1^\prime=\varepsilon+\varepsilon_1$
we have
\begin{align}\nn
\delta \mathcal{Z}=&
\frac1{T}\mathrm{e}^{-(\varepsilon-\eF)/T}(1-n)
\mathrm{e}^{-(\varepsilon_1-\eF)/T}(1-n_1)
\label{delta Z}\\&
\times (1-n^\prime)
(1-n_1^\prime) 
\left[ 
\varphi^\prime+\varphi_1^\prime-\varphi-\varphi_1
\right].
\end{align}
Using again \Eqref{n/(1-n)}
we finally arrive in linearized approximation
\begin{align}
\delta \mathcal{Z}=&
\frac1{T} \, \overline n\, 
\overline n_1
(1-\overline n^\prime)
(1-\overline n_1^\prime) 
(\varphi^\prime+\varphi_1^\prime-\varphi-\varphi_1).
\label{final delta Z}
\end{align}
And linearized Stoss-integral \cite[Eq.~(74.24)]{LL10}
\begin{align}
\mathrm{St}(\mathbf{p})
=\frac1{T}\int &w\, \overline n\, \overline n_1 \,
(1-\overline n^\prime)\,(1-\overline n_1^\prime)\,
(\varphi^\prime+\varphi_1^\prime-\varphi-\varphi_1)\nn\\&
\times
\delta(\varepsilon^\prime+\varepsilon_1^\prime -\varepsilon-\varepsilon_1)\,
\frac{\md^D p_1\, \md^D p^\prime}{(2\pi)^{2D}}.
\label{74.24}
\end{align}
Where argument of the Stoss-integral $\mathbf{p}$ is fixed
and $\overline n=n^{(0)}(\varepsilon_\mathbf{p})$
can be written before the integral.
According momentum conservation in \Eqref{St1}
$\mathbf{p}_1^\prime=\mathbf{p}+\mathbf{p}_1-\mathbf{p}^\prime$
and 
$n_1^\prime=n^{(0)}(\varepsilon_{\mathbf{p}+\mathbf{p}_1-\mathbf{p}^\prime})$
Integration on two dimensional momentum space 
we will perform as integration on the 
Constant Energy Curves (CEC)
followed by the energy integration
\be
\int\limits_\mathbf{p}\dots \, \md^2 p
=
\int \md \varepsilon
\oint\limits_{\varepsilon_\mathbf{p}=\varepsilon}
\dots \frac{\md l}{v(\mathbf{p})},\qquad
\md l=\md p_l,
\label{energy_layer}
\ee
where $l \equiv p_l$ is the length in $\mathbf{p}$-space along the CEC.
We will apply this formula for the integration with respect of
$\mathbf{p}_1$ and $\mathbf{p}^\prime$.
Starting with 
$\md^2p_1=\md l_1 \md\varepsilon_1/v_1$
the integration 
of $\delta$-function in the integrant of \Eqref{74.24} gives
with respect of $\varepsilon_1$
gives
\begin{align}&
\varepsilon_1=\varepsilon^\prime+\varepsilon_1^\prime-\varepsilon,\\&
\overline n_1=n^{(0)}(\varepsilon^\prime+\varepsilon_1^\prime-\varepsilon)
=\frac1{\exp((\varepsilon^\prime+\varepsilon_1^\prime-\varepsilon-\eF)/T)+1}.\nn
\end{align}
and this energy argument is supposed in the next representation 
of the Stoss-integral
\begin{align}
\mathrm{St}(\mathbf{p})
=\frac{\overline n(\varepsilon(\mathbf{p}))}{T}\!\!&
\oint\limits_{l_1}
\int\limits_{\varepsilon^\prime}
\oint\limits_{l^\prime}
w\, \overline n_1 \,
(1-\overline n^\prime)\,(1-\overline n_1^\prime)\,
\\&\qquad
\times
\left[\varphi^\prime+\varphi_1^\prime-\varphi-\varphi_1\right]\,
\frac{\md l_1\, \md\varepsilon^\prime \,\md l^\prime}{(2\pi)^{4}\,v_1v^\prime}.
\nn
\label{next_step}
\end{align}
For numerical implementation we need to use the function
$\mathbf{p}^\prime=\mathbf{p}^\prime(\varepsilon^\prime, l^\prime)$
and 
$\mathbf{p}_1=\mathbf{p}_1(\varepsilon_1,l_1)$
which express the momentum $(p_x,p_y)$
as function of the energy of CEC $\varepsilon$ and the length $p_l$ along it.
Then in the integrant we calculate
\begin{align}&
\mathbf{p}_1^\prime=\mathbf{p}+\mathbf{p}_1-\mathbf{p}^\prime,\quad
v_1=v_{_\mathrm{F}}(\mathbf{p}_1),\quad
v^\prime=v_{_\mathrm{F}}(\mathbf{p}^\prime),\\&
\varepsilon_1^\prime=\varepsilon(\mathbf{p}_1^\prime), \quad
\varepsilon_1=\varepsilon(\mathbf{p}_1),\quad
\varepsilon^\prime=\varepsilon(\mathbf{p}^\prime),
\end{align}
equilibrium distributions from \Eqref{Fermi_distribution}
and $w$ from \Eqref{centrum}.

For the static homogeneous electric field the kinetic equation
\Eqref{kinetic_equation_0}
writes
\be 
-\mathbf{F}\cdot\mathbf{v_p}=T\,\mathrm{St}(\mathbf{p}),\qquad
\mathbf{F}\equiv \frac{e a_0\mathbf{E}}{\hbar}\, (1-\overline n)\,\overline{n}.
\ee
So introduced variable $\mathbf{F}$ has dimension of frequency
and $\mathbf{v_p}$ has dimension of energy;
$\mathbf{V}_\mathbf{\!p}=a_0\mathbf{v_p}/\hbar$ has dimension of velocity.
For the energy shift $\varphi(\mathbf{p})$
the kinetic equation is actually a system of in-homogeneous linear equations.
For this system we apply a relaxation scheme 
\begin{align}
\varphi = &
\ddfrac{
\mathbf{F}\cdot\mathbf{v_p} + 
\overline n \int w\, \overline n_1 
(1-\overline n^\prime)(1-\overline n_1^\prime)
\left[\varphi^\prime+\varphi_1^\prime-\varphi_1\right]\mathcal{D}
}
{\overline n  \int w\, \overline n_1 \,
(1-\overline n^\prime)\,(1-\overline n_1^\prime)\mathcal{D}
},\nn
\label{relax}\\&
\mathcal{D}\equiv 
\frac{\md l_1\, \md\varepsilon^\prime \,\md l^\prime}{(2\pi)^{4}\,v_1v^\prime},
\qquad\qquad
\int\equiv\oint\limits_{l_1}
\int\limits_{\varepsilon^\prime}
\oint\limits_{l^\prime}.
\end{align}
As $p_x,\,p_y\in(0,\, 2\pi)$ are actually dimensionless phases,
$\mathcal{D}$ has dimension of 1/energy;
the vector $\mathbf{P}=\hbar\mathbf{p}/a_0$ has dimension of momentum.

In the right side of \Eqref{relax} we apply the old iteration and explicitly we obtain the new
one $\varphi^{\mathrm{(new})}.$
It is also possible to apply successive over-relaxation 
\be
\varphi^{\mathrm{(sor})}=\varphi+
\left(\varphi^{\mathrm{(new})}-\varphi\right)\omega^{\mathrm{(sor)}},
\quad 1<\omega^{\mathrm{(sor)}}<2.
\ee
We can take say $N_\varepsilon=10$1 CEC in the energy interval
$(\eF-5T,\,\eF+5T)$ and $N_p=100$ points along every CEC.
Having this solution we can express the current \Eqref{current} as 
\be
\mathbf{j}=2 e\int \left(\frac{a_0}{\hbar}\mathbf{v_p}\right)
\left[(1-\overline n)\,\overline n\,\frac{\varphi}{T}\right]
\frac{\md^2 p}{(2\pi a_0)^2},
\ee
where in parentheses and brackets we recognize sequentially $\mathbf{V_p}$
and $\delta n$ and multiplier 2 comes from spin summation.
At low temperatures according \Eqref{energy_derivative}
and \Eqref{energy_derivative} we have
\be
\mathbf{j}=2 e\oint\limits_{\varepsilon_\mathbf{p}=\eF} \left(\frac{a_0}{\hbar}\frac{\mathbf{v_p}}{v_\mathbf{p}}\right)
\varphi\,
\frac{\md l}{(2\pi a_0)^2},\qquad
\varphi=\mathbf{v_p}\cdot\mathbf{Q}.
\ee
The conductivity can be determined as ratio $\sigma=j/E$ 
and due to symmetry of the quadratic lattice the orientation of $\mathbf{E}$
is irrelevant.
Finally linear regression $\varrho_{_\Omega}=1/\sigma$ versus $T^2$ gives the searched coefficient
$A_\Omega$ which is the fingerprint of the applicability of the
usual Landau Fermi liquid theory. 
The numerical implementation is described in the next sub-section.
Where for illustration we neglect consideration of the real energy of the quasi-particles,
cf. Ref.~\cite[Eqs.~(74.17) and (74.21)]{LL10}.
The application of so derived kinetic equation to 
derivation of kinetics coefficient will be subject of a different study.

\section{Re-scaling of energy}

\begin{figure}[ht]
\centering
\includegraphics[scale=0.45]{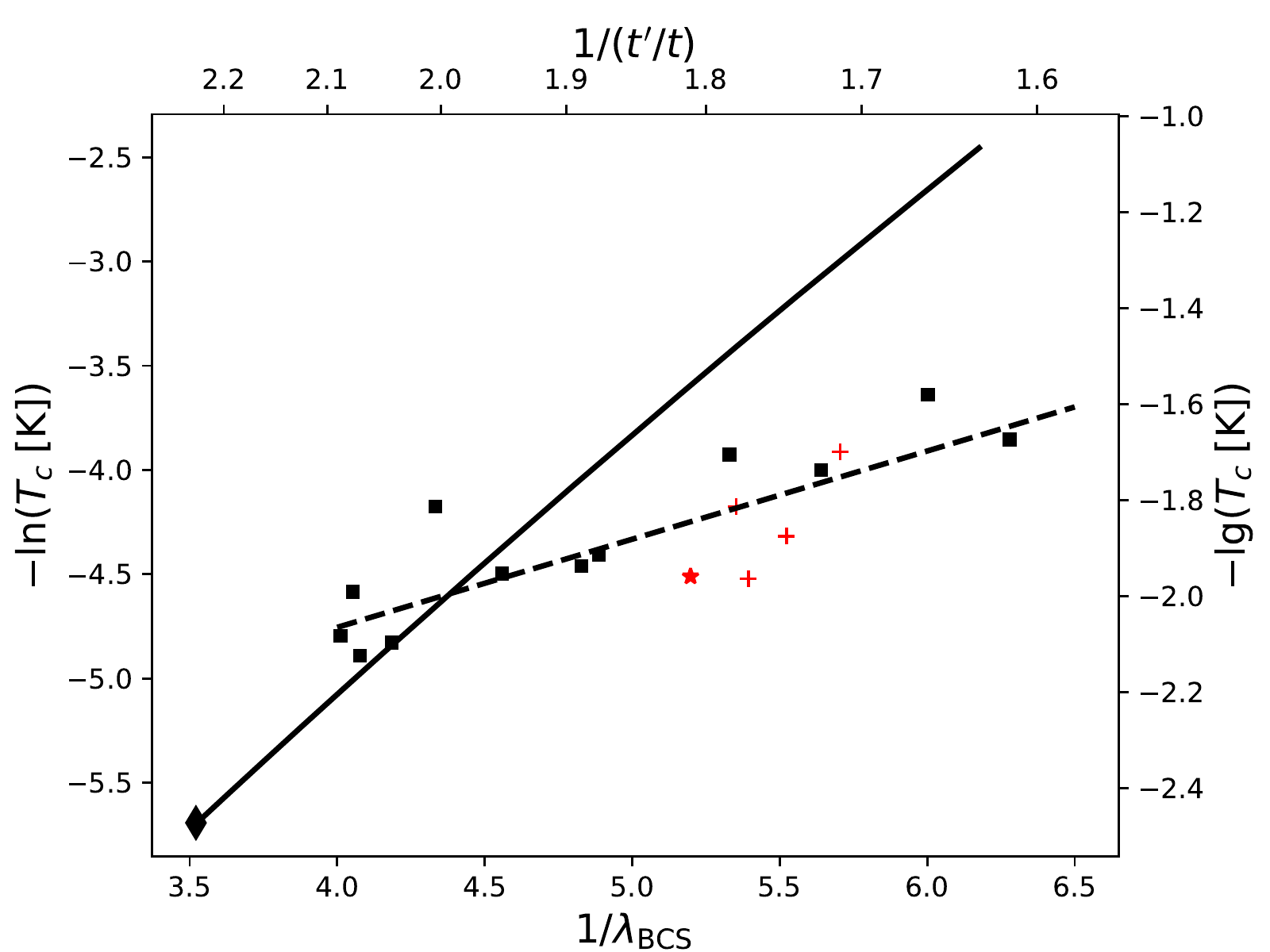}
\caption{
BCS correlation between the critical temperature $-\ln(T_c)$
and coupling constant $\lambda$.
According to \Eqref{TcBCS} this should be in initial approximation a straight line.
The high correlation coefficient $\rho>0.82$ is a hint that nature wishes to tell
us something.
The straight line is derived 
supposing that for different cuprates only 
position of  4$s$ level  it is $\epsilon_s$ is different 
and all other parameters of the Hamiltonian are constant, and $J_{sd}=2776$~meV.
Within this approximation the coupling constant $\lambda$ 
is  a linear function of $t^\prime/t$.
The slope of the linear regression of the experimental data (dashed line)
from 
Pavarini \textit{et al.}~\cite{Pavarini:01} ($_\blacksquare$),
Kaminski \textit{et. al.}~\cite{Kaminski:05} for underdoped Bi-2212 ({\color{red}\textbf{+}}) and
ARPES data \cite{Vishik:10} ({\color{red}$\mathbf{\star}$}) (same as in \Fref{Fig:tptr})
is lower than the $1/\lambda$ line (continuous) line.
We interpret this difference as a linear dependence of $1/J_{sd}$ 
as function of $\epsilon_s$.
}
\label{Fig:lmbTc}
\end{figure}

Rougly speaking, the science begins with simplicity and 
perhaps linear dependence is the simplest possibility.
In \Fref{Fig:lmbTc} is depicted a correlation between
the experimentally measured $\ln T_c$ and the theoretically 
calculated reciprocal coupling constant $1/\lambda$.
In spite of many approximations the LCAO method for the band theory
and common value of the exchange integral $J_{sd}$ for all cuprates
we arrive to the conclusion that linear regression can be used 
for prediction of the critical temperature $T_c$ of some new cuprate.
In our opinion it is a good step in our quantitative description of the theory
of high-$T_c$ superconductivity which starts with calculation of $T_c$.
In our LCAO analysis we follow the detailed works by
Andersen \textit{et al.}~\cite{Andersen:95,Andersen:96,Pavarini:01}.
However, it is well known that electron band calculations give 
significantly broader conductivity band than observed by photoemission data.
In order to solve this contradiction we make a compromise re-normalizing
all energy parameters given by the group of Andersen by a common
energy denominator: energy/=$Z_\varepsilon$. 
We choose $Z_\varepsilon=3.34$ in order for the Fermi velocity 
$V_\mathrm{_F}$ for optimally doped
Bi$_2$Sr$_2$Ca$_1$Cu$_2$O$_8$
to be close to the ARPES data  by 
Zhou~\textit{et al.}~\cite[Chap.~3, Fig.~3.21 and Fig.~1]{Handbook,Lanzara:01} $V_\mathrm{F}= 37$~km/s. 
See also \cite[Fig.~1.1, K.~A.~M\"uller]{Handbook,Shen:02}.
We use Fermi velocity $V_\mathrm{F}$ for $\eF -\varepsilon>\hbar\omega_\mathrm{D}$,
for energy difference lower than the Debye frequency $\hbar\omega_\mathrm{D}$
due to polaronic effects the slope of energy dispersion is smaller \cite[Eqs.~(65.12-13)]{LL9} 
\be 
V_\mathrm{F}^{(0)}=\left.\frac{\partial \varepsilon}{\partial P}
\right|_{(\eF -\varepsilon)<\hbar\omega_\mathrm{D}}.
\ee
This change of the slope is evident in the ARPES data we use not only for Bi2212 but for all studied materials~\cite[Chap.~3, Fig.~3.21 and Fig.~1]{Handbook,Lanzara:01}.

The corresponding re-normalized parameters are represented in Table~\ref{tbl:norm-in_energy}.
The re-normalized optical mass $m_\mathrm{opt}\approx 3$ is already an acceptable
3\% agreement with the effective mass of Cooper pairs (per particle) 
$m_\mathrm{eff}\approx 5.1\times 0.6=3.06$
if we take into account that 40\% of the mass can be attributed 
to the grain boundaries~\cite{Fiory:91}.

\begin{table}[ht]
\centering
\begin{tabular}{c c c c c  c c}	
		\hline \hline
		&  \\ [-1em]
		$\epsilon_s$  & $\epsilon_p$  & $\epsilon_d$  &
		$t_{sp}$  & $t_{pp}$~\cite{Mishonov:96} & $t_{pd}$   \\ 
			&  \\ [-1em] 
			1.2	& -0.27 & 0.0 & 0.6 & 0.06 & 0.45 & \\
\hline \hline
\end{tabular}
\caption{
Re-normalized single site energies $\epsilon$, hopping amplitudes $t$ in eV
and effective masses
from Table~\ref{tbl:in_energy} taken from 
Refs.~\cite{Andersen:95,Pavarini:01}
with the energy denominator $Z_\varepsilon=3.34$;
$m_\mathrm{opt}^\mathrm{(ren)}=Z_\varepsilon m_\mathrm{opt}=
3.34 \times 0.89 = 2.97\approx 3$ (from Table~\ref{tbl:out_energy}).
}
\label{tbl:norm-in_energy}
\end{table}
Now $J_{sd}$ has the more acceptable value of 2.776~eV, too.
The re-scaling of the energy is somehow in agreement with atomic and ionic spectra
because for all atoms and single charge ions the energies of
$4s^23d^n\leftrightarrow 4s^13d^{n+1}$ never exceed 3~eV.
Another hint for significant energy re-scaling gives the infrared absorption of
cuprates in energy $\sim$1~eV, for a reference if IR spectra see the monograph by
Plakida~\cite{Plakida:10}.
The simplest possible explanation is that we observe absorption between 
filled $3d$ conduction band and completely empty $4s$ band.
The corresponding theory is to simple to attract attention
for article publication.

The high correlation coefficient $\rho>0.82$ reveals that less 
20\% deviation from the straight line can be ascribed to the 
material science of accessorizes related to the specific 
properties of the crystal structure.

No doubts Pavarini~\textit{et. al.}~\cite{Pavarini:01} $T_c$-$r$ correlation is perhaps 
the most significant hint revealing the mechanism of high-$T_c$ superconductivity;
this article received more than 555 citations for the last 20 years.
Alas we do not find theoretical attempts to explain this remarkable correlation.
Moreover it is proclaimed as \textit{empirical correlation}
\cite[P.~A.~Lee, Chap.~14, Page 531]{Handbook}.
and \textit{Who is afraid of Virginia Woolf} the authors
of Ref.~\cite{Pavarini:01} are reluctant to consider the fundamental grounds.
According to our traditional interpretation we are witnesses of standard
BCS correlation $-\ln T_c$ versus $1/\lambda$ for which we have to apply
Pokrovsky theory \cite{Pokr:61,PokrRiv:63} for anisotropic superconductors.
It is a property of LCAO model applied to \textit{s-d} exchange interaction that
BCS coupling constant $\lambda$ is linear function of the ratio $t^\prime/t$
parameterizing the shape of the Fermi contour at optimal doping. 
Using $t^\prime/t=0.3$  in their numerical calculation \cite[K.~A.~M\"uller, Fig.~1.7]{Handbook,Kohen:03} have shown that $\Delta_\mathrm{max}\propto T_c$.
See also the study by Kugler \textit{et al.} \cite{Kugler:01} 
reproduced by Kirtley and Tafuri \cite[Fig.~2.31]{Handbook}
where $2\Delta_\mathrm{max}/T_c\approx 4.3$
which is in acceptable 5\% agreement with the BCS value of 4.116 from Table~\ref{tbl:out_energy}
according to Pokrovsky's theory \cite{Pokr:61},
calculated as
\begin{align}
\frac{2\Delta_\mathrm{max}}{T_c}=\frac{2\pi}{\gamma}
\dfrac{\mathrm{max} \left|\Delta_\mathbf{p}\right|}
{\exp\left\{\dfrac{\langle\Delta_\mathbf{p}^2\ln|\Delta_\mathbf{p}|\rangle}{\langle\Delta_\mathbf{p}^2\rangle}\right\}},\quad
\Delta_\mathbf{p}(T)=\tilde\Xi(T)\tilde\chi_\mathbf{p}
\end{align}
or
\begin{align}
\frac{2}{T_c}\left.\exp\left(\langle\Delta_\mathbf{p}^2\ln|\Delta_\mathbf{p}|\rangle/\langle\Delta_\mathbf{p}^2\rangle\right)\right\vert_{_{T=0}}=\frac{2\pi}{\gamma}
\approx 3.53 .
\label{Trivialized_by_Pokrovsky}
\end{align}
We have applied a 60 years old theory to a 30 years old experimental problem
and have obtained acceptable understanding of the thermodynamics of
high-$T_c$ cuprates.
The shape of the Fermi contour and the \textit{s-d} hybridization of the conduction band 
are  determined by the position of Cu4$s$ level with respect to Cu3$d$ level.
Here we have to give an indispensable credit to the intuition of Roehler~\cite{Roehler:00,Roehler:00a}
who has emphasized the importance of  Cu4$s$ long time ago.
We have to repeat the banal phrase that physics is an experimental science because
in the early attempts to build the foundation of the theory of high-$T_c$ superconductivity 
the  Cu4$s$ state has been ignored in the sophisticated lattice models 
\cite{Anderson:87,Emery:87,Varma:87,Lee:06} and initial set of dogmas~\cite{Anderson_dogmas}.
According to the best of our knowledge the Cu4$s$ state was never incorporated as an  indispensable ingredient
of the mechanism of high-$T_c$ superconductivity.

\begin{figure}[ht]
\centering
\includegraphics[scale=0.55]{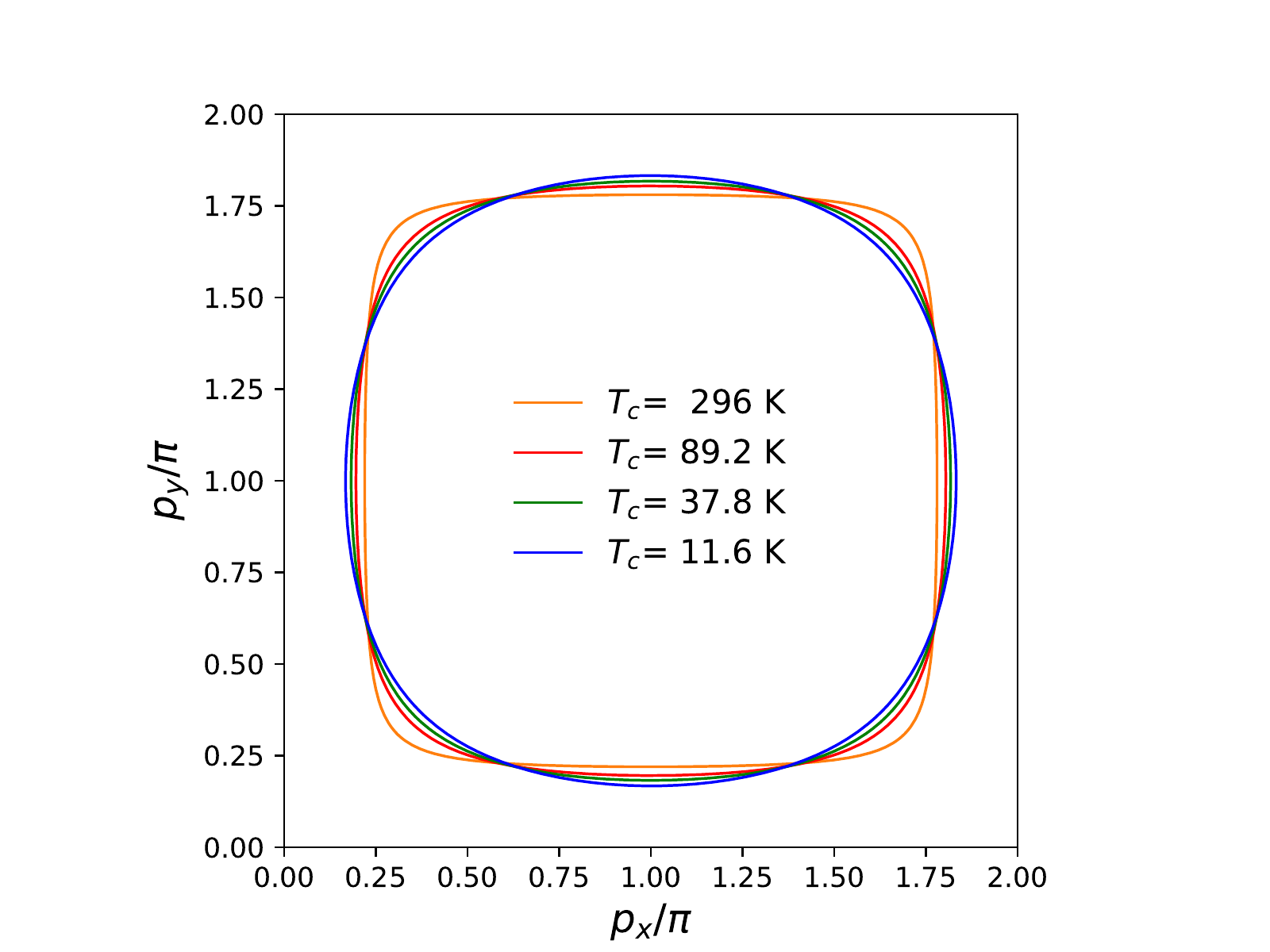}
\caption{Three rounded-square Fermi contours 
for optimally doped concentration $f=0.58$
with different critical temperatures $T_c$.
In the used LCAO approximation the shape of the contours
is determined by the equation
$-2t\,[\cos (p_x)+\cos (p_y)]+4t^\prime\cos (p_x)\cos (p_y)=\mathrm{const}$.
The contour with lowest $T_c$ is the most rounded one, while
the highest $T_c$ contour with higher $t^\prime/t$ ratio according to
\Fref{Fig:tpt-lambda} and \Fref{Fig:lmbTc}
corresponds to maximal curvature at the Brillouin zone diagonals.
The difference in the shape is subtle,
but subtle is the Lord.
In order to reach room $T_c$
\textcolor{blue}{(blue dream of physicists)}
by a meta-stable cuprate layer
according to our linear extrapolation in \Fref{Fig:lmbTc},
it is necessary the Cu4$s$ level $\epsilon_s$ to be only slightly 
above the Fermi level $\eF$ using an appropriate apex structure ($\epsilon_s - \eF > 2 T_c$).
\textit{Raffiniert is der Her Got aber boshaft is er nicht}
-- it is a doable task for MBE (Molecular Beam Epitaxy) technologists.
The basic parameters corresponding to these Fermi contours are given in Table~\ref{tbl:FC}.
}
\label{Fig:triptych}
\end{figure}

Last but not least, we wish to explain why so simple ingredients of the theory have to wait
quarter of century to be incorporated in a text-book compilation explaining main properties of
high-$T_c$ superconductivity. 
As a simple example we have to consider how stubborn was one of the authors of the present study (TMM):
Long time ago the late E.H.~Brand explained that in his MPI
there is a guy who only looking on the Fermi surface can immediately guess whether the
cuprate $T_c$ is high or low.
The stupid contra-argument that it is impossible was base on the conclusion
that $T_c$ is determined by particle-particle interaction and
the Fermi surface 
\textit{``Thou comest in such a questionable shape''}~\cite[Shakespeare, Hamlet, Act.~1, Scene~4, motto of  Chap.~9]{Ziman:64}
is irrelevant.
However for cuprates Cu4$s$ state determines simultaneously
\textit{s-d} hybridization with corresponding coupling constant and
the shape of the Fermi contour.
As a training for the eye we draw in \Fref{Fig:triptych} the Fermi contours for the cuprates with 3 different critical temperatures. 
\begin{table}[ht]
\centering
\begin{tabular}{r@{\hspace{0.5cm}} r@{\hspace{0.5cm}} r@{\hspace{0.5cm}} 
r@{\hspace{0.5cm}} r@{\hspace{0.5cm}} }	
		\hline \hline
		&  \\ [-1em]
		$\epsilon_s$~[eV]  & $\eF$~[eV]  & $t^\prime/t$  & $\lambda$  &  $T_c$~[K] \\ [0.1cm]
			\hline
			 0.569 & 0.493 & 0.623 & 0.284 & 296.75 \\ 
			 1.237 & 0.557 & 0.542 & 0.224 & 89.22 \\
			 1.619 & 0.582 & 0.500 & 0.194 & 37.81 \\
			 2.096 & 0.607 & 0.454 & 0.162 & 11.65 \\
\hline \hline
\end{tabular}
\caption{Parameters of the Fermi contours drawn in \Fref{Fig:triptych}.
The energy of Cu4$s$ level $\epsilon_s$ (the first column) and the Fermi level
$\eF$ (the second column) are taken into account from Cu3$d_{x^2-y2}$ atomic level.
These energies are re-scaled by a multiplier $Z_\varepsilon=3.34$ chosen to describe
the Fermi velocity according to ARPES data. 
The BCS coupling constant $\lambda$ is calculated supposing that the
exchange integral $J_{sd} = 2.776$~eV has almost one and the same value for all CuO$_2$
planes.
The dimensionless $t^\prime/t$ parameter is widely used in many studies.
The critical temperature, actually its logarithm is well described 
by the linear correlation in  \Fref{Fig:lmbTc} and
its high correlation coefficient encourages us to represent in the first row 
of the table a linear extrapolation of 15\% difference for the $t^\prime/t$,
from 0.542 to 0.623 for 50\% decrease of $\epsilon_s$.
This extrapolation is marked by $\blacklozenge$ in the lower left corner of  \Fref{Fig:lmbTc}.
Whether chemistry allows existence of such metastable structure
-- this is the question.
}
\label{tbl:FC}
\end{table}
Analyzing the subtle difference in the shape we can evaluate the professionalism
of the colleagues working in the electron-band calculation.

Considering \textit{why high temperature superconductivity is difficult}
Kivelson and Fradkin in Ref.~\cite[Chap. 15]{Handbook} cited
the well known article \textit{more is different}
(quantity is itself a quality :-) but an \textit{enemy of working class}
was a great trivialisator.
We repeat that linear temperature dependence of resistivity 
and strong anisotropy of the lifetime can be understand not  
in Fermi liquid but in gaseous approach.
Simultaneous experimentally confirmed relation between gap anisotropy
and critical temperature \Eqref{Trivialized_by_Pokrovsky}
are hints for conventional behavior of cuprates.
\textit{We fools of nature} (Act.~1, Scene~4) 
have only to 
to put in the agenda calculation of $J_{sd}$ and to 
explain the antiferromagnetic sign 
of the exchange interaction between itinerant electrons in the conduction band. 
The spin-singlet pairing is most likely to arise on the border of 
antiferromagnetism, cf.~\cite[Sec.~16.5]{Handbook}.

Let us repeat and clarify the basic idea. 
For cuprate superconductors the most typical Kondo interaction substitutes 
the electron-phonon interaction as pairing mechanism.
The applicability of the standard  BCS scheme is checked by the fit of parameters.
The comparison with the experiment convincingly demonstrates that 
ratio of critical temperature and exchange amplitude 
$T_c/J_{sd}$ see Table~\ref{tbl:out_energy}, 
is so small that Eliashberg type corrections taking into account retardation effects 
are negligible.
Real interaction amplitude is actually not $J_{sd}$ but 
the eigenvalue of the pairing kernel $V_0$ \Eqref{EigenValue},
for which $T_c/V_0$ is also small enough.
One can say the same for the Euler-Mascheroni energy $E_\mathrm{C}$
which substitutes the Debye frequency in the formula for the critical temperature
\Eqref{TcBCS}.
Here we have to emphasize one important difference between phonon superconductors 
and the exchange mediated cuprates.
For phonon superconductors the applicability of the Pokrovsky theory
is a consequence of the weak coupling limit.  
However for \textit{s-d} exchange superconductors with a single transition atom
in the elementary cell the pairing interaction is separable.
This property of the Hamiltonian can be checked by 
one A4 sheet of handwriting pen calculation.
In other words the weak coupling Pokrovsky approximation is an
exact result for the \textit{s-d} exchange applied to LCAO 
electron band functions.
In this sense cuprate superconductivity mediated by \textit{s-d}
exchange interaction is more conventional than the phonon superconductivity.
We can use BCS trial function even in the case when  BCS coupling constant 
$\lambda$ is scarcely small Table~\ref{tbl:out_energy},
simply because the pairing is exactly separable and we do not need to take into account 
any corrections.
That is why for cuprates calculated within \textit{s-d} LCAO approximation
$\Delta_\mathrm{max}/T_c$,
see \Eqref{Trivialized_by_Pokrovsky},
is within high precision of 5\% agreement with weak coupling Pokrovsky theory for the anisotropic superconductors;
the ARPES data are not directly involved in this agreement;
see also model calculations \cite{MishKlenov:05} for parabolic dispersion
where by chance we have numerical coincidence with the tunneling experiment.
Here we have to mention that for MgB$_2$ we can see the the same good work for the week coupling approximation \cite{MishPokr:05}.
In other words the separability of the pairing interaction extends the applicability 
of the weak coupling theory.
It is not necessary for the coupling itself to be  small.

Our final remark is that for the relative jump of heat capacity for cuprates \cite{MishKlenov:05} and MgB$_2$ \cite{MishPokr:05}
we can see the same good agreement with the Pokrovsky theory for anisotropic
BCS superconductors 
\be
\frac{\Delta C}{C_\mathrm{N}(T_c)}
=\frac{12}{7\zeta(3)}\frac{\langle\Delta_\mathbf{p}^2\rangle^2}{\langle\Delta_\mathbf{p}^4\rangle}.
\ee
This result can be easily re-derived by minimization of the 
free energy \cite{MishonovPenev:02};
for applications see also the comments \cite{Comment:02,Comment:03}.

\chapter{Discussion and conclusions}
\label{Discussion and conclusions}

\section{Psychoanalysis of the phenomenology}

Analyzing the zone-diagonal-dominated transport in high-$T_c$ cuprates 
Ioffe and Millis~\cite{Ioffe:98}
pointed out that angular dependence of the Fermi-liquid scattering rate 
is reminiscent of $d_{x^2-y^2}$ superconducting gap
and proposed that the life time is caused
by interaction of electrons with nearly singular 
$d_{x^2-y^2}$ pairing fluctuations.
Led by religious arguments, here we have to insert 
only a minor correction to their consideration:
both the pairing fluctuations
and the scattering rate in the normal phase
has to be derived from one and the same 
interaction Hamiltonian.

\section{``In the beginning was the Hamiltonian, and the $\hat H$ was by the God, 
and $\hat H$ was the God.'' Saint~John (citation by memory)}

When Allah wrote the Hamiltonian the Universe blew up;
something are so serious that about them we can speak only by joke;
\textit{Nota bene} by Niels Bohr.
Popularizing this idea, St.~John emphasized (citing by memory)
that in the beginning was the Hamiltonian.
We conclude that one and the same Shubin-Kondo-Zener \textit{s-d} exchange Hamiltonian
creates the pairing in the superconducting phase of CuO$_2$ high-$T_c$ superconductors
and the scattering rate of the charge carriers in the normal phase.
In such a way the best investigated high-$T_c$ materials 
have a common basic Hamiltonian single electron hopping
between Cu$3d_{x^2-y^2}$, O$2p_x$, O$2p_y$, and Cu$4s$,
and the electron exchange with antiferromagnetic sign 
between Cu$4s$ and Cu$3d_{x^2-y^2}$ orbitals.
For every cuprate to this generic Hamiltonian, 
accessories describing double planes, chains, apex oxygen etc. have to be added.
In the present work we demonstrate that the main phenomenological properties 
of the normal charge carriers scattering time
can be at least qualitatively derived from the $s$-$d$ pairing exchange Hamiltonian.
That is why the $s$-$d$ exchange Hamiltonian can be put into the agenda to be 
treated by standard methods of the statistical mechanics which can
explain the complete set of phenomena of the normal state of high-$T_c$ cuprates.
Definitely high-$T_c$ is not a mystery -- all details of its theory can be found in the textbooks 
written long time ago before Bednorz and Mueller to discover superconductivity in cuprates.
We strongly believe that the approach we use interaction projected on LCAO basis is applicable for other transition metal perovskites
and zero-sound propagating along the cold spot direction is 
a new phenomenon which we can predict if the $s$-$d$ interaction has ferromagnetic sign.
We suppose that charge neutral zero sound oscillations can be detected when they are converted
in Tera-Hertz hyper-sound in the opposite sing of the transition metal perovskite.
Excitation can be made by nonspecific rough impulse in the exciting side of the layered perovskite crystal.
The sample has to be cut in [110] plane.

Returning to the consideration of cuprates
the Pavarini \textit{et al.}~\cite{Pavarini:01} relation
reveals also that exchange amplitude $J_{sd}$ is a common constant for all cuprates 
and the difference in $T_{c,\,\mathrm{max}}$ is related to different band structure.
Band structure calculations have low social rank, 
the specialists in these numerical calculations are not considered as theorists
midst high level science fiction authors.
But honest work is nevertheless \textit{modus vivendi} at least at surviving level.
Band calculators have to be proud that 
their noble efforts revealed which 
parameter is most important for 
determination of $T_c$
which reveals the mechanism of high-$T_c$.

The band calculations can give a reliable set of LCAO parameters:
transfer integrals and single side energies which together
with $s$-$d$ exchange integral completely determine the lattice Hamiltonian.
Then calculation of kinetic properties is already a technical task
of the statistical physics without the freedom to change the 
Hamiltonian and the rule of the game.

In the present work we qualitatively trace only the initial path
which can be extended to the high-way of layered cuprate physics.
And the developed methods can be useful for many other materials
for which the exchange interaction is essential. 

\section{Small quantum of history}
Analyzing only plane dimpling in YBa$_2$Cu$_3$O$_{7-\delta}$ even in 2000
R\"ohler~\cite{Roehler:00,Roehler:00a} emphasized that 
the Cu$4s$-3$d_{x^2-y^2}$ hybridization seems to be the crucial quantum chemical parameter controlling related electronic degree of freedom.
We appreciate this early insight which becomes the precursor
of the detailed electron band studies and microscopic
investigation of the influence of \textit{s-d} exchange originally
suggested by Shubin~\cite{Shubin} on the statistical properties of the 
cuprates.

Few words we have to add also to the history of 2-electron correlations.
Soon after discovery of the electron J.~J.~Thompson~\cite{Thomson} suggested that
electric current is created by \textit{electron doublets}.
Later on in the beginning of quantum physics N.~Bohr~\cite{Bohr}
considered that two electrons in helium are moving with opposite momenta
$\mathbf{P}_1=-\mathbf{P}_2$,
this possibility for two \textit{s}-electrons was experimentally 
observed in double Rydberg states of noble gas atoms, 
see the review by Read~\cite{Read:82}.
In this strongly correlated states two electrons with zero angular momentum 
fall simultaneously to the nucleus like resurrecting kamikaze.

The history of self-consistent approximation starts from 19$^\mathrm{th}$ century 
and the first work on collective phenomena is the consideration by J.-C.~Maxwell~\cite{Maxwell:Sat}
that Saturn rings cannot be rigid discs but consist of self-consistent motion
of gravitating particles. 
This idea was developed in the atomic physics bay Hartree and Fock, and works by 
Bardeen, Cooper and Schrieffer~\cite{BCS} and Bogolyubov~\cite{Bogolyubov} 
develop for the physics of superconductivity the same idea of free particles
moving in a self-consistent field created by the interaction Hamiltonian.
We consider that Hubbard $U_d$, $U_s$ and $U_p$ have to be taken into a self-consistent 
way in the single site energies $\epsilon_d$, $\epsilon_s$ and $\epsilon_p$
while the Schubin~\cite{Shubin} \textit{s-d} exchange is considered as the pairing interaction in the standard BCS scheme. 
The \textit{s-d} exchange parameter $J_{sd}$ is actually the main amplitude
determining many phenomena with transition ion compounds;
for a review of strong correlations and exchange phenomena see the 
monograph by Anisimov and Izyumov~\cite{Izyumov:10}.

Having an unified scenario is indispensable, we open the Pandora box
of the necessity of making compromises between researches in different areas.
For example, an optical mass calculated according to \textit{ab initio} band calculation
exceeds almost $2\pi$ times
the same determined by electrostatic modulation of the kinetic inductance.
With such energy reduction the unexplained maximum of the mid infrared absorption
can be explained as a direct inter-band absorption caused by electron transitions between
conduction band and completely empty Cu4$s$ band. 
This is however only an example which type of disagreement can create 
a trial for unified description of the electron properties of the CuO$_2$ plane.

We finish with one unresolved problem.
What is the explanation of the anti-ferromagnetic sign of the Kondo
\textit{s-d} exchange in Cu transition ion $J_{sd}$?
The two electron exchange is a correlation,
and words ``strongly correlated'' is repeated as mantra already
33 years (the age of Jesus Christ)  in the physics of high-$T_c$ superconductivity.
The present work is not an exception.
This anti-ferromagnetic sign is against the Hund rule from the atomic physics and 
indispensably requires consideration of strong correlations in the simplest
cluster CuO$_2$ which plays an important fundamental role in the physics of 
cuprates.
Multiplet splitting of energy levels of a transition ion
surrounded by non-innocent ligands has been a
fundamental problem of the quantum chemistry for decades.
We hope that the development of the physics cuprates can stimulate the
satisfactory solution of this old problem.

\section{Results}

This investigation of the electronic properties of a generic CuO$_2$ plane
in the framework of the Shubin-Kondo-Zener \textit{s-d} exchange interaction
simultaneously describes two phenomena in cuprates.
Using the Pokrovsky theory for the
anisotropic gap BCS superconductors
and a microscopic model to calculate a separable interaction,
the following phenomena in the superconducting regime are quantitatively explained:
1) The dependency between the superconducting critical temperature $T_c$
and the Cu4\textit{s} energy level;
2) The scattering rate and the superconducting gap anisotropy, as
our theoretical approach reproduces the phenomenological analysis  Refs.~\cite{Hlubina:95,Ioffe:98}
performed to describe data from ARPES experiments on cuprates.
Simply stated the hot/cold spots phenomenology
is derived from a microscopic Hamiltonian
describing the superconducting spectrum of optimally doped and overdoped cuprates.
Moreover,  this theory is  also applicable to underdoped cuprates
if additional features like glassy nematicity is included
as accomplished by Lee, Kivelson and Kim~\cite{Lee:16}.
We can conclude that the electric charge fluctuations should be analyzed
in the framework of the standard theory of electromagnetic
fluctuations in continuous media~\cite[Chap.~8]{LL9} and
\cite[Chap.~6]{AbrGorDzya}.

Another result of this study is the prediction for propagation of zero sound in
non-super-conducting cuprates.
For anti-ferromagnetic sign of the of the $s$-$d$ interaction $J_{sd}>0$
we have tendency for superconductivity,
while for ferromagnetic sign, $J_{sd} < 0$, we expect zero sound to be observed.
Superconductivity via the $s$-$d$ exchange interaction can be reached
only for anti-ferromagnetic sign meaning that
in the normal phase of high-$T_c$ cuprates propagation of zero sound is impossible,
it is a dissipation mode.
The thermal excitation of charge density fluctuation modes creates 
Ohmic resistivity and intensive scattering,
and the strong angular dependence of this scattering rate causes the so called 
``hot spots'' phenomenologically postulated for the interpretation of the experimental data~\cite{Hlubina:95}.
In addition, it should be also noted that the
thermal fluctuations of plasmons could contribute to the
hot spots along the Fermi contour~\cite{Greco:19}.

Now it is difficult to determine 
the zero sound propagation for different layered compounds;
which material is more or less technologically appropriate
for production of a clean $ac$-surface.
This theoretical study is just the beginning,
now it is turn for the experiments.
As it seems the most probable solution would be provided by a thin layer geometry.

Let us repeat the main results obtained in our short study.
We have derived a well-known and working phenomenology of the hot/cold spots 
along the Fermi contour of layered high-$T_c$ cuprates.
We use the most typical Kondo-Zener exchange interaction
incorporated in the LCAO approach.
Then we perform two standard reductions of one and the same Hamiltonian.
The BCS one describes the well-known properties of the superconducting phase of the
overdoped cuprates,
while the Fermi liquid reduction explains the hot/cold spots phenomenology,
which is our main result belonging to the physics of normal metals.

The next step will be the derivation of these result within some alternative
approach
and the main problem we set in the agenda is the derivation of $J_{sd}$
exchange amplitude starting with LCAO approximation and Coulomb repulsion.

To summarize, our main result is a
qualitative physical explanation rather than
some technical details related to derivation of the necessary formulae.
The new result is that the kernel of the Landau zero sound theory coincides with the kernel
of the BCS coupling.
This is a property of the \textit{s-d} exchange interaction and is also a
hint of the importance of this interaction.
%

\section*{Acknowledgments}

Authors are thankful to Patrick~Lee for pointing out significant works on 
$T$-linearity of conductivity. 
The authors  are also  thankful to Davide~Valentinis for the interest to the present study
and pointing out for recently appeared related works on kinetic theories for the 
electrodynamic response of Fermi liquids and anisotropic metals 
\cite{Valentinis:21,Valentinis:22, Baker:22}.
Critical remarks by Valery~Pokrovsky are also highly appreciated.
Considerations with Mihail~Mishonov and Evgeni~Penev
in the early stages of this study are highly appreciated.
This work is supported by grant KP-06-N58/1
from 15.11.2021 of the Bulgarian National Science
Fund.

Eleven years ago
a preliminary fragment of the present work was represented at the conference
in memoriam of Matey~Mateev.
When proximity of the three levels in CuO$_2$ plane was juxtaposed to the sequence
of three holidays in the spring Alvaro~De~Rujula said -- you are right, congratulations
happy Easter to the CuO$_2$ plane.
One of the authors of the present work (TMM)
received big impetus to finalize the initial idea.

The authors are also thankful to Hassan~Chamati for the interest to the present study, 
creative atmosphere in ISSP, BAS,
critical reading of the manuscript, stimulating discussions and many suggested amendments and references.
The initial version of this work was presented in a conference devoted to the 90$^\mathrm{th}$
anniversary of VLP. 
Different our results from this  manuscript were reported in conferences on physics of superconductivity and magnetism in Bodrum and  Madrid.
We are thankful to colleagues for the comments and reviews.
\setsecnumdepth{none}
\maxsecnumdepth{none}
\maxsecnumdepth{subsubsection}
\setsecnumdepth{subsubsection}
\backmatter

\bibliographystyle{unsrturl}
\bibliography{Pokrovsky}

\end{document}